\pdfoutput=1
\documentclass[%
 reprint,
 superscriptaddress,
 amsmath,amssymb,
 aps
]{revtex4-2}

\usepackage{array}
\usepackage{braket}
\usepackage{amsmath}
\usepackage{mathtools}
\usepackage{here}
\mathtoolsset{showonlyrefs} 
\usepackage{latexsym}
\usepackage{amsfonts}
\usepackage{amsthm}
\usepackage{mathrsfs}
\usepackage{graphicx}
\usepackage{xcolor}
\usepackage{comment}
\usepackage{bm}
\usepackage{physics}
\usepackage{hyperref}
\hypersetup{
    colorlinks=true,
    linkcolor=blue,
    filecolor=magenta,   
    citecolor=magenta,
    urlcolor=cyan
    }
\usepackage{url}
\usepackage{amssymb}
\usepackage{qcircuit}
\usepackage{afterpage}
\usepackage{qcircuit}
\usepackage[ruled,linesnumbered]{algorithm2e}

\theoremstyle{plain}
\newtheorem{thm}{Theorem}

\newcommand{\beb}{\begin{itembox}}
\newcommand{\enb}{\end{itembox}}

\newcommand{\rhohat}{\hat{\rho}}
\newcommand{\zhat}{\hat{Z}}

\begin{document}

\title{STAR-Magic Mutation: Even More Efficient Analog Rotation Gates\\
for Early Fault-Tolerant Quantum Computer}

\author{Riki Toshio}
 \email{toshio.riki@fujitsu.com}

\affiliation{
Quantum Laboratory, Fujitsu Research, Fujitsu Limited,
4-1-1 Kawasaki, Kanagawa 211-8588, Japan
}
\affiliation{
Fujitsu Quantum Computing Joint Research Division,
Center for Quantum Information and Quantum Biology, The University of Osaka, 1-2 Machikaneyama, Toyonaka, Osaka, 565-8531, Japan
}

\author{Shota Kanasugi}
\affiliation{
Quantum Laboratory, Fujitsu Research, Fujitsu Limited,
4-1-1 Kawasaki, Kanagawa 211-8588, Japan
}
\affiliation{
Fujitsu Quantum Computing Joint Research Division,
Center for Quantum Information and Quantum Biology, The University of Osaka, 1-2 Machikaneyama, Toyonaka, Osaka, 565-8531, Japan
}

\author{Jun Fujisaki}
\affiliation{
Quantum Laboratory, Fujitsu Research, Fujitsu Limited,
4-1-1 Kawasaki, Kanagawa 211-8588, Japan
}
\affiliation{
Fujitsu Quantum Computing Joint Research Division,
Center for Quantum Information and Quantum Biology, The University of Osaka, 1-2 Machikaneyama, Toyonaka, Osaka, 565-8531, Japan
}

\author{Hirotaka Oshima}
\affiliation{
Quantum Laboratory, Fujitsu Research, Fujitsu Limited,
4-1-1 Kawasaki, Kanagawa 211-8588, Japan
}
\affiliation{
Fujitsu Quantum Computing Joint Research Division,
Center for Quantum Information and Quantum Biology, The University of Osaka, 1-2 Machikaneyama, Toyonaka, Osaka, 565-8531, Japan
}

\author{Shintaro Sato}
\affiliation{
Quantum Laboratory, Fujitsu Research, Fujitsu Limited,
4-1-1 Kawasaki, Kanagawa 211-8588, Japan
}
\affiliation{
Fujitsu Quantum Computing Joint Research Division,
Center for Quantum Information and Quantum Biology, The University of Osaka, 1-2 Machikaneyama, Toyonaka, Osaka, 565-8531, Japan
}

\author{Keisuke Fujii}
\affiliation{
Fujitsu Quantum Computing Joint Research Division,
Center for Quantum Information and Quantum Biology, The University of Osaka, 1-2 Machikaneyama, Toyonaka, Osaka, 565-8531, Japan
}

\affiliation{
Graduate School of Engineering Science, The University of Osaka,
1-3 Machikaneyama, Toyonaka, Osaka, 560-8531, Japan
}
\affiliation{
Center for Quantum Information and Quantum Biology, The University of Osaka, 560-0043, Japan
}
\affiliation{
RIKEN Center for Quantum Computing (RQC), Wako Saitama 351-0198, Japan
}

\date{\today}

\begin{abstract}
We introduce {\it STAR-magic mutation}, an efficient protocol for implementing a logical analog rotation gate on early fault-tolerant quantum computers. This protocol judiciously combines two of the latest state preparation protocols: transversal multi-rotation protocol and magic state cultivation.  It achieves a logical analog rotation gate with a favorable error scaling of $\mathcal{O}(\theta_L^{2(1-\Theta(1/d))}p_{\text{ph}})$, while requiring only the ancillary space of a single surface code patch.
Here, $\theta_L$ is the logical rotation angle, $p_{\text{ph}}$ is the physical error rate, and $d$ is the code distance. 
This scaling marks a significant improvement over the previous state-of-the-art, $\mathcal{O}(\theta_L p_{\text{ph}})$, making our protocol particularly powerful for implementing a sequence of small-angle rotation gates, like Trotter-based circuits.
Notably, for $\theta_L \lesssim 10^{-5}$, our protocol achieves a two-order-of-magnitude reduction in both the execution time and the error rate of analog rotation gates compared to the standard $T$-gate synthesis using cultivated magic states.
Building upon this protocol, we also propose a novel quantum computing architecture designed for early fault-tolerant quantum computers, dubbed ``STAR ver.~3". This architecture employs a refined circuit compilation strategy based on Clifford+$T$+$\phi$ gate set, rather than the conventional Clifford+$T$ or Clifford+$\phi$ gate sets. We establish a theoretical bound on the feasible circuit size on this architecture and illustrate its capabilities by analyzing the spacetime costs for simulating the dynamics of quantum many-body systems. Specifically, we demonstrate that our architecture can simulate biologically-relevant molecules or lattice models at scales beyond the reach of exact classical simulation, with only a few hundred thousand physical qubits, even assuming a realistic error rate of $p_{\text{ph}}=10^{-3}$.
\end{abstract}

\maketitle


\section{Introduction}

Full-fledged quantum computers are widely believed to offer exponential speedups in various applications, ranging from materials simulation~\cite{Lloyd1996,Abrams1999,Aspuru-Guzik2005} to cryptography~\cite{Shor1994,Shor1999, Gidney2021RSA, Gidney2025RSA}.
However, in real quantum devices, interactions with the environment
always disrupt the state of qubits, thereby preventing us from realizing the full benefits of quantum advantages.
Fortunately, the quantum error correction (QEC) techniques enable exponential suppression of idling errors in logical qubits by encoding quantum information using a polynomial number of physical qubits~\cite{Shor1996, Aharonov1997, Kitaev1997}.

However, protecting quantum information with QEC codes alone is insufficient to achieve fault-tolerant quantum computing (FTQC). It is essential to develop techniques enabling a universal logical gate set on QEC codes in a fault-tolerant manner.
For typical QEC codes, like the surface code~\cite{Dennis2002,Kitaev2003}, logical Clifford gates are relatively easy to perform by using well-known techniques such as transversal gates~\cite{Nielsen2000} or lattice surgery~\cite{Horsman2012,Litinski2019}.
In contrast, non-Clifford gates like $T$-gates generally cannot be implemented using the above techniques, except for some special cases such as [[15,1,3]] quantum Reed-Muller code~\cite{Knill1996, Bravyi2005}.
To resolve this issue, typical FTQC architectures prepare an ancillary non-Clifford state, especially a magic state $\ket{m}_L\equiv T\ket{+}_L$, and consume it to execute the gate teleportation scheme~\cite{Zhou2000}.
Magic state distillation (MSD)~\cite{Bravyi2005, Fowler2012, Gidney2019, Litinski2019magic} is a common approach to prepare such non-Clifford states, often leveraging transversal non-Clifford gates~\cite{Bravyi2005, Bravyi2012} or transversal Clifford measurements~\cite{Knill1998,Knill1998_2,Meiher2013,Jones2013multilevel} by embedding faulty logical qubits onto specific QEC codes that permit these operations.

Unfortunately, MSD incurs substantial spacetime overhead compared to logical Clifford operations.  For example, Ref.~\cite{Litinski2019magic} demonstrated that preparing a clean magic state with an error rate of $p_L < 10^{-10}$ requires a spacetime cost exceeding $30d^3$ qubit-cycles, where $d$ is the code distance of the surface code. This is substantially larger than the cost of Clifford gates, which typically require from $d^3$ to $3d^3$ qubit-cycles via lattice surgery~\cite{Litinski2019}. 
In particular, to sustain a supply rate of one magic state per clock (= $d$ cycles), a significant fraction of physical qubits must be dedicated to MSD factories, which would be prohibitive in early FTQC devices with limited physical qubits. Consequently, an alternative approach is highly desired to avoid the substantial overhead of MSD in the coming early-FTQC era.

A series of recent works~\cite{Chamberland2020, Itogawa2024distillation,Hirano2024zerolevel, Gidney2024cultivation, Hirano2025cultivation} has achieved a significant breakthrough on this problem by introducing novel techniques referred to as {\it zero-level distillation}~\cite{Itogawa2024distillation,Hirano2024zerolevel} or {\it magic state cultivation} (MSC)~\cite{Gidney2024cultivation, Hirano2025cultivation, Hetenyi2026MSC}.
In particular, MSC enables the preparation of moderately clean magic states with $p_L\gtrsim 10^{-9}$ using only a single surface code patch as ancillary space. Using these cultivated states as input for MSD, the spacetime overhead to achieve $p_L < 10^{-10}$ can also be reduced~\cite{Hirano2024zerolevel, Gidney2025RSA}. In addition, we can leverage the locality of the cultivation technique to realize locality-aware Pauli-based computation on early-FTQC devices~\cite{Hirano2025localityaware}.

Despite these advances in addressing the overhead of logical $T$-gates, another challenge 
remains in the fault-tolerant implementation of arbitrary-angle analog rotation gates.
In conventional FTQC, an analog rotation gate must be decomposed into a long sequence of single-qubit gates such as $\{T,S,H\}$~\cite{Selinger2015, Ross2016,Paetznick2014rus,Bocharov2015rus,Yoshioka2025crafting}, incurring significant time overhead. 
Notably, analog rotation gates play a predominant role in circuits for typical quantum algorithms such as Trotter simulation~\cite{Trotter1959, Suzuki1990, Suzuki1991, Lloyd1996, Abrams1997} and variational quantum algorithms~\cite{Peruzzo2014, McClean2016, Cerezo2021review}.
For example, a single Trotter step for simulating molecules with around 100 orbitals includes more than $10^{6}$ analog rotations gates with small angles $\theta\lesssim 10^{-6}$ to achieve the chemical accuracy~\cite{Wecker2014}. 

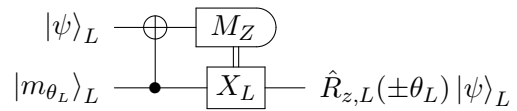
\begin{figure}[t]
\hspace{-10mm}
\centering
\fontsize{11pt}{11pt}\selectfont
  \mbox{
  \Qcircuit @C=1em @R=.7em {
    \lstick{\ket{\psi}_L}                      & \targ     & \measureD{M_{Z}}  &  \\
    \lstick{\ket{m_{\theta_L}}_L} & \ctrl{-1} & \gate{X_L} \cwx  & \rstick{\hat{R}_{z,L}(\pm\theta_L) \ket{\psi}_L } \qw 
    }
  }
  \caption{Quantum circuit for implementing an analog $Z$-rotation gate $\hat{R}_{z,L}(\theta_L)$ on a single logical qubit. $M_Z$ denotes a destructive $Z_L$-measurement on an encoded qubit. The sign of the resulting rotation angle is randomly determined with equal probability, depending on the measurement outcome.}
  \label{fig:GT_circ}
\end{figure}

To tackle these issues, Ref.~\cite{Akahoshi2023} has introduced a {\it partially fault-tolerant} quantum computing architecture, called {\it Space-Time efficient Analog Rotation quantum computing (STAR) architecture}.
The core idea is to decompose any quantum circuit with the Clifford+$\phi$ gate set by directly implementing analog rotation gates, rather than synthesizing them with the Clifford+$T$ gate set. Here, the Clifford+$\phi$ gate set consists of any Clifford gate and arbitrary analog rotation gates. Specifically, in the STAR architecture, an ancillary non-Clifford state $\ket{m_{\theta_L}}_L\equiv \hat{R}_{z,L}(\theta_L)\ket{+}_L$ is prepared in some non-fault-tolerant manner, and consumed for executing a Pauli-$Z$ rotation gate $\hat{R}_{z,L}(\theta_L)\equiv e^{i\theta_L \hat{Z}_L}$~\cite{comment} via gate teleportation (see Fig.~\ref{fig:GT_circ}). 


In the past few years, several promising approaches to prepare the $\ket{m_{\theta_L}}_L$ state has been proposed in  Refs.~\cite{Akahoshi2023,Choi2023,Toshio2024}. These proposals share a common important feature: they allow the execution of $\ket{m_{\theta_L}}_L$ state preparation within a single surface code patch, just like MSC.  
Particularly, the approaches proposed in Refs.~\cite{Choi2023,Toshio2024} can prepare a $\ket{m_{\theta_L}}_L$ state with an infidelity of $\mathcal{O}(\theta_L^{2(1-\Theta(1/d))}p_{\text{ph}})$, making them especially useful for implementing small-angle rotation gates in Trotter-based circuits. In our paper, we will refer to these approaches collectively as the {\it transversal multi-rotation (TMR) protocol}.
Then, Refs.~\cite{Toshio2024,Akahoshi2024} have demonstrated that the TMR protocol enables practical quantum speedups on the STAR architecture, even though achieving such speedups requires an optimistic physical error rate of $p_{\text{ph}}=10^{-4}$.

Here it is noteworthy that, as clarified in Ref.~\cite{Toshio2024}, the low infidelity of $\ket{m_{\theta_L}}_L$ states in the TMR protocol can inherently deteriorate before accomplishing the gate-teleportation process for the following reason: When executing the gate-teleportation circuit in Fig.~\ref{fig:GT_circ}, we encounter a failure event with a probability of 1/2, resulting in an inverse rotation. Therefore, to achieve the target rotation angle, we have to repeatedly perform feedback analog rotations while doubling the feedback rotation angle until we succeed in the teleportation. These feedback processes are often referred to as the {\it repeat-until-success (RUS) process}.
In this process, unlucky events where the feedback angle (referred to as the RUS angle) reaches the order of unity occur with a probability of $\mathcal{O}(\theta_L)$ and the corresponding feedback gate has an error rate of $\mathcal{O}(p_{\text{ph}})$. These events dominate the average error rate of the resulting rotation gate, leading to the error scaling of $\mathcal{O}(\theta_L p_{\text{ph}})$.

{\renewcommand{\arraystretch}{1.5}
\begin{table*}
    \centering
    \caption{Comparison of typical FTQC architectures~\cite{Litinski2019,Gidney2025RSA}, previous versions of STAR architecture (STAR ver.~1/ver.~2)~\cite{Akahoshi2023,Toshio2024}, and our proposal (STAR ver.~3). The columns ``$T$-gate" and ``Analog Rotation" represent the approaches used for implementing logical $T$-gates and logical analog rotation gates, respectively. ``MSD" and ``MSC" represent magic state distillation and magic state cultivation, respectively. ``[[4,1,1,2]] protocol" denotes the injection protocol based on [[4,1,1,2]] subsystem code, proposed in Ref.~\cite{Akahoshi2023}.
    $P_L(\theta_L)$ denotes an effective error rate of logical rotation gate $R_{P,L}(\theta_L)$ ($\theta_L\ll 1$) for each approach.}
    \begin{tabular}{p{3.5cm}p{0.3cm}p{2.3cm}p{0.3cm}p{3cm}p{0.3cm}p{3.5cm}p{0.3cm}p{3cm}p{0.1cm}}
    \hline\hline
        Architecture && gate set && $T$-gate && Analog Rotation && Scaling of $P_L(\theta_L)$ &\\
        \hline
        FTQC~\cite{Litinski2019,Gidney2025RSA} && Clifford+$T$ or Clifford+Toffoli && MSD or/and MSC && Gate synthesis && --- &\\ 
        STAR ver.~1~\cite{Akahoshi2023} && Clifford+$\phi$ && --- && [[4,1,1,2]] protocol && $\mathcal{O}(p_{\text{ph}})$ &\\ 
        STAR ver.~2~\cite{Toshio2024} && Clifford+$\phi$ && --- && TMR protocol and [[4,1,1,2]] protocol && $\mathcal{O}(\theta_L p_{\text{ph}})$ &\\ 
        STAR ver.~3 (this work) && Clifford+$T$+$\phi$ &&  MSD or/and MSC && STAR-magic mutation && $\mathcal{O}(\theta_L^{2(1-\Theta(1/d))}p_{\text{ph}})$ &\\ 
        \hline\hline
    \end{tabular}
    \label{tab:compare STAR}
\end{table*}
}

In this work, we introduce {\it STAR-magic mutation} (SMM), a novel protocol for implementing a logical rotation gate with high fidelity. The protocol enables an efficient implementation of an analog rotation gate with an effective error rate of $\mathcal{O}(\theta_L^{2(1-\Theta(1/d))}p_{\text{ph}})$, while requiring an ancillary space of only a single surface code patch. This scaling improves over the previous state-of-the-art, $\mathcal{O}(\theta_L p_{\text{ph}})$, reported in Ref.~\cite{Toshio2024}, and is particularly advantageous for executing small-angle rotation-based circuits. 

A key idea of SMM is to combine two of the latest state preparation protocols---the TMR protocol~\cite{Toshio2024, Choi2023} and magic state cultivation~\cite{Gidney2024cultivation}---within the RUS process for gate teleportation.
Specifically, as long as the RUS angle $\theta_{\text{RUS}}$ remains below a threshold $\theta_{\text{th}}$, the TMR protocol is employed to prepare a $\ket{m_{\theta_{\text{RUS}}}}$ state with an infidelity of $\mathcal{O}(\theta_{\text{RUS}}^{2(1-\Theta(1/d))}p_{\text{ph}})$.
Once the RUS angle reaches the threshold ($\theta_{\text{RUS}}\geq\theta_{\text{th}}$), the subsequent rotation gate $R_{P,L}(\theta_{\text{RUS}})$ is executed using a standard $T$-gate synthesis scheme~\cite{Ross2016}, where each $T$-gate is supplied by MSC or MSD. 
This adaptive strategy suppresses error accumulation during the RUS process, yielding the favorable error-rate scaling of SMM.

One might expect that the partial use of $T$-gate synthesis would substantially increase the time overhead of SMM, thereby negating the speed advantage of the STAR-based approach.
However, as we show below, RUS events in which $\theta_{\text{RUS}}$ exceeds $\theta_{\text{th}}$ occur only with a probability of $\order{\theta_{L}/ \theta_{\text{th}}}$. 
Consequently, by choosing $\theta_{\text{th}}$ sufficiently large, SMM achieves high-fidelity logical rotations at a speed comparable to an implementation that entirely avoids 
$T$-gate synthesis. Indeed, we find that, compared to an approach that relies solely on cultivation, SMM reduces the execution time of analog rotation gates by two orders of magnitude while achieving error rates that are several orders of magnitude smaller. 
Moreover, relative to the TMR-only approach in the previous STAR architecture~\cite{Toshio2024}, SMM reduces error rates by more than an order of magnitude, while keeping the increase in execution time to only a few tens of percent for $\theta_L\lesssim 10^{-4}$. Such small-angle rotation gates naturally appear in Trotter circuits for typical molecular systems~\cite{Wecker2014, Kanasugi2026}.


When performing each $T$-gate in gate synthesis, we primarily use the MSC protocol to prepare a magic state locally within an ancilla surface-code patch.
This allows us to perform a high-fidelity $T$-gate locally without requiring additional spatial overhead, in contrast to MSD~\cite{Fowler2012, Gidney2019, Litinski2019magic}.
Consequently, we can even perform multiple analog rotation gates in parallel by properly allocating local ancilla spaces for each SMM process.
Such locality-preserving parallelism is particularly advantageous for accelerating computations via locality-aware circuit compilation~\cite{Hirano2025localityaware, Sethi2025, Akahoshi2024}.

We next consider how SMM and MSC can be incorporated into earlier STAR architectures~\cite{Akahoshi2023, Toshio2024, Akahoshi2024} to further improve their performance. We refer to the resulting early-FTQC architecture as ``STAR ver.~3", distinguishing it from the first and second versions of the STAR architecture reported in Ref.~\cite{Akahoshi2023} and Ref.~\cite{Toshio2024}, respectively (see also TABLE~\ref{tab:compare STAR}). In STAR ver.~3, arbitrary quantum circuits are compiled into gate sequences over the Clifford+$T$+$\phi$ gate set, instead of Clifford+$T$ or Clifford+$\phi$ gate sets.
Here, Clifford+$T$+$\phi$ gate set consists of (i) any Clifford gates, (ii) digital rotation gates $\hat{R}_P(\pm\pi/8)\equiv e^{\pm i(\pi/8) \hat{P}}$, and (iii) analog rotation gates $\hat{R}_P(\theta)\equiv e^{i\theta \hat{P}}$ ($\theta\in \mathbb{R}$), where $\hat{P}$ denotes an arbitrary Pauli-string operator.
In our architecture, each analog rotation gate is implemented via SMM,
while digital rotation gates (i.e., $T$-gates) are executed using MSC (or MSD). 
In TABLE~\ref{tab:compare STAR}, we summarize the key differences from previous FTQC or STAR architectures.

Moreover, we clarify the theoretical bound on the maximum feasible circuit size for STAR ver.~3. This bound is determined by the cost of error mitigation and is closely related to the L1-norm of the target Hamiltonian in Trotter-based circuits.
Finally, as a promising application, we analyze the spacetime costs of simulating quantum many-body dynamics in molecules and lattice models using the Time Evolution by Probabilistic Angle Interpolation (TE-PAI) algorithm~\cite{Kiumi2024}. We show that STAR ver.~3 can simulate the dynamics of a variety of many-body systems, including biologically-relevant molecules, at scales beyond the reach of exact classical simulation, even assuming a realistic physical error rate of $p_{\text{ph}}=10^{-3}$. 
For example, we estimate that it enables Hamiltonian simulation of the $[\text{4Fe-4S}]$ cluster (72 spin orbitals)~\cite{Lee2023evaluating} up to $T=10$ [a.u.] within a week using $1.9\times 10^{5}$ physical qubits.
This contrasts with previous studies on the STAR architecture~\cite{Toshio2024,Akahoshi2024}, which inevitably required an optimistic error rate of $p_{\text{ph}}=10^{-4}$ to address comparable tasks.

This paper is organized as follows: In Sec.~\ref{sec:Preliminary}, we overview the basic concepts and the recent developments of the STAR architecture, including the details of the TMR protocol.
Then, in Sec.~\ref{sec:main_results}, we introduce our main proposal, STAR-magic mutation. Sec.~\ref{sec:cultivation} presents an overview of MSC. Sec.~\ref{sec:higher-order} formulates the method to suppress the higher-order coherent errors in the resource states.
Sec.~\ref{sec:STAR-magic mutation} details the mechanism of STAR-magic mutation and the theoretical scaling of the effective error rate of the resulting rotation gate. In Sec.~\ref{sec:numerical_results}, we present numerical results demonstrating the superior performance of SMM compared to the standard FTQC approach. Building upon these proposals, in Sec.~\ref{sec:STARv3}, we define a novel early-FTQC architecture, ``STAR ver.~3", and discuss its error mitigation strategies and the theoretical bounds on feasible circuit size. Finally, in Sec.~\ref{sec:practical applications}, we provide detailed resource estimates for simulating quantum many-body dynamics on STAR ver.~3, specifically focusing on the TE-PAI algorithm and its application to molecules and lattice models.


\section{Preliminary: STAR architecture}
\label{sec:Preliminary}

One of the promising applications of STAR-magic mutation is the substantial refinement of previous STAR architectures~\cite{Akahoshi2023, Toshio2024}.
Our primary goal in this context is to significantly improve the performance of this architecture by replacing its original gadgets for implementing analog rotations and its circuit compilation with more sophisticated alternatives.
To this end, in this section, we provide preliminary details of the latest version of the STAR architecture (for details, see Ref.~\cite{Toshio2024}).

\subsection{Definition}

The original concept of the STAR architecture was proposed in Ref.~\cite{Akahoshi2023} as a promising framework for early-FTQC devices.
In the original construction, each logical qubit is encoded on the rotated planar surface code~\cite{Horsman2012}, and the quantum computation running on it is implemented with two types of operations: (i) fault-tolerant Clifford operations with lattice surgery~\cite{Litinski2019} and (ii) noisy analog rotation gates produced via non-fault-tolerant ancilla state preparation with [[4,1,1,2]] subsystem code~\cite{Bacon2006}.
Thanks to the second operation, STAR architecture can avoid the costly procedures of magic state distillation~\cite{Bravyi2005, Fowler2012, Gidney2019, Litinski2019magic} and Solovay-Kitaev decomposition~\cite{Ross2016} when performing a logical rotation gate. This significantly reduces the number of physical qubits and execution time required for each non-Clifford gate. In this paper, we refer to this original architecture as ``STAR ver.~1". Recently, similar partially fault-tolerant architectures were successfully demonstrated on the Quantinuum H2-2~\cite{Yamamoto2025} and Helios~\cite{Dasu2026Quantinuum} trapped-ion quantum processors with [[7,1,3]] color code and (two-level) iceberg codes, respectively, providing compelling evidence for the utility of partial fault-tolerance in the early FTQC era.

After the original proposal, Ref.~\cite{Toshio2024} has significantly improved the computational capability of the STAR architecture by introducing a more sophisticated framework to perform a high-fidelity small-angle rotation gate (see Sec.~\ref{sec:STARv2} for details). This enables the achievement of quantum speedups on STAR even in practical tasks, including quantum phase estimation for the fermionic Hubbard model~\cite{Toshio2024, Akahoshi2024}. Importantly, the proposed framework also made it easier to extend the STAR architecture to generic QEC codes such as quantum low-density parity-check codes, as detailed in Ref.~\cite{Ismail2025}. In this paper, we refer to this refined architecture as ``STAR ver.~2".

To put these developments in perspective, Ref.~\cite{Toshio2024} redefined the concept of the STAR architecture as a class of quantum computing architectures based on the following design principles:

\begin{itemize}
    \item {\bf Partial fault-tolerance}: Quantum information is encoded on some error-correcting codes, and arbitrary Clifford operations are performed on it in a fault-tolerant manner.
    \item {\bf Clifford + $\phi$ gate set}: Most of logical operations are performed by synthesizing a gate set composed of the Clifford gates and analog rotation gates.
    \item {\bf Noisy analog rotation gates}: Analog rotation gates are implemented using a non-fault-tolerant resource state preparation protocol followed by the gate-teleportation of the prepared states. 
    \item {\bf Error mitigation}: Some error mitigation strategies (e.g., probabilistic error cancellation (PEC)~\cite{Temme2017,Endo2018}) are employed to suppress quantum errors occurring in analog rotation gates.
\end{itemize}

In this paper, we attempt to construct a further refined early-FTQC architecture named ``STAR ver.~3", which is built upon SMM proposed herein.
From the above perspectives, our achievements can be summarized as follows:

\begin{itemize}
    \item {\bf Further improvement of non-fault-tolerant analog rotation gates:} By introducing the STAR-magic mutation, we improve the error rate of analog rotation gates by more than one order of magnitude.
    \item {\bf Extension to the Clifford+$T$+${\bm \phi}$ gate set:} By incorporating MSC into the STAR architecture, we enable the architecture to execute a broader class of quantum circuits.
\end{itemize}
Here, it is noteworthy that previous studies on STAR ver.~2~\cite{Toshio2024,Akahoshi2024} reluctantly required an optimistic value of physical error rate (i.e., $p_{\text{ph}}=10^{-4}$) in order to demonstrate practical quantum advantages on STAR.
In contrast, our architecture can alleviate this challenging requirement by improving the logical error rate of analog rotation gates by at least one digit, enabling practical quantum advantages on STAR even under $p_{\text{ph}}=10^{-3}$, a value within reach of near-term experiments. This represents a significant step towards the practical use of early-stage FTQC devices in the near future.

In what follows, we summarize the existing framework proposed in Refs.~\cite{Akahoshi2023,Toshio2024} to make it easy to understand our results.

\begin{figure}[tbp]
  \centering
  \fontsize{11pt}{11pt}\selectfont
  \mbox
  {
  \Qcircuit @C=1.3em @R=1.0em {
    & \multigate{5}{M_{P \otimes Z}} & \multigate{4}{P} & \qw \\
    & \ghost{M_{P \otimes Z}} & \ghost{P} & \qw & \\
    \lstick{\ket{\psi}_L} & \ghost{M_{P \otimes Z}} & \ghost{P} &   \rstick{\hat{R}_{P,L}(\pm\theta_L)\ket{\psi}_L} \qw \\
    & \ghost{M_{P \otimes Z}} & \ghost{P} & \qw \\
    & \ghost{M_{P \otimes Z}} & \ghost{P} & \qw \\
    \lstick{\ket{m_{\theta_L}}_L} & \ghost{M_{P \otimes Z}} & \measureD{M_X} \cwx &
  }
  }
  \caption{Quantum circuit for implementing an analog multi-Pauli rotation gate $\hat{R}_{P,L}(\theta)$ on any encoded state $\ket{\psi}_L$. In this setup, we can choose any Pauli string operator $\hat{P}$. The sign of the resulting rotation angle is randomly determined with equal probability, depending on the outcome of the destructive $X$-measurement.}
  \label{fig:mprot_circ}
\end{figure}
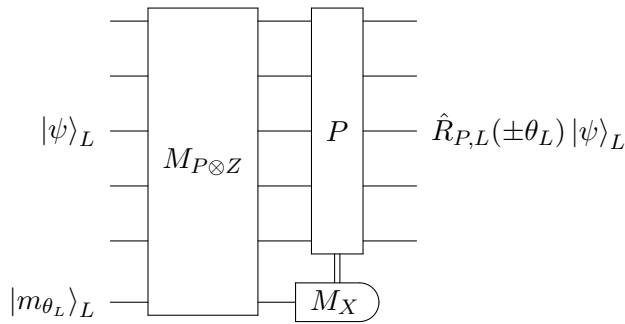

\subsection{Non-fault-tolerant implementation of analog rotation gates}
\label{sec:STARv2}

In this subsection, we will illustrate how to implement analog rotation gates in STAR ver.~2~\cite{Toshio2024}. First, we define the {\it resource state}, which is consumed to implement an analog multi-Pauli rotation gate via gate-teleportation circuit.
Next, we explain the transversal multi-rotation (TMR) protocol~\cite{Toshio2024}, which can prepare a resource state with 
a low infidelity of $\mathcal{O}(\theta^{2(1-\Theta(1/d))}p_{\text{ph}})$. 
We then illustrate that applying a probabilistic inverse rotation can improve the worst-case error rate of analog rotation gates. Finally, we comment on the limitations of the non-fault-tolerant approach used in the STAR architecture.

\subsubsection{Repeat-until-success implementation of analog rotation gate}
\label{sec:RUS}

In the STAR architecture, we exploit the following type of ancillary non-Clifford states instead of magic states~\cite{Bravyi2005}:
\begin{equation}
\label{eq:resource_state}
    \ket{m_{\theta_L}}_L \equiv \hat{R}_{z,L}(\theta_L)\ket{+}_L = \cos\theta_L \ket{+}_L +i\sin\theta_L \ket{-}_L,
\end{equation}
where encoded quantum states are denoted as $\ket{\cdots}_L$ and a logical Pauli-$Z$ rotation gate as $\hat{R}_{z,L}(\theta_L)=e^{i\theta_L \hat{Z}_L}$~\cite{comment}. The rotation angle $\theta_L$ can be arbitrarily chosen. In what follows, we will refer to this type of non-Clifford state as a {\it resource state}.
As illustrated in Fig.~\ref{fig:GT_circ}, we can execute $\hat{R}_{z,L}(\theta_L)$ gate on any target state $\ket{\psi}_L$ by entangling it with a resource state $\ket{m_\theta}_L$ through the gate-teleportation circuit~\cite{Zhou2000}.
In the circuit, the output state becomes a correctly rotated state $\hat{R}_{z,L}(\theta)\ket{\psi}_L$ if
the measurement outcome is $+1$ with a probability of $1/2$; otherwise, the
output becomes an inversely rotated state $\hat{R}_{z,L}(-\theta)\ket{\psi}_L$.
If the inversely rotated state is obtained, we need to apply the teleportation process again to correct its rotation direction while doubling the rotation angle of the input resource state. 
This procedure is repeated until we obtain a measurement outcome of $+1$, thereby yielding the desired state $\hat{R}_{z,L}(\theta)\ket{\psi}_L$~\cite{Jones2012}. We will refer to this procedure as {\it repeat-until-success (RUS) process}.
As is easily checked, this procedure succeeds in two trials on average.

More generally, we can implement any multi-Pauli rotation gates $\hat{R}_{P,L}(\theta)$ via the quantum circuit in Fig.~\ref{fig:mprot_circ} by consuming a single resource state $\ket{m_\theta}_L$. 
This circuit is based on a multi-Pauli measurement operation, instead of applying multiple $CNOT$ gates. This is preferable for an efficient implementation of multi-Pauli rotations via the lattice surgery techniques~\cite{Litinski2019}. Therefore, we usually assume the circuit in Fig.~\ref{fig:mprot_circ} for implementing these rotation gates.

\subsubsection{Transversal multi-rotation protocol}
\label{sec:TMR protocol}

\begin{figure}[t]
    \centering
    \includegraphics[width=1.0\linewidth]{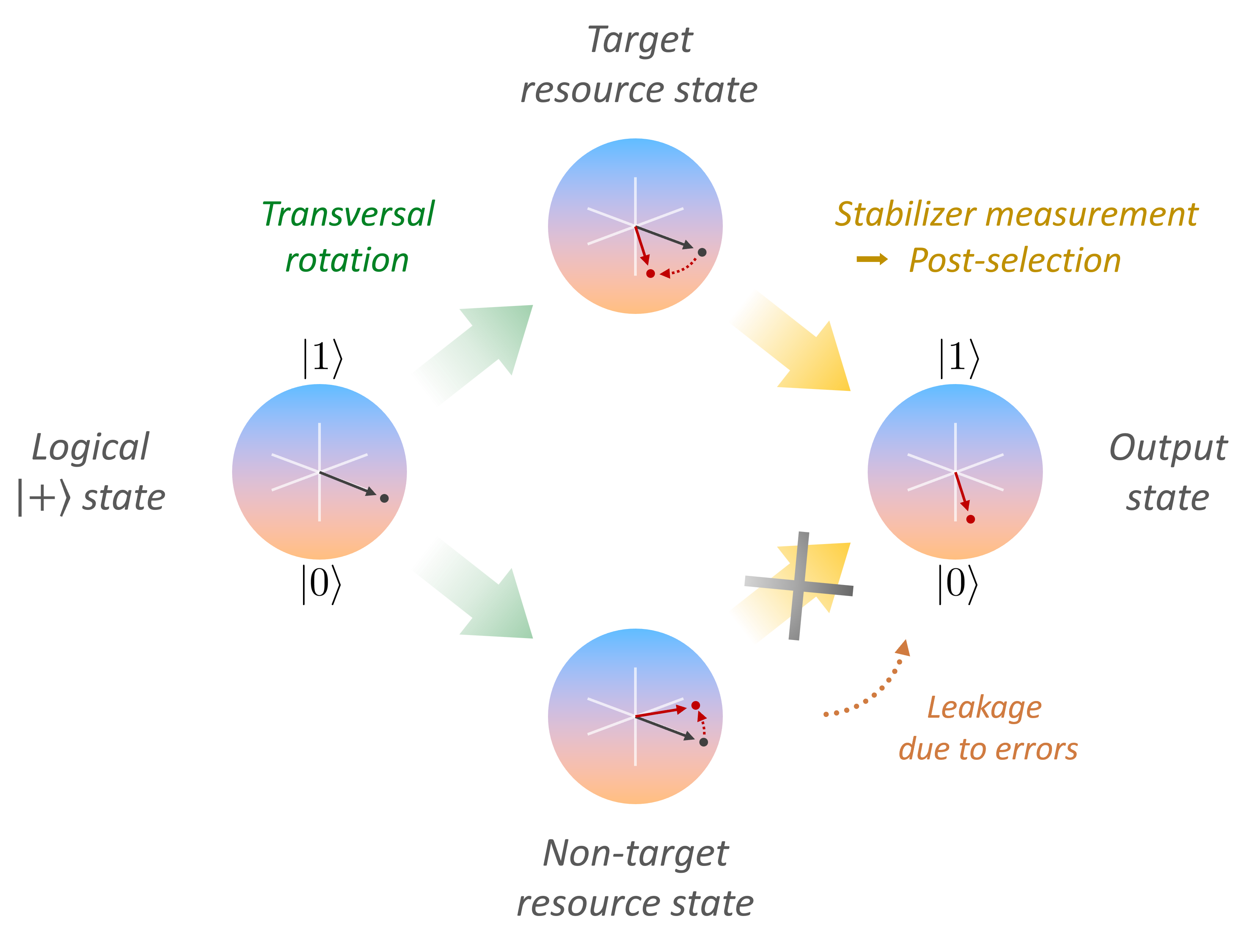}
    \caption{Intuitive picture of the transversal multi-rotation (TMR) protocol. The transversal multi-Pauli rotation gate (green) produces superpositions of logical resource states with different logical angles and different stabilizer eigenvalues. By performing stabilizer measurement (yellow), we can post-select only the target resource state, discarding other erroneous resource states. However, under physical noises, erroneous resource states rarely appear in the output state (orange), leading to a finite error rate of the TMR protocol.}
    \label{fig:intuitive picture}
\end{figure}

A key component of the STAR architecture is a space-time efficient preparation protocol for a resource state $\ket{m_{\theta_L}}_L$. Especially, as formulated in Ref.~\cite{Toshio2024}, the latest version of STAR architecture (STAR ver.~2) is based on the so-called {\it transversal multi-rotation  (TMR) protocol}. This protocol was originally inspired by the proposal in Ref.~\cite{Choi2023} and improved it in terms of success probability and logical error rate. In what follows, we briefly overview the TMR protocol (see also Fig.~\ref{fig:intuitive picture}).

\begin{figure}[t]
    \centering
    \includegraphics[width=0.9\linewidth]{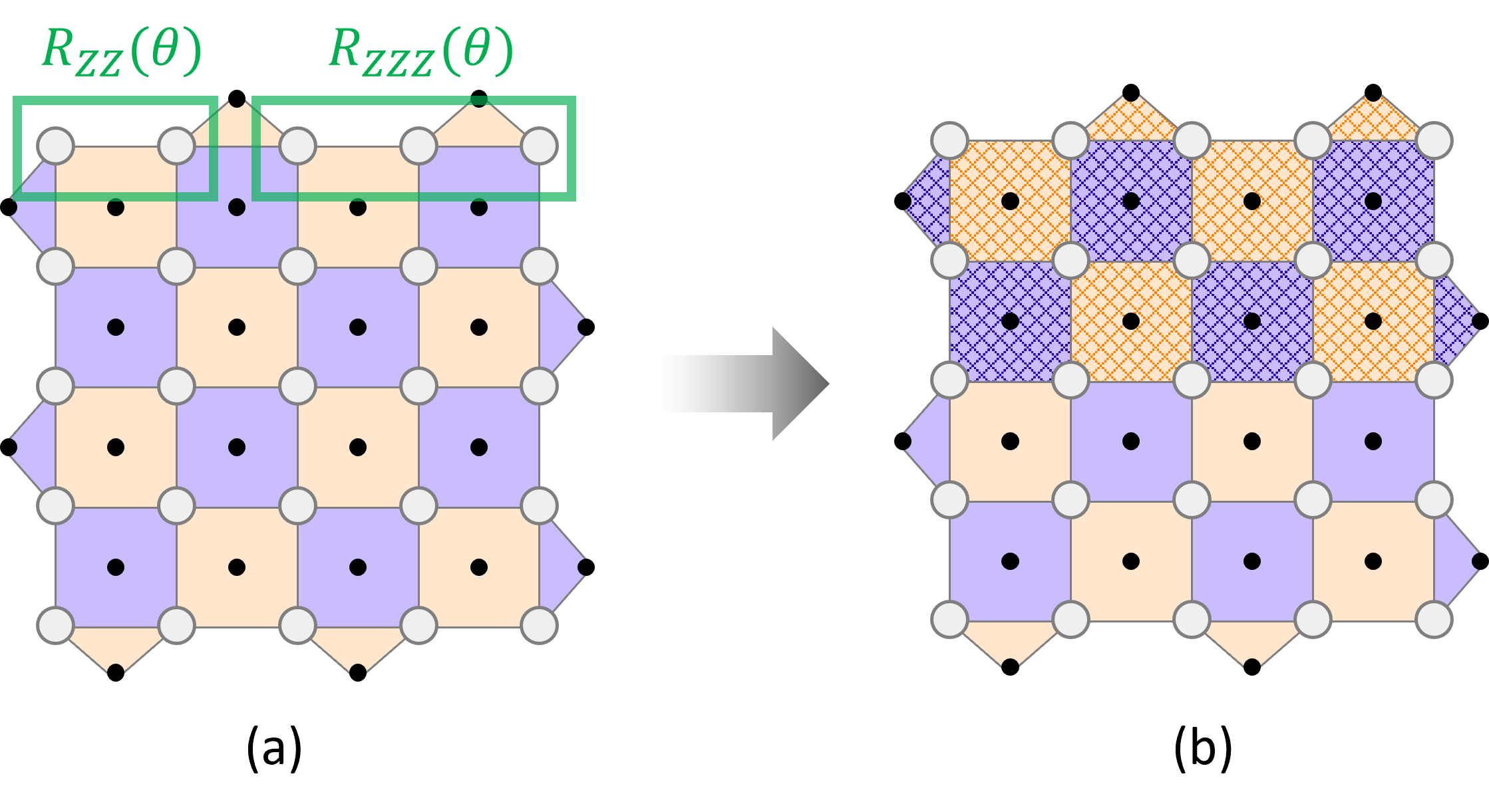}
    \caption{Schematic of the TMR protocol on a rotated surface code with code distance $d=5$.
    (a) First, the ancilla state is initialized in $\ket{+}_L$, then multiple rotation gates are applied along the logical $Z_L$ operator. Each multi-$Z$ rotation gate can be performed by combining nearest-neighbor CNOT gates and single-qubit $Z$-rotation gates.
    (b) After applying the transversal multi-Pauli rotation gates, we perform a post-selection on the hatched stabilizers to remove critical errors that bring leading logical errors. The remaining errors detected by other stabilizers can be corrected in the subsequent quantum error correction procedure.}
    \label{fig:injection}
\end{figure}

First, let us assume that we prepare a logical state $\ket{+}_L$ of an arbitrary $[[n,1,d]]$ error-correcting code.
Then, we apply a {\it transversal multi-Pauli rotation gate} defined as follows: 
\begin{equation}
    \prod_{i=1}^{k} \hat{R}_{Z_{Q_i}}(\theta) = \prod_{i=1}^{k} \left[ \cos \theta\cdot  \hat{I}_{Q_i} + i \sin \theta \cdot \hat{Z}_{Q_i} \right],  
\end{equation}
where $\{Q_i\}_{i=1,\cdots,k}$ is a collection of mutually disjoint subsets of physical qubits that satisfies $\bigcup_i Q_i = Q_{Z_L}$. Here we define $Q_{Z_L}$ as a subset of physical qubits that support a logical-$Z$ operator $\hat{Z}_L$ such that  $\prod_{j \in Q_{Z_L}} \hat{Z}_j = \hat{Z}_L$. 
$\hat{I}_{Q_i}$ and $\hat{Z}_{Q_i}$ denote an identity gate and a Pauli-$Z$ string operator acting on the subset $Q_i$, respectively. The parameter $k$ is an integer that determines how to divide $Q_{Z_L}$ into $\{Q_i\}_{i=1,\cdots,k}$. 
A simple case of $k=2$ is shown in Fig.~\ref{fig:injection} (a). In what follows, we assume $\theta>0$ for simplicity.

Applying an appropriate logical Clifford correction $\hat{C}\propto R_{x,L}(j\pi/4)$ ($j\equiv k-1\ (\text{mod}\ 4)$), the resultant state becomes a superposition of a resource state with the target angle $\theta_L$ and other states as follows: 
\begin{equation} \label{eq:ver2_state}
    \hat{C}\prod_{i=1}^{k} \hat{R}_{Z_{Q_i}}(\theta) \ket{+}_L = \sqrt{p_{\rm ideal}} \ket{m_{\theta_L}}_L 
    + \ket{\text{out}}, 
\end{equation}
where $p_{\rm ideal}$ is defined as 
\begin{equation}
    p_{\rm ideal}(\theta, k) = \sin^{2k} \theta + \cos^{2k} \theta, 
\end{equation}
and the logical-level angle $\theta_L$ is related to the physical-level angle $\theta$ as 
\begin{equation}
    \theta_L = \sin^{-1} \left( \frac{1}{\sqrt{p_{\rm ideal}}} \sin^k (\theta) \right) \approx \theta^k + {\mathcal O}(\theta^{k+2}).
\end{equation}
The second term $\ket{\text{out}}$ is a superposition of perturbed $\ket+$ states on which various Pauli-$Z$ string operators act, 
\begin{equation}
\label{eq:noisy terms}
    \ket{\text{out}} = \sum_{b\neq 00\cdots0, 11\cdots1} u_{|b|} \hat{C} \hat{Z}^b \ket{+}_L  =  \sum_{0<|b|\leq k/2}\hat{C}\ket{\psi_b},
\end{equation}
where $b= b_1b_2\cdots b_k\in \{0,1\}^k$ is a $k$-bit string and $|b|$ denotes the Hamming weight of the bit-string $b$. We define $u_{j} \equiv i^j\sin^{j}\theta \cos^{k-j}\theta$, $\ket{\psi_b}\equiv(u_{|b|}  \hat{Z}^b + u_{|\bar{b}|}\hat{Z}^{\bar{b}}) \ket{+}_L$, and $\hat{Z}^b \equiv \prod_{i:b_i=1} \hat{Z}_i$. 

The $\ket{\text{out}}$ state exists outside of the logical space of the QEC code, since non-logical $Z$ operators acting on $\ket{+}_L$ flips at least one of stabilizer eigenvalues.
Consequently, we can extract the target ancilla state $\ket{m_{\theta_L}}_L$ by post-selecting only the case where there are no error syndromes detected from stabilizer measurements. The value of $p_{\rm ideal}$ denotes the success probability of the post-selection process in the ideal limit of $p_{\text{ph}}\to 0$.

However, in actual quantum devices, quantum noises on $\ket{\text{out}}$ cause finite overlap between this state and the logical subspace (see Fig.~\ref{fig:intuitive picture}). Consequently, the output state of the TMR protocol is represented as the following density matrix:
\begin{equation}
\label{eq:final output state}
\begin{aligned}
 \hat{\rho}_{\text{output}}\simeq \  &\frac{1}{p_{\text{suc}}} \Big[ q_0 \ket{m_{\theta_L}}\bra{m_{\theta_L}}_L  \\
    & \qquad +\ q_1 \ket{m_{\theta_{\text{error}}}}\bra{m_{\theta_{\text{error}}}}_L\Big] + {\mathcal{O}(\theta_L p_{\text{ph}}^2)},\\
\end{aligned}
\end{equation}
where $\ket{m_{\theta_{\text{error}}}}_L$ is the leading error term that has a different rotation angle
\begin{equation}
\label{eq:error angle}
\begin{aligned}
    \theta_{\text{error}}  \  \equiv\  -\sin^{-1}\left(
        \frac{\sin^{k-2}\theta}{\sqrt{\sin^{2k-4}\theta+\cos^{2k-4}\theta}} \right)
\end{aligned}
\end{equation}
The coefficients $q_0$ and $q_1$ are related to the probability that each corresponding state is post-selected via stabilizer measurements, and $p_{\text{suc}}\equiv q_0+q_1+\mathcal{O}(p_{\text{ph}}^2)$ represents the
success rate of the TMR protocol. Importantly, the second coefficient $q_1$ scales as $\mathcal{O}(\theta_L^{2/k}p_{\text{ph}})$.
Thus, the second term in Eq.\eqref{eq:final output state} leads to a finite infidelity of the output state, which scales as $\mathcal{O}(\theta_L^{2(1-1/k)}p_{\text{ph}})$. 
Furthermore, Ref.~\cite{Zeng2025star} has recently suggested that, in specific hardware setups, we can reduce the magnitude of $q_1$ into $\mathcal{O}(\theta_L^{2/k}p_{\text{ph}}^2)$ by leveraging realistic dissipative noise processes like dispersive coupling between transmons. This modification improves the infidelity of the TMR protocol into $\mathcal{O}(\theta_L^{2(1-1/k)}p_{\text{ph}}^2)$. This technique can be easily combined with our proposal in this paper.

In practice, the post-selection process should be optimized to maximize its success probability by combining the QEC mode. 
Specifically, the stabilizer operators checked in the post-selection can be limited to the shaded region shown in Fig.~\ref{fig:injection} (b). 
Detected errors outside the post-selection region can be corrected by subsequent QEC procedures and do not contribute to leading logical errors. This hybrid approach significantly improves the success probability of the TMR protocol by several orders. 
For example, assuming $p_{\text{ph}}=10^{-3}$ and the rotated surface code with $d=11$, the optimized TMR protocol can prepare a resource state with $\theta_L=10^{-3}$ within about 1.3 clocks. This is much faster than magic state preparation via MSC~\cite{Gidney2024cultivation}, which requires around 10 clocks for similar setups. Furthermore, Ref.~\cite{Chung2026} has recently discussed a patch-growth technique to improve the success rate of the TMR protocol even for larger code distances.

Most notably, the TMR protocol is executed on a single ancilla QEC code block, eliminating the need for magic state factories in usual FTQC~\cite{Fowler2012, Gidney2019, Litinski2019magic}. This compactness and locality of the TMR protocol also enable us to execute multiple rotation gates in parallel by leveraging ancilla space for lattice surgery, as discussed in Refs.~\cite{Hirano2025localityaware, Sethi2025, Akahoshi2024}.
This ``easy-to-parallelize" feature of the STAR architecture clearly contrasts with usual FTQC architectures based on MSD, where parallel $T$ gates require many more physical qubits to parallelize the MSD processes for achieving a sufficient supply rate of magic states.

\subsubsection{Effective quantum channel and probabilistic inverse rotation}
\label{sec:inverse rotation}

We can perform a noisy rotation gate by employing the state in Eq.~\eqref{eq:final output state} as an input of the gate-teleportation circuit in Fig.~\ref{fig:mprot_circ}.
Theoretically, this noisy gate is described with the following quantum channel when the gate-teleportation succeeds in the first trial:
\begin{equation}
\label{eq:noisy rotation channel in the first trial}
\begin{aligned}
    \mathcal{N}_{\theta_L}(\hat{\rho}) =&\  \mathcal{E}_{\theta_L} \circ \mathcal{R}_{\theta_L}(\hat{\rho})\\
    =& \ (1-\bar{q}_1) \cdot \mathcal{R}_{\theta_L}(\hat{\rho}) \\
    & \qquad + \bar{q}_1 \cdot \mathcal{R}_{\theta_{\text{error}}}(\hat{\rho}) + \order{\theta_L p_{\text{ph}}^2},
\end{aligned}
\end{equation}
where $\mathcal{R}_{\theta_L}$ denotes the quantum channel that corresponds to an ideal rotation gate $\hat{R}_{z,L}(\theta_L)$, and $\mathcal{E}_{\theta_L}$ denotes the following stochastic channel:
\begin{equation}
\label{eq:our error channel}
\begin{aligned}
    \mathcal{E}_{\theta_L}(\hat{\rho}) &= (1-\bar{q}_1)\cdot\hat{\rho}\\
    & \qquad  + \bar{q}_1\cdot \mathcal{R}_{\Delta_{\theta_L}}(\hat{\rho})
+ \order{\theta_L p_{\text{ph}}^2}. 
\end{aligned}
\end{equation}
Here we define the logical error rate $\bar{q}_1\equiv q_1/p_{\text{suc}}$ and the over-rotation angle $\Delta_{\theta_L}\equiv \theta_{\text{error}}- \theta_L$.
From a simple analysis, we find that the coherent error term of $\mathcal{R}_{\Delta_{\theta_L}}$ plays the most dominant role in the channel $\mathcal{N}_{\theta_L}$, leading to the worst-case error rate of $\order{\theta_L p_{\text{ph}}}$. 
More generally, in each RUS trial for gate-teleportation, $\mathcal{N}_{\theta_{\text{RUS}}}$ or $\mathcal{N}_{-\theta_{\text{RUS}}}$ occurs with a probability of 1/2, respectively. By completing the RUS process, we can obtain a noisy quantum channel $\tilde{\mathcal{N}}_{\theta_L}$, which has a worst-case error rate of $\tilde{\mathcal{O}}(\theta_L p_{\text{ph}})$.

To remove the coherent term in $\mathcal{N}_{\theta_L}$, Ref.~\cite{Toshio2024} introduced a post-processing method called the {\it probabilistic coherent error cancellation} (PCEC). 
In this method, we use the following stochastic channel,
\begin{equation}
\label{eq:cancellation channel}
\begin{aligned}
    \mathcal{C}_{\theta_L}(\hat{\rho}) &\equiv (1-\bar{q}_1)\cdot\hat{\rho} + \bar{q}_1 \cdot \tilde{\mathcal{N}}_{-\Delta_{\theta_L}}(\hat{\rho})  \\
    & \simeq (1-\bar{q}_1)\cdot\hat{\rho} + \bar{q}_1 \cdot \mathcal{R}_{-\Delta_{\theta_L}}(\hat{\rho})
    + \tilde{\mathcal{O}}(\theta_L p_{\text{ph}}^2).
\end{aligned}
\end{equation}
In the last equality, we used the fact that $\tilde{\mathcal{N}}_{-\Delta_{\theta_L}}\simeq  \mathcal{R}_{-\Delta_{\theta_L}}(\hat{\rho}) + \tilde{\mathcal{O}}(\theta_L^{1-2/k} p_{\text{ph}})$ and  $\bar{q}_1 \simeq \order{\theta_L^{2/k}p_{\text{ph}}}$.

By applying this channel after the noisy rotation channel $\mathcal{N}_{\theta_L}$, we obtain a composed quantum channel as follows:
\begin{equation}
    \mathcal{N}^{c}_{\theta_L}(\hat{\rho})\ \equiv\  \mathcal{C}_{\theta_L} \circ \mathcal{N}_{\theta_L}(\hat{\rho}) \ =\  \mathcal{E}_{\theta_L}^{c} \circ \mathcal{R}_{\theta_L}(\hat{\rho}),
\end{equation}
where an effective error channel for $\mathcal{N}^{c}_{\theta_L}$ is defined as $\mathcal{E}_{\theta_L}^{c} \equiv \mathcal{C}_{\theta_L} \circ \mathcal{E}_{\theta_L}$.
Then, we can explicitly decompose this error channel as 
\begin{equation}
\label{eq:canceled rotation gate}
\begin{aligned}
    \mathcal{E}_{\theta_L}^{c}(\hat{\rho})\ &\simeq \ 
    [1-2\bar{q}_1\sin^2(\Delta_{\theta_L})]\cdot \hat{\rho}  \\
    & \qquad + 2\bar{q}_1 \sin^2(\Delta_{\theta_L})\cdot \hat{Z}\hat{\rho} \hat{Z} + \tilde{\mathcal{O}}(\theta_L p_{\text{ph}}^2).
\end{aligned}
\end{equation}
This formula leads to the worst-case error rate of $\mathcal{E}_{\theta_L}^{c}$ as $2\bar{q}_1 \sin^2(\Delta_{\theta_L}) \simeq \mathcal{O}(\theta_L^{2(1-1/k)}p_{\text{ph}})$, when we can neglect the term of $\tilde{\mathcal{O}}(\theta_L p_{\text{ph}}^2)$. In our work, the $\tilde{\mathcal{O}}(\theta_L p_{\text{ph}}^2)$ term is not necessarily negligible since the target angle $\theta_L$ could be smaller than $p_{\text{ph}}$ and the RUS process is interrupted before the RUS angle grows to $\mathcal{O}(1)$. Therefore, we modify the above PCEC method to be able to cancel the higher-order errors in Sec.~\ref{sec:higher-order}.

Unfortunately, even with the above technique, the worst-case error rate of the resulting rotation gate via the RUS process still falls into the order of $\mathcal{O}(\theta_L p_{\text{ph}})$.
This is because, in the RUS process, unlucky events where the RUS angle reaches the order of unity occur with a probability of $\mathcal{O}(\theta_L)$ and the corresponding feedback gate has an error rate of $\mathcal{O}(p_{\text{ph}})$. These events dominate the average error rate of the resulting rotation gate, leading to the error scaling of $\mathcal{O}(\theta_L p_{\text{ph}})$. This bottleneck is the central challenge our study aims to resolve.

\subsubsection{Resilience to control errors}
\label{sec:control errors}

In the previous subsections, we implicitly assumed that quantum errors are well described by stochastic Pauli error channels, which are usually adopted in standard QEC simulations. 
However, in actual devices, we can never neglect the presence of control errors in physical gates due to imperfect pulse calibrations, misalignment of the quantization axes, crosstalk between qubits, and so on.

As detailed in Ref.~\cite{Toshio2024}, the TMR protocol is relatively weak against certain types of coherent errors, particularly over-rotation errors. 
More specifically, assuming the transversal multi-rotation gate is slightly shifted by over-rotation errors as
\begin{equation}
\label{eq:over-rotation model}
    \prod_{i=1}^{k} \hat{R}_{Z_{Q_i}}(\theta)\ \ \to \ \ \prod_{i=1}^{k} \hat{R}_{Z_{Q_i}}(\theta+\phi_i),
\end{equation}
we readily find that the target angle $\theta_L$ is also shifted as follows:
\begin{equation}
     \theta_L\ \ \to \ \ \theta_L\cdot\left(1+\sum_i \phi_i/\theta\right)
\end{equation}
Here $\phi_i$ $(\ll \theta\ll 1)$ denotes a small shift angle in the $i$-th rotation gate due to over-rotation errors.
This result means that the relative error on the output logical angle is proportional to that of the input physical angles. Therefore, compared to the usual injection protocols~\cite{Li2015magic, Lao2022} including the [[4,1,1,2]] protocol in STAR ver.~1~\cite{Akahoshi2023}, the TMR protocol is relatively robust to these coherent errors because the over-rotation $\phi$ directly shifts the output angle as $\theta_L\to \theta_L +\phi$ in the former approaches.

Fortunately, we can further cancel out the leading coherent errors by randomizing the direction of each rotation in the transversal multi-rotation gate~\cite{Toshio2024}, assuming the use of the virtual $Z$-gate~\cite{Mckay2017}. 
This approach suppresses the absolute amplitude of the over-rotation error from $\order{\theta_L \phi_i/\theta}$ to $\order{\theta_L \phi_i^2}$ at the logical level. 

Furthermore, the residual over-rotation error at the logical level could be calibrated by measuring the resultant rotation angle via quantum phase estimation before performing quantum computation. 
However, in general, error parameters in realistic devices will systematically change in time, preventing us from removing control errors perfectly. This problem would be alleviated by performing the above QPE-based calibration within a predefined auxiliary space during quantum computations.
By following these prescriptions, we believe that our approach will satisfy at least a transient demand of simulating classically-intractable quantum circuits within a limited qubit budget in the early-FTQC era.

\section{STAR-magic mutation: High-fidelity logical rotation gate with a small angle}
\label{sec:main_results}

In this section, we formulate SMM, a novel method for implementing logical analog rotation gates in early fault-tolerant quantum computers.
This method is based on the TMR protocol discussed in Sec.~\ref{sec:TMR protocol} and MSC. The latter is a novel type of magic state preparation protocol recently proposed by C. Gidney {\it et al.}~\cite{Gidney2024cultivation}. 
SMM enables implementing efficient and high-fidelity analog rotation gates by combining the above methods in a way that they mutually compensate for each other's disadvantages.

In what follows, we first overview the technique of MSC~\cite{Gidney2024cultivation}. Then, we modify the PCEC method to cancel the higher-order errors in the TMR protocol. This modification is required to formulate the error scaling of SMM.  
In Sec.~\ref{sec:STAR-magic mutation}, we present our main claims, namely the detailed procedures of SMM and a theorem on the logical error rate of the resulting rotation gates. Finally, we show numerical data for SMM and provide a discussion on the tradeoff between accuracy and speed in SMM. 

\subsection{Magic state cultivation}
\label{sec:cultivation}

In this subsection, we briefly overview MSC. It is a method for efficiently preparing a high-fidelity magic state within a surface code patch. This technique gradually grows the size and reliability of a magic state encoded in a small color code with a limited number of physical gates and qubits. As detailed below, the cultivation process is divided into three stages: injection, cultivation, and escape. 

The cultivation process begins with the injection stage, where a faulty $\ket{m}$ state is encoded in a distance-3 color code. Subsequently, the cultivation stage iteratively increases the fault distance of the encoded $\ket{m}$ state through a check-grow-stabilize cycle. The ``check" step enhances the fault distance by verifying the encoded value of the code, while the ``grow" step expands the size of the color code to facilitate further increases in fault distance. The ``stabilize" step measures the stabilizers of the code to detect errors and prevent correlated errors between check steps. This cultivation stage uses a ``double-check" approach, where we perform two checks, with the second check being the time reverse of the first. This streamlined circuit reduces the cost of checking the logical state.
Importantly, in each step, we have to discard the generated state when we encounter unwanted measurement outcomes to maximize the final fidelity of the output state. This post-selection process results in a high probability of failure in this cultivation process.

The final process is the escape stage, which addresses how to embbed the generated $\ket{m}$ state into a larger QEC code to maintain its high fidelity during subsequent logical operations.
In this stage, we rapidly expand the code distance of the code by grafting a large surface code onto a small color code. The resulting QEC code, called the grafted code, is stabilized by measuring its stabilizers several times until and finally transformed into a fully matchable code. Finally, by calculating the complementary gap of the escape stage, we determine whether or not to keep the obtained state.
More recently, Ref.~\cite{Hirano2025cultivation} has suggested that the spacetime cost of this escape state can be significantly reduced by teleporting the cultivated
state from the color code to the rotated surface code, instead of the grafted code, and by enabling early rejection based on a preset lookup table.

End-to-end simulations in Ref.~\cite{Gidney2024cultivation} demonstrate that cultivation can achieve logical error rates as low as $p_m = 2 \times 10^{-9}$ under a uniform depolarizing circuit noise with $p_{\text{ph}}=10^{-3}$, assuming $d=15$ for the escape stage.
However, this process has an end-to-end discard rate of $99\%$, resulting in an expected spacetime cost of $6.6\times 10^4$ [qubit$\cdot$rounds]. 
For example, given an ancillary space of $d=15$ surface code, this suggests that MSC requires roughly $t_m = 10$ [clocks] (= $10d$ [rounds]) on average to generate a single magic state with an ancillary surface code patch. Here it should be noted that Ref.~\cite{Gidney2024cultivation} ignores how to pack multiple cultivation processes in a limited space and therefore, the above estimation of spacetime cost is somewhat underestimated.

In what follows, we use the above parameters, $p_m = 2 \times 10^{-9}$ and $t_m = 10$ [clocks], to characterize the cultivation process. Strictly speaking, the value of $t_m$ depends on the code distance of ancillary spaces and the packing efficiency. We will not delve into the details of these aspects in this paper, since we do not expect them to significantly alter our results.

\subsection{Higher-order probabilistic coherent error cancellation}
\label{sec:higher-order}

In this subsection, we modify the PCEC protocol in Sec.~\ref{sec:inverse rotation} to enable the cancellation of higher-order errors in the TMR protocol. This modification is required for the theoretical formulation of our proposal.

Generally, the output state of the TMR protocol is described with the following density matrix:
\begin{equation}
\label{eq:general fom of output state}
 \hat{\rho}_{\text{output}}\simeq \  \frac{1}{p_{\text{suc}}}  \sum_{j=0}^{k} q_j  \ket{m_{\theta_j}}\bra{m_{\theta_j}}_L,
\end{equation}
where $p_{\text{suc}}=\sum_{j=0}^k q_j$ represents the success probability of the post-selection process in the TMR protocol.
Here, $\ket{m_{\theta_0}}_L$ equals to the target resource state $\ket{m_{\theta_L}}_L$, and $\ket{m_{\theta_j}}_L$ ($j=1,2,\cdots,k$) denotes a non-target resource state that arises due to errors in order $\order*{p_{\text{ph}}^j}$. Their angles are specified as 
\begin{equation}
\label{eq: general fom of angle}
\begin{aligned}
    \theta_j   &\equiv (-1)^j  \sin^{-1}\left(
    \frac{|u_{k-j}|}{\sqrt{|u_{j}|^2+|u_{k-j}|^2}} \right) \simeq  \order{\theta^{k-j-n_j}},   
\end{aligned}
\end{equation}
where $u_{j} \equiv i^j\sin^{j}\theta \cos^{k-j}\theta$ and $n_j\equiv \min(j, k-j)$.
In particular, the term of $j=1$ equals to the leading error term in Eq.~\eqref{eq:general fom of output state}, namely, $\theta_1=\theta_{\text{error}}$.

The coefficients $\{q_j\}_{j=0,1,\cdots,k}$ in Eq.~\eqref{eq:general fom of output state} represent the probabilities that the non-Clifford state $\ket{m_{\theta_{j}}}_L$ is post-selected via stabilizer measurements. 
These coefficients can be decomposed into two types of the contributions as $q_j=q_j^{\text{sample}}\cdot q_j^{\text{pass}}$. Here, $q_j^{\text{sample}}$ denotes the sum of the probability amplitude of $\hat{C}\ket{\psi_b}$ with $|b|=j$ or $|\bar{b}|=j$ in Eq.~\eqref{eq:noisy terms}. This is explicitly given as $q_j^{\text{sample}}\equiv  {}_kC_j(|u_{j}|^2+|u_{k-j}|^2)\simeq \order{\theta^{2n_j}}$.
Whereas, $q_j^{\text{pass}}$ denotes the sum of the probability that the normalized state of $\hat{C}\ket{\psi_b}$ with $|b|=j$ or $|\bar{b}|=j$ is projected to $\ket{m_{\theta_j}}_L$ after the stabilizer measurements of the TMR protocol under physical noises. This factor is independent of the input angle $\theta$ (or the target angle $\theta_L$) and in the order of $\mathcal{O}(p_{\text{ph}}^{j})$. The concrete value of $q_j^{\text{pass}}$ can be determined via numerical approaches illustrated in Refs.~\cite{Choi2023, Toshio2024}, if the explicit noise model is specified. In conclusion, the coefficient $q_j$ scales as $\mathcal{O}(\theta^{2n_j}p_{\text{ph}}^{j})$.

Based on the formula in Eq.~\eqref{eq:general fom of output state}, we can modify the quantum channel in Eq.~\eqref{eq:noisy rotation channel in the first trial} to include higher-order errors as follows:
\begin{equation}
\label{eq:}
\begin{aligned}
    \mathcal{N}_{\theta_L}(\hat{\rho}) =&\  \mathcal{E}_{\theta_L} \circ \mathcal{R}_{\theta_L}(\hat{\rho})
    = \sum_{j=0}^k \bar{q}_j \cdot \mathcal{R}_{\theta_j}(\hat{\rho}) ,
\end{aligned}
\end{equation}
where $\bar{q}_j=q_j/p_{\text{suc}}$ ($\sum_{j=0}^k \bar{q}_j=1$) and $\mathcal{E}_{\theta_L}$ denotes the following stochastic error channel:
\begin{equation}
\label{eq:}
\begin{aligned}
    \mathcal{E}_{\theta_L}(\hat{\rho}) = \left(1-\sum_{j=1}^k \bar{q}_j \right)\cdot\hat{\rho} + \sum_{j=1}^k \bar{q}_j \cdot \mathcal{R}_{\Delta_{j}}(\hat{\rho}).
\end{aligned}
\end{equation}
Here, $\Delta_{j}(\theta_L)= \theta_j(\theta_L) - \theta_L$ represents the angle of the over-rotation error that arises due to physical errors in order of $\mathcal{O}(p_{\text{ph}}^j)$. If we neglect the higher-order terms of $j>2$, these equations result in Eq.~\eqref{eq:noisy rotation channel in the first trial} and Eq.~\eqref{eq:our error channel}.

Next, to remove the higher-order coherent terms in $\mathcal{N}_{\theta_L}$, we introduce the following stochastic operation:
\begin{equation}
\label{eq:higher-order cancellation channel}
\begin{aligned}
    \mathcal{B}_{\theta_L}(\hat{\rho}) &\equiv \left(1-\sum_{j=1}^k \bar{q}_j \right)\cdot\hat{\rho} + \sum_{j=1}^k \bar{q}_j \cdot \mathcal{R}_{-\Delta_{j}}(\hat{\rho}).
\end{aligned}
\end{equation}
Importantly, unlike in Eq.~\eqref{eq:cancellation channel}, this channel directly uses an ideal rotation channels $\{ \mathcal{R}_{-\Delta_{j}} \}_{j=1,2,\cdots,k}$ for avoiding the sub-leading error in order of
$\tilde{\mathcal{O}}(\theta_L p_{\text{ph}}^2)$ in Eq.~\eqref{eq:cancellation channel}. This term is not essential in the formulation of Ref.~\cite{Toshio2024} because the RUS trials with relatively large angles dominate the errors accumulated during the RUS process. However, in our proposal, we can avoid such RUS trials and thereby $\theta_{\text{RUS}}$ can be smaller than $p_{\text{ph}}$. This makes the terms of $\tilde{\mathcal{O}}(\theta_L p_{\text{ph}}^2)$ non-negligible compared to the terms of $\mathcal{O}(\theta_L^{2(1-1/k)}p_{\text{ph}})$.
In this work, we always assume that the $\mathcal{R}_{-\Delta_{j}}$ in Eq.~\eqref{eq:higher-order cancellation channel} channel is performed with $T$-gate synthesis by using high-fidelity magic states generated via MSC or MSD.

By applying the $ \mathcal{B}_{\theta_L}$ channel after the noisy rotation channel $\mathcal{N}_{\theta_L}$, we obtain a composed quantum channel as follows:
\begin{equation}
    \mathcal{N}^{b}_{\theta_L}(\hat{\rho})\ \equiv\  \mathcal{B}_{\theta_L} \circ \mathcal{N}_{\theta_L}(\hat{\rho}) \ =\  \mathcal{E}_{\theta_L}^{b} \circ \mathcal{R}_{\theta_L}(\hat{\rho}),
\end{equation}
where an effective error channel for $\mathcal{N}^{b}_{\theta_L}$ is defined as $\mathcal{E}_{\theta_L}^{b} \equiv \mathcal{B}_{\theta_L} \circ \mathcal{E}_{\theta_L}$.
Then, we can explicitly decompose this error channel as 
\begin{equation}
\label{eq:canceled rotation gate for STARv3}
\begin{aligned}
    \mathcal{E}_{\theta_L}^{b}(\hat{\rho})\ &\simeq \ 
    \left(1-2\sum_{j=1}^k\bar{q}_j\sin^2(\Delta_{j}) \right)\cdot \hat{\rho}  \\
    & \qquad + 2\sum_{j=1}^k\bar{q}_j\sin^2(\Delta_{j})\cdot \hat{Z}\hat{\rho} \hat{Z} + \mathcal{O}(\theta_L^2 p_{\text{ph}}^2).
\end{aligned}
\end{equation}
Using the fact that $\theta_j \simeq \order{\theta^{k-j-n_j}}$ and $\bar{q_j}\simeq\mathcal{O}(\theta^{2n_j}p_{\text{ph}}^{j})$, the coefficient of the second terms are evaluated as 
\begin{equation}
    2\bar{q}_j\sin^2(\Delta_{j}) \simeq 2\bar{q}_j\Delta_{j}^2 \simeq   \mathcal{O}(\theta^{2(k-j)}p_{\text{ph}}^{j}) \simeq \mathcal{O}(\theta_L^{2(1-j/k)}p_{\text{ph}}^{j}).
\end{equation}
Therefore, as long as $\theta_L \gg p_{\text{ph}}^{k/2}$ is satisfied, we can regard the worst-case error rate of $\mathcal{E}_{\theta_L}^{b}$ to be $\mathcal{O}(\theta_L^{2(1-1/k)}p_{\text{ph}})$.

\subsection{STAR-magic mutation}
\label{sec:STAR-magic mutation}

\begin{algorithm}[tb]
\SetKwComment{Comment}{/* }{ */}
\SetKwInOut{Input}{Input}
\SetKwInOut{Output}{Output}
\caption{STAR-magic mutation}
\label{alg:STAR-magic mutation}
\Input{
\begin{itemize}
\renewcommand{\labelitemi}{}
    \item $\rho \gets$ Target quantum state
    \item $\theta_L \gets$ Target rotation angle
    \item $\theta_{\text{th}} \gets$ Threshold of rotation angle
    \item $\epsilon \gets$ Accuracy of $T$-gate synthesis
    \item $\{ \bar{q_j}(\theta_{\text{RUS}})\} \gets$ Probability functions of over-rotation errors
\end{itemize}
}
\Output{
\begin{itemize}
\renewcommand{\labelitemi}{}
    \item $\rho_{\mathrm{out}}\  \simeq \ \hat{R}_{z,L}(\theta_L)\rho\hat{R}_{z,L}^\dagger(\theta_L)$
\end{itemize}
}
\BlankLine
Set the RUS angle and output state: $\theta_{\text{RUS}}=\theta_L$, $\rho_{\mathrm{out}}=\rho$\;
\While{$\theta_{\mathrm{RUS}}<\theta_{\mathrm{th}}$}{
Prepare a resource state $\ket{m_{\theta_{\mathrm{RUS}}}}$ via the TMR protocol\;
Input $\rho_{\mathrm{out}}$ and $\ket{m_{\theta_{\mathrm{RUS}}}}$ into the gate-teleportation circuit: $\rho_{\mathrm{out}}\gets \hat{R}_{z,L}(\pm\theta_{\mathrm{RUS}})\rho_{\mathrm{out}}\hat{R}_{z,L}^\dagger(\pm\theta_{\mathrm{RUS}})$\;
\If{the gate-teleportation succeeded}{
Sample an over-rotation channel $\mathcal{R}_{-\Delta_j(\theta_{\text{RUS}})}$ according to the probabilities $\{ \bar{q_j}(\theta_{\text{RUS}})\}$ and apply it to $\rho_{\mathrm{out}}$: $\rho_{\mathrm{out}}\gets \hat{R}_{z,L}(-\Delta_j)\rho_{\mathrm{out}}\hat{R}_{z,L}^\dagger(-\Delta_j)$\;
\Return $\rho_{\mathrm{out}}$
}
Sample an over-rotation channel $\mathcal{R}_{\Delta_j(\theta_{\text{RUS}})}$ according to the probabilities $\{ \bar{q_j}(\theta_{\text{RUS}})\}$ and apply it to $\rho_{\mathrm{out}}$: $\rho_{\mathrm{out}}\gets \hat{R}_{z,L}(\Delta_j)\rho_{\mathrm{out}}\hat{R}_{z,L}^\dagger(\Delta_j)$\;
Update the RUS angle: $\theta_{\text{RUS}} \gets 2\theta_{\text{RUS}}$\;
}
Decompose $\hat{R}(\theta_{\text{RUS}})$ into a gate series $\{G_i\}_{i=1,\cdots,N_g}$\;
\For{$G_i$ in $\{G_i\}_{i=1,\cdots,N_g}$}{
\If{$G_i$ is Clifford}{
Perform $G_i$ with lattice surgery: $\rho_{\mathrm{out}}\gets G_i\rho_{\mathrm{out}}G_i^\dagger$\;
}
\If{$G_i$ is $R_P(\pi/8)$}{
Prepare a magic state $\ket{m}$ via MSC or MSD\;
Input $\rho_{\mathrm{out}}$ and $\ket{m}$ into the gate-teleportation circuit: $\rho_{\mathrm{out}}\gets G_i\rho_{\mathrm{out}}G_i^\dagger$\;
}
}
\Return $\rho_{\mathrm{out}}$
\end{algorithm}

\begin{figure*}
    \centering
    \includegraphics[width=0.8\linewidth]{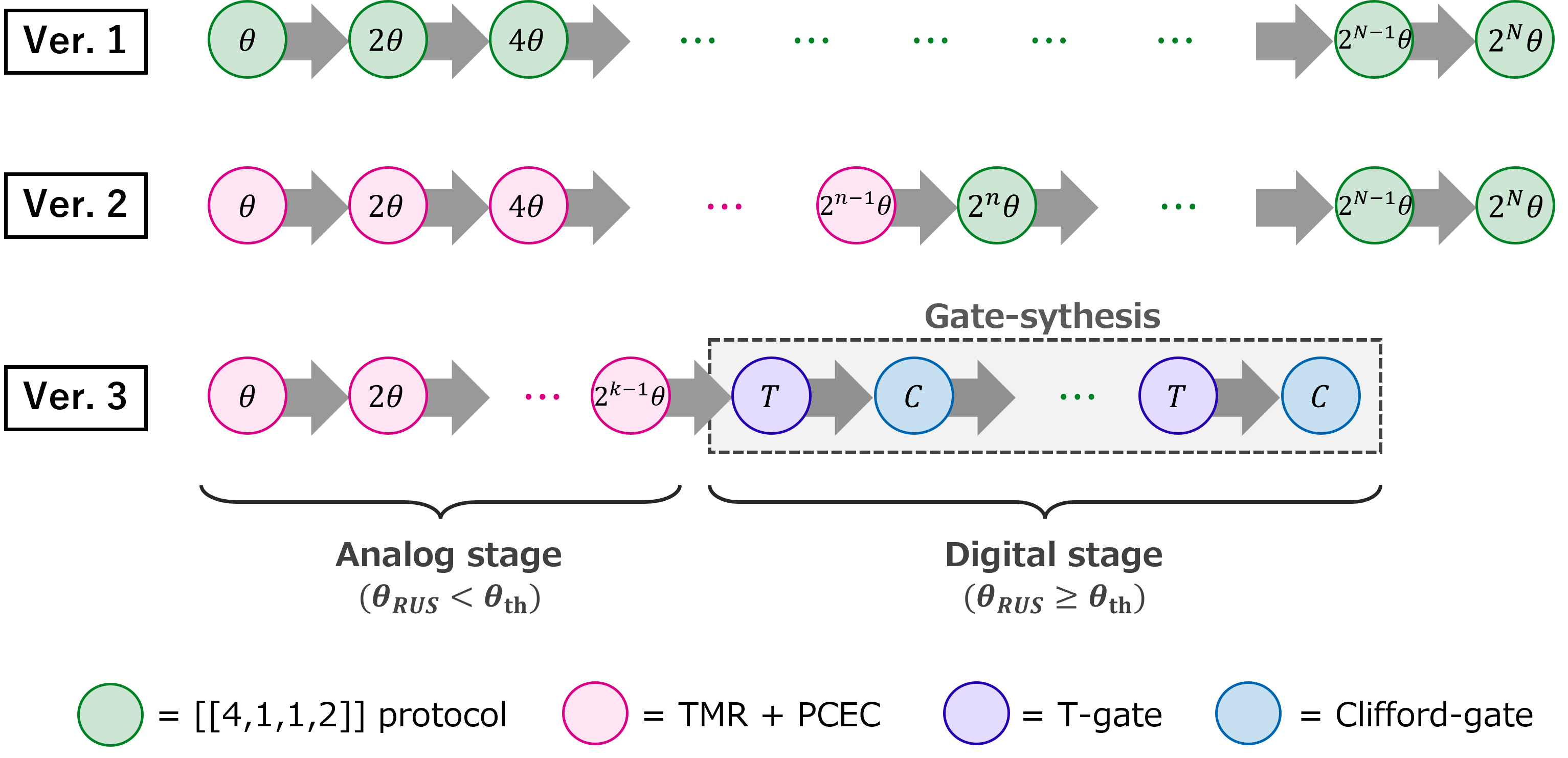}
    \caption{STAR-magic mutation and the previous approaches~\cite{Akahoshi2023, Toshio2024} for directly executing a logical rotation gate with an arbitrary angle. In the original proposal of the STAR architecture (STAR ver.~1)~\cite{Akahoshi2023}, any analog rotation gate is executed using an injection protocol based on the [[4,1,1,2]] subsystem code (green circle). Whereas, STAR ver.~2 employs a more refined approach based on the TMR protocol~\cite{Toshio2024}. In this approach, the initial trials of RUS are executed using the TMR and PCEC protocols (pink circle). Once the RUS angle exceeds a specific threshold, the subsequent RUS trials are executed by using the injection protocol in STAR ver.~1. On the other hand, STAR-magic mutation (or STAR ver.~3) adopts another type of adaptive switching during the RUS process. In this protocol, as with STAR ver.~2, we perform the RUS trials using the TMR and PCEC protocols as long as the RUS angle is smaller than a predefined threshold $\theta_{\text{th}}$ (analog stage). However, in STAR ver.~3, PCEC is modified to suppress the higher-order errors in resource states as discussed in Sec.~\ref{sec:higher-order}.  Then, once the RUS angle $\theta_{\text{RUS}}$ gets greater than or equal to the threshold angle $\theta_{\text{th}}$ (digital stage), we perform the next rotation gate using a standard $T$-gate synthesis (purple and blue circles). In this stage, each $T$-gate is executed by preparing a magic state via MSC or MSD.}
    \label{fig:STAR_history}
\end{figure*}

Eventually, we are ready to introduce {\it STAR-magic mutation} (SMM), a novel protocol for implementing a logical analog rotation gate with a small spacetime overhead. This protocol consists of the following two stages:
\begin{itemize}
    \item {\bf Analog stage}: First execute the RUS process with the TMR protocol as long as the RUS angle $\theta_{\text{RUS}}$ remains below the threshold angle $\theta_{\text{th}}$ ($\theta_{\text{RUS}}<\theta_{\text{th}}$). Here, $\theta_{\text{th}}$ is a tunable parameter and controls the trade-off between the output fidelity and time overhead of SMM. After each RUS trial is completed, apply the $\mathcal{B}_{\theta_{\pm\text{RUS}}}$ channel for PCEC. 
    \item {\bf Digital stage}: Once the RUS angle $\theta_{\text{RUS}}$ gets greater than or equal to the threshold angle $\theta_{\text{th}}$ ($\theta_{\text{RUS}}\geq\theta_{\text{th}}$), we deterministically perform the next RUS trial of analog rotation gate with $T$-gate synthesis. Each $T$-gate in the synthesis is executed by preparing magic states via MSC or MSD.
\end{itemize}
The more detailed description of SMM is shown in Algorithm~\ref{alg:STAR-magic mutation}. A similar two-stage approach is employed in Ref.~\cite{Toshio2024}, where the injection protocol in Ref.~\cite{Akahoshi2023} is used for the large-angle region in the RUS process, instead of $T$-gate synthesis. Importantly, this previous approach can reduce the total error rate of the resulting rotation gate, but cannot improve the scaling with $\theta_L$.
In Fig.~\ref{fig:STAR_history}, we summarize these recent developments of the non-fault-tolerant implementation of analog rotation gates.

It should be noted here that the digital stage is rarely required in this protocol, as the RUS process typically succeeds before the RUS angle reaches the threshold.
The probability that we proceed to the digital stage is readily evaluated as $p_{\mathrm{switch}}=2^{-\lceil\log_2(\theta_{\mathrm{th}}/\theta_L)\rceil}\simeq \order{\theta_{\mathrm{th}}/\theta_L}$. 
Consequently, SMM can effectively mitigate the time overhead associated with $T$-gate synthesis and remain viable even when magic states are supplied at a low rate.
As mentioned in the previous section, MSC requires roughly 10 clocks under $p=10^{-3}$ to supply a magic state on a single patch. This is relatively slower than that of the TMR protocol, which requires 2 or 3 clocks to supply a resource state with $\theta_L\lesssim10^{-3}$ on average~\cite{Toshio2024}.

When we employ the MSC in the digital stage, we can carry out SMM with only an ancillary space of a single surface code patch. 
This property enables us to perform a locality-aware Pauli-based computation as discussed in Ref.~\cite{Hirano2024zerolevel, Akahoshi2024}. Namely, we can execute multiple rotation gates in parallel by leveraging the ability to prepare resource or magic states locally. For example, in Ref.~\cite{Akahoshi2024}, the time-overhead for Trotter simulation of the Hubbard model is substantially reduced from $\mathcal{O}(N)$ to $\mathcal{O}(\sqrt{N})$ by combining the locality of resource state preparation and the fermionic swap network technique~\cite{Babbush2018fswap,Kivlichan2018}.

In contrast, as an alternative to cultivation, we can also utilize standard MSD techniques~\cite{Bravyi2005,Litinski2019magic} during the digital stage. In this case, while we can no longer perform the STAR-magic mutation locally, we can arbitrarily improve the fidelity and the supply rate of magic states by scaling up the size of magic state factories. Assuming such a situation makes the error analysis of SMM easier, enabling us to show the following theorem:

\begin{thm}[Asymptotic behavior of STAR-magic mutation under the supply of noiseless magic states]
\label{thm:scaling}
  Let us assume an ideal case where we can supply noiseless magic states in the digital stage $(p_m=0)$. In this case, as long as $\theta_L \gg p_{\text{ph}}^{k/2}$, the STAR-magic mutation achieves an analog rotation gate $R_{z,L}(\theta_L)$ with the following scaling of the worst-case error rate:
  \begin{equation}
      P_L(\theta_L)\simeq \mathcal{O}(\theta_L^{2(1-1/k)}p_{\mathrm{ph}})\ \ \ \ \ \ \  (r:\mathrm{fixed}; \ \theta_L\to 0)
  \end{equation}
  \begin{equation}
      P_L(\theta_L)\simeq \mathcal{O}(\theta_L p_{\rm{ph}})\ \ \ \ \ \ \  (\theta_{\mathrm{th}}:\mathrm{fixed};\  \theta_L\to 0)
  \end{equation}
  where $r=\theta_{\mathrm{th}}/\theta_L$ is the ratio between the threshold angle $\theta_{\mathrm{th}}$ and the target angle $\theta_L$.  
\end{thm}

\noindent
{\it --- Proof.} 
Given a supply of noiseless magic states, we only need to analyze the accumulation of errors during the analog stage.
According to Sec.~\ref{sec:higher-order}, the worst-case error rate of a single RUS trial is determined as $2\bar{q}_1(\theta_{\text{RUS}})\sin^2(\Delta_{1}(\theta_{\text{RUS}}))\simeq \mathcal{O}(\theta_{\text{RUS}}^{2(1-1/k)}p_{\text{ph}})$ in the leading order as long as $\theta_L \gg p_j^{k/2}$.
Consequently, the effective error rate during the analog stage $P_L$ is generally determined as follows:
\begin{equation}
\label{eq:effective error rate}
    P_L(\theta_L) = \sum_{m=1}^{N_{\text{RUS}}} \left( \frac{1}{2}\right)^m P_L^{(m)}(\theta_L)
\end{equation}
\begin{equation}
    P_L^{(m)}(\theta_L) = \sum_{i=0}^{m-1}  2\bar{q}_1(2^i\theta_L)\sin^2(\Delta_{1}(2^i\theta_L))
\end{equation}
where $N_{\text{RUS}}=\lceil\log_2(r)\rceil$ and $P_L^{(m)}(\theta_L)$ denotes the accumulated error when the RUS process succeeds at the $m$-th trial. Here, we implicitly neglect the approximation error of $T$-gate synthesis for simplicity, assuming a sufficiently large number of $T$-gates are consumed.

When the ratio $r$ is fixed, $N_{\text{RUS}}$ does not depend on $\theta_L$. As a result, $P_L(\theta_L)$ scales as $\mathcal{O}(\theta_L^{2(1-1/k)}p_{\mathrm{ph}})$ while it involves a large prefactor.
Meanwhile, when the threshold value $\theta_{\mathrm{th}}$ is fixed, $N_{\text{RUS}}$ scales as $\mathcal{O}(\log_2(1/\theta_{\mathrm{th}}))$. In this case, the term of $m=N_{\text{RUS}}$ becomes most dominant in Eq.~\eqref{eq:effective error rate}. Considering that $(1/2)^{N_{\text{RUS}}}\simeq \mathcal{O}(\theta_{\mathrm{th}})$ and $P_L^{(N_{\text{RUS}})}(\theta_L)\simeq \mathcal{O}(\theta_{\text{th}}^{2(1-1/k)}p_{\mathrm{ph}})$, we find that $P_L(\theta_L) \simeq \mathcal{O}(\theta_L\theta_{\text{th}}^{2(1-1/k)}p_{\mathrm{ph}})$. Since $\theta_{\text{th}}$ is independent on $\theta_L$, we can conclude that $P_L(\theta_L)\simeq \mathcal{O}(\theta_L p_{\mathrm{ph}})$. $\square$

This result represents a significant improvement over the scaling achieved in the prior work~\cite{Toshio2024}, where the worst-case error rate of analog rotation gates is given in the order of $\mathcal{O}(\theta_L p_{\rm{ph}})$. As shown in the next subsection, assuming a finite error rate of magic states ($p_m\neq0$), the ideal scaling of $\mathcal{O}(\theta_{\text{RUS}}^{2(1-1/k)}p_{\text{ph}})$ in Theorem.~\ref{thm:scaling} is violated when the target angle $\theta_L$ falls below a specific value. 
Nonetheless, STAR-magic mutation can achieve a significantly smaller error rate for analog rotation gates compared to the approach used in STAR ver.~2.

\begin{figure*}
    \centering
  \begin{tabular}{lll}
{\normalsize (a)} & {\normalsize (b)} & {\normalsize (c)} \\
   \includegraphics[width=5.8cm]{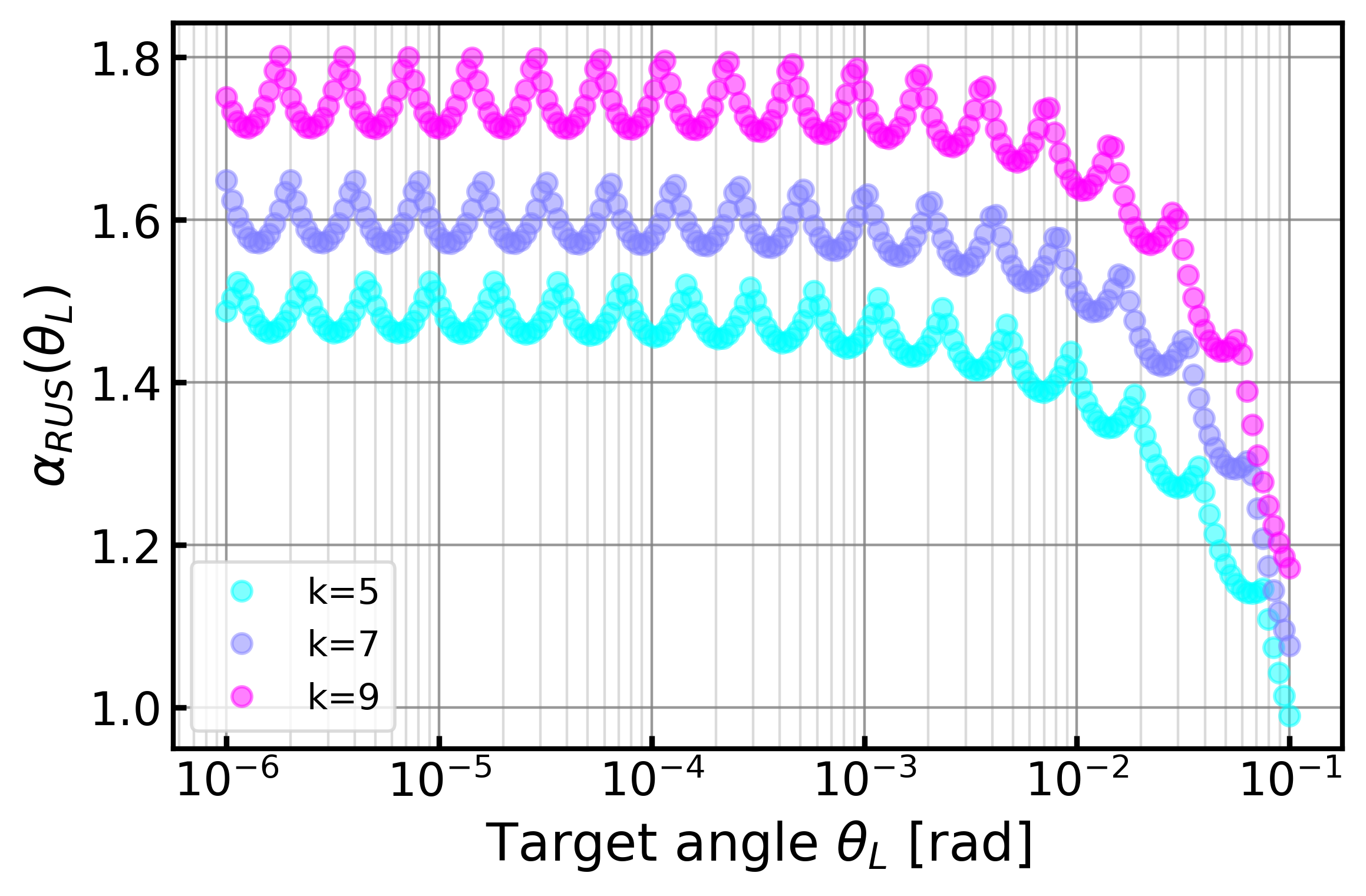}&
   \includegraphics[width=5.8cm]{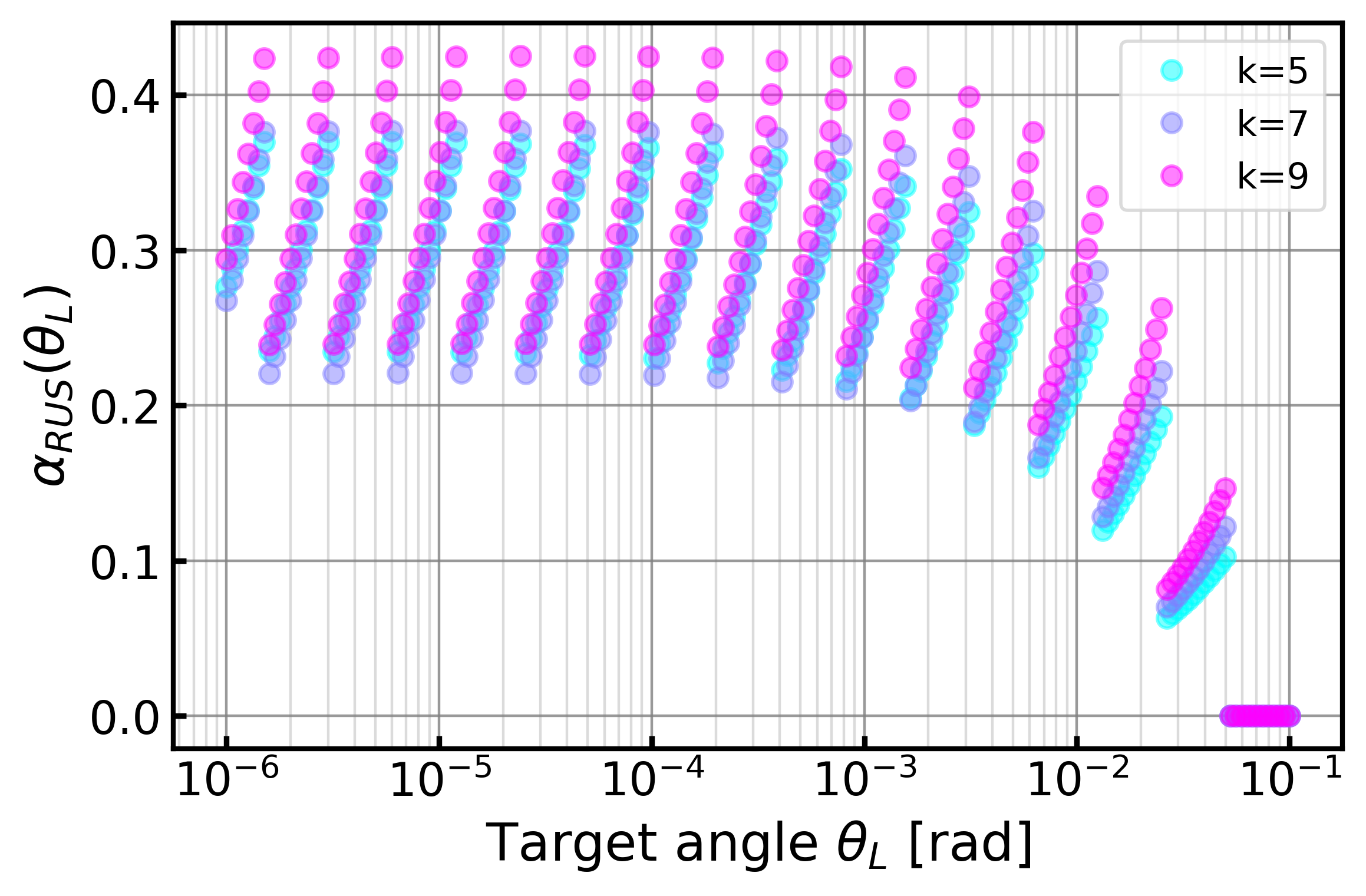}&
   \includegraphics[width=5.8cm]{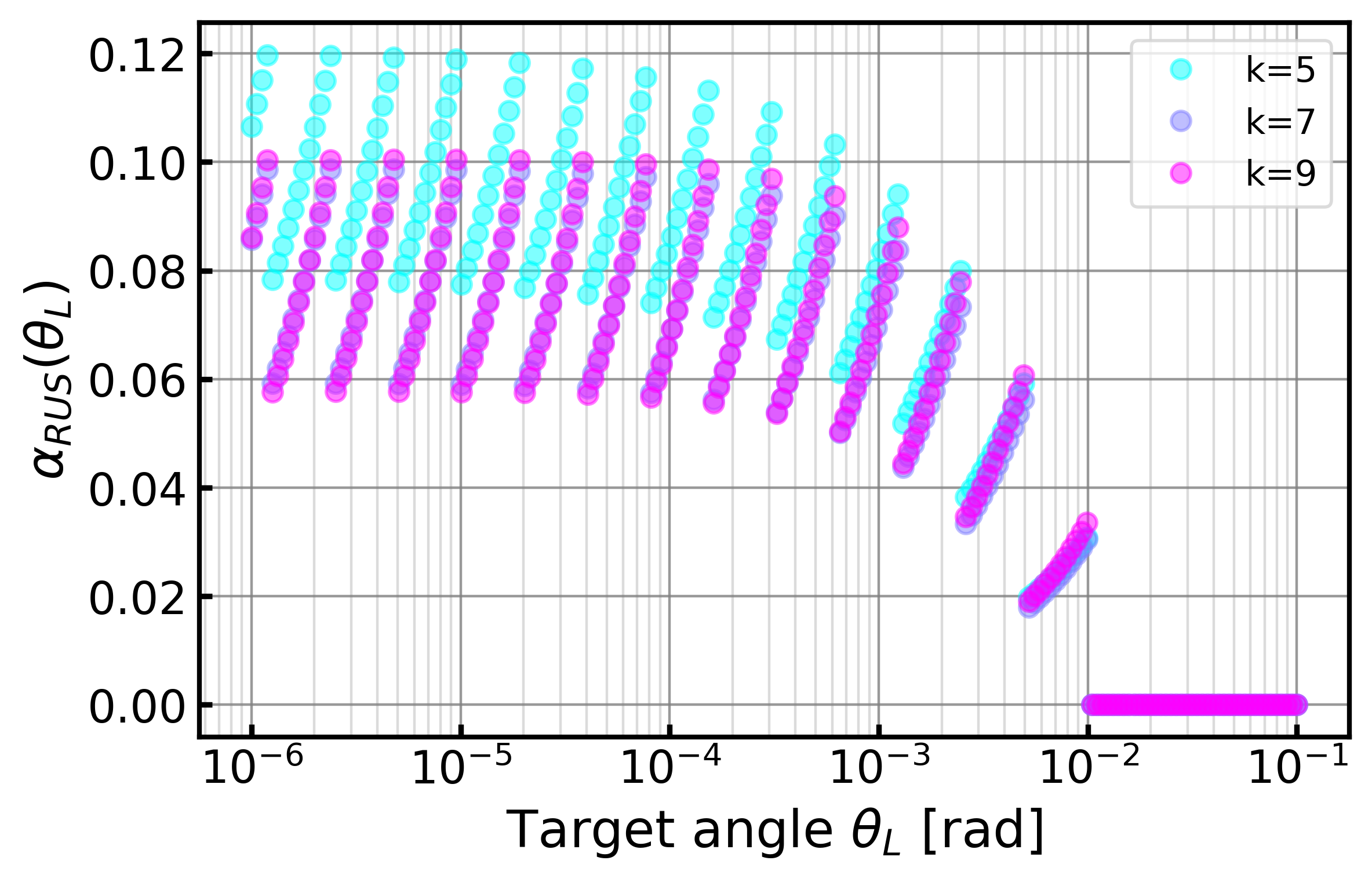}\\
{\normalsize (d)} & {\normalsize (e)} & {\normalsize (f)} \\
   \includegraphics[width=5.8cm]{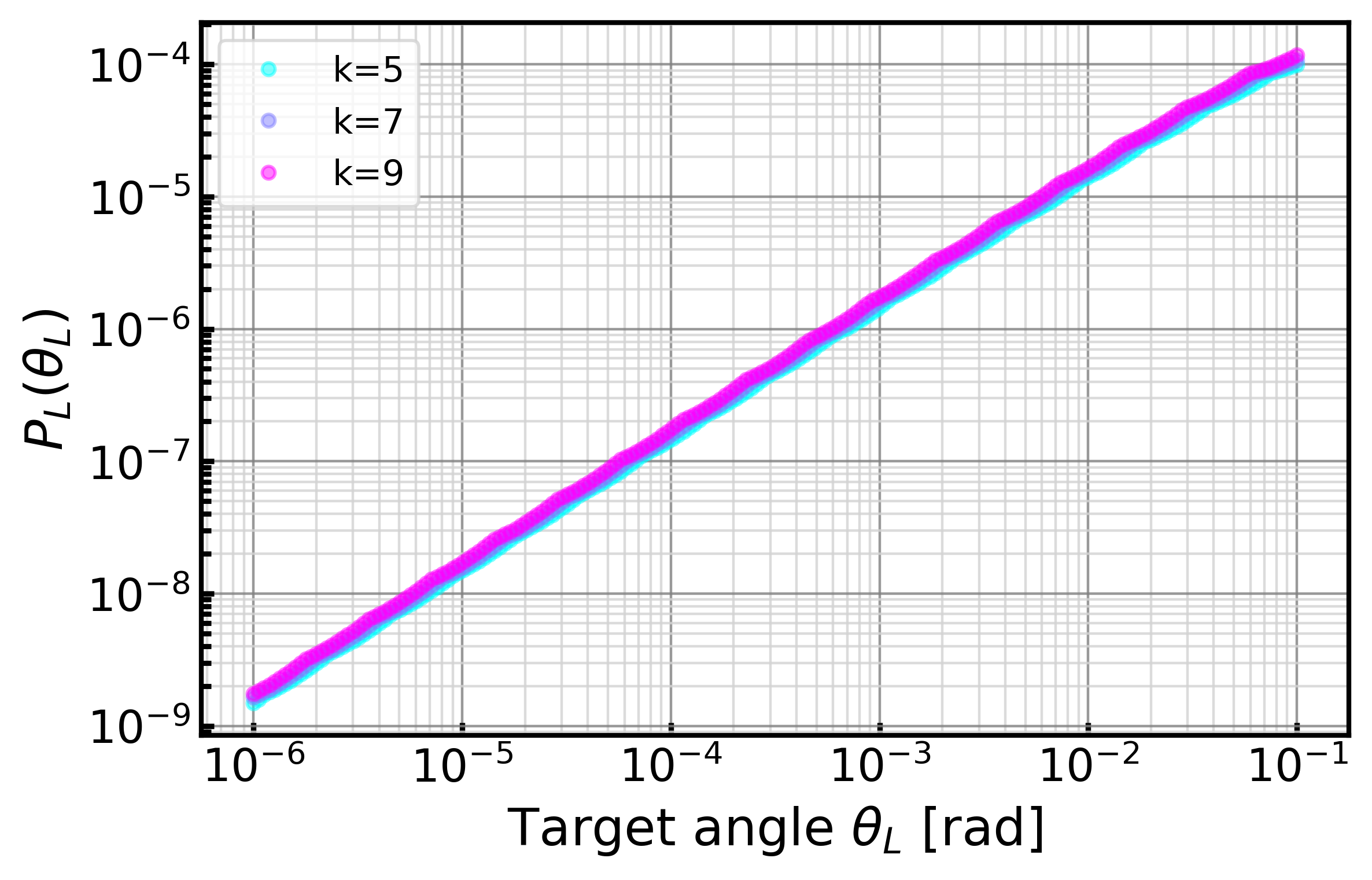}&
   \includegraphics[width=5.8cm]{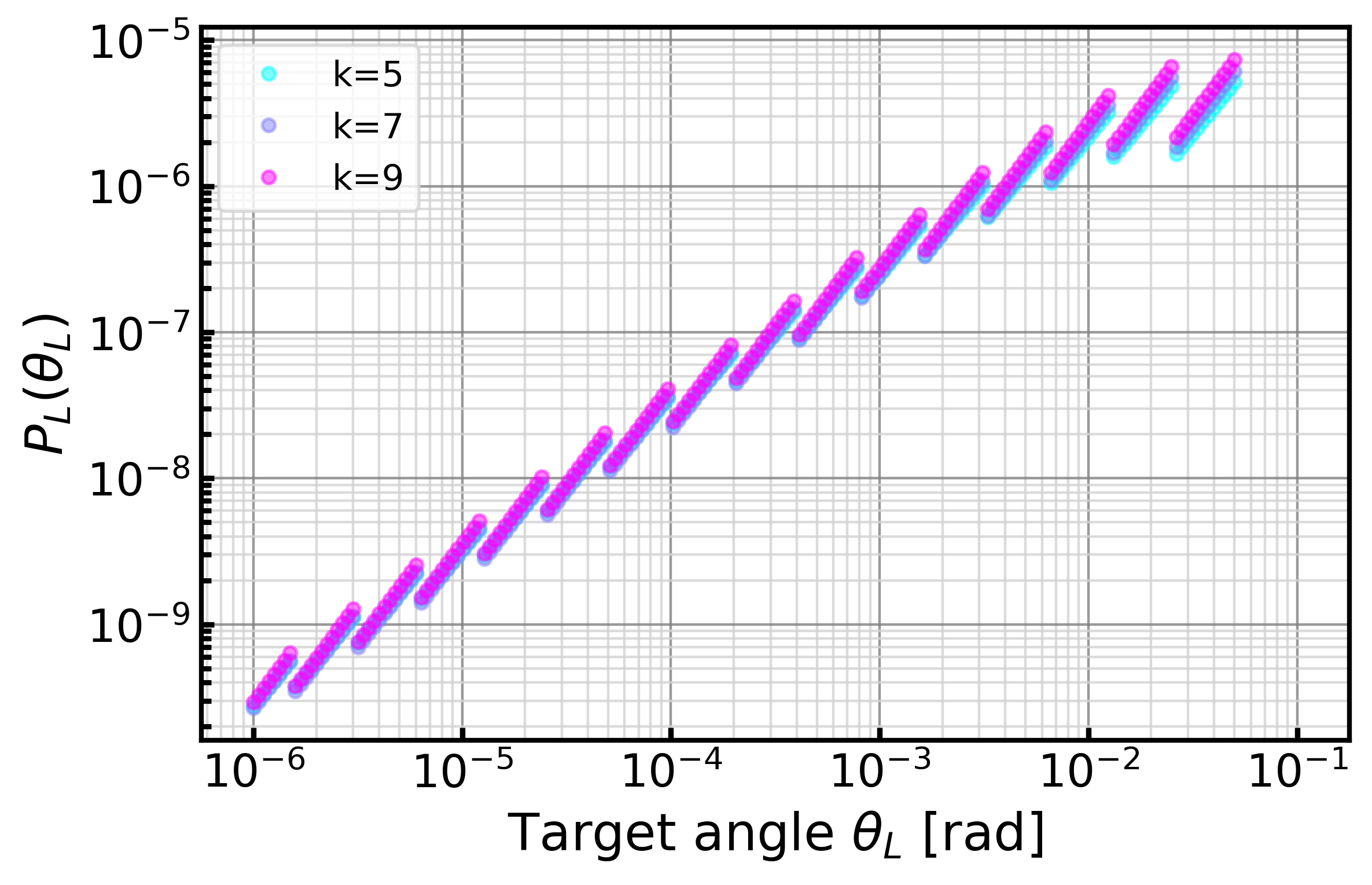}&
   \includegraphics[width=5.8cm]{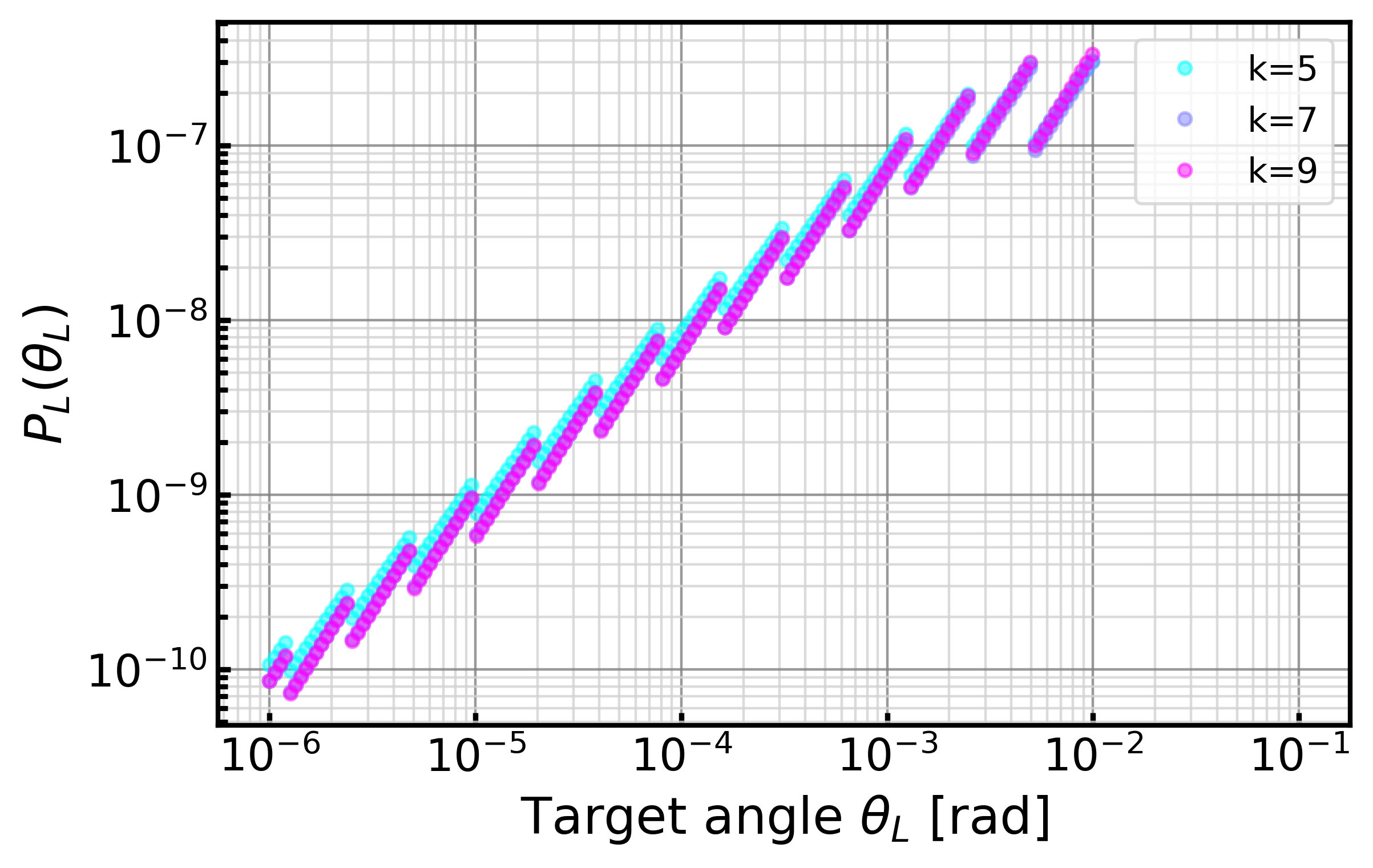}\\
  \end{tabular}
    \caption{The RUS factor $\alpha_{\text{RUS}}$ and the effective error rate $P_L$ of STAR-magic mutation under $p_{\text{ph}}=10^{-3}$ and $p_m=0$. {\bf (a-c)} The RUS factor for (a) the previous approach (STAR ver.~2)~\cite{Toshio2024} or STAR-magic mutation with (b) $\theta_{\mathrm{th}}=0.05$ or (c) $\theta_{\mathrm{th}}=0.01$. {\bf (d-f)} The effective error rate for the same setups with (a-c), respectively.}
    \label{fig:error_rate_w_fixed_threshold}
\end{figure*}

\begin{figure*}
    \centering
  \begin{tabular}{lll}
{\normalsize (a)} & {\normalsize (b)} & {\normalsize (c)} \\
   \includegraphics[width=5.8cm]{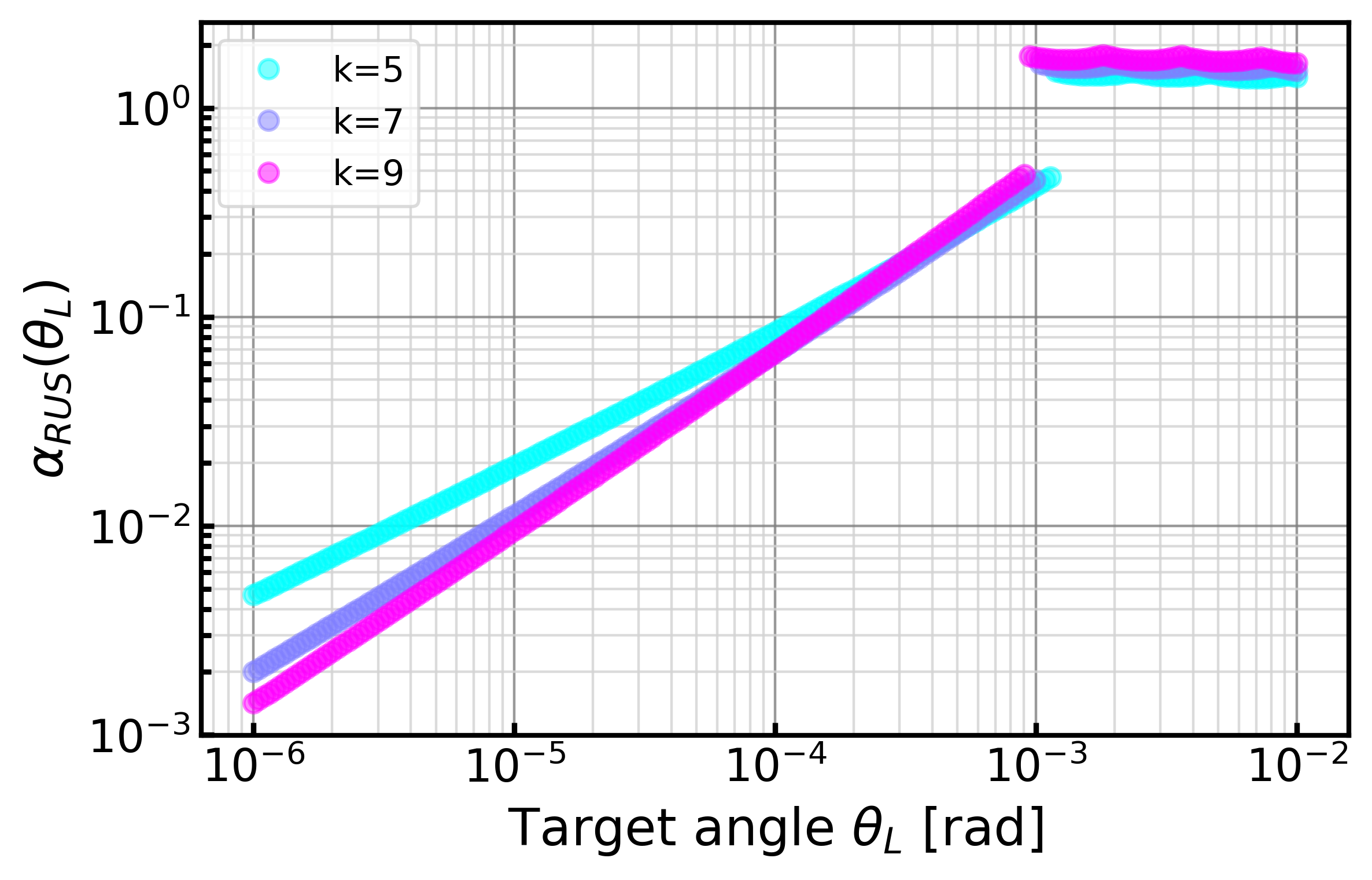}&
   \includegraphics[width=5.8cm]{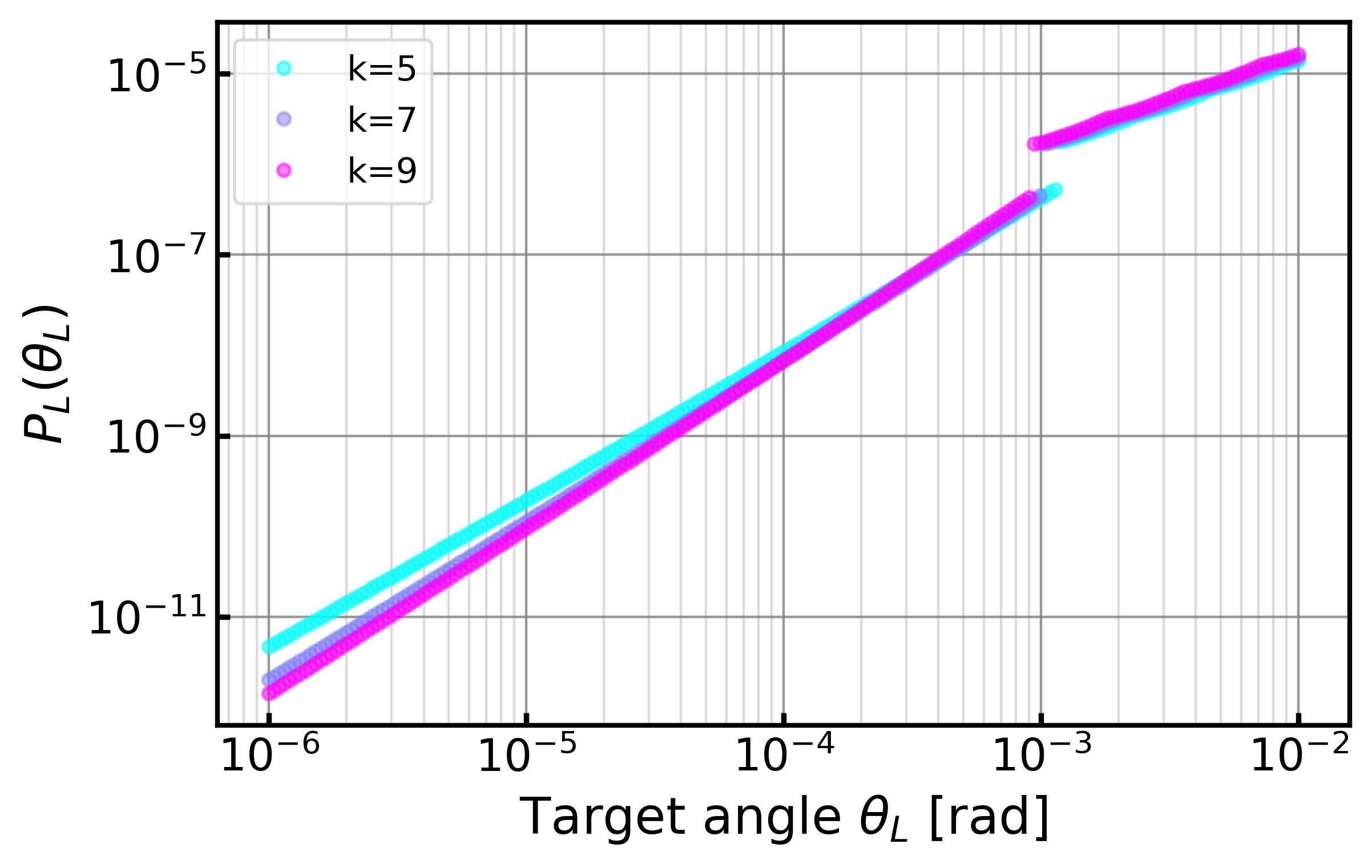}&
   \includegraphics[width=5.8cm]{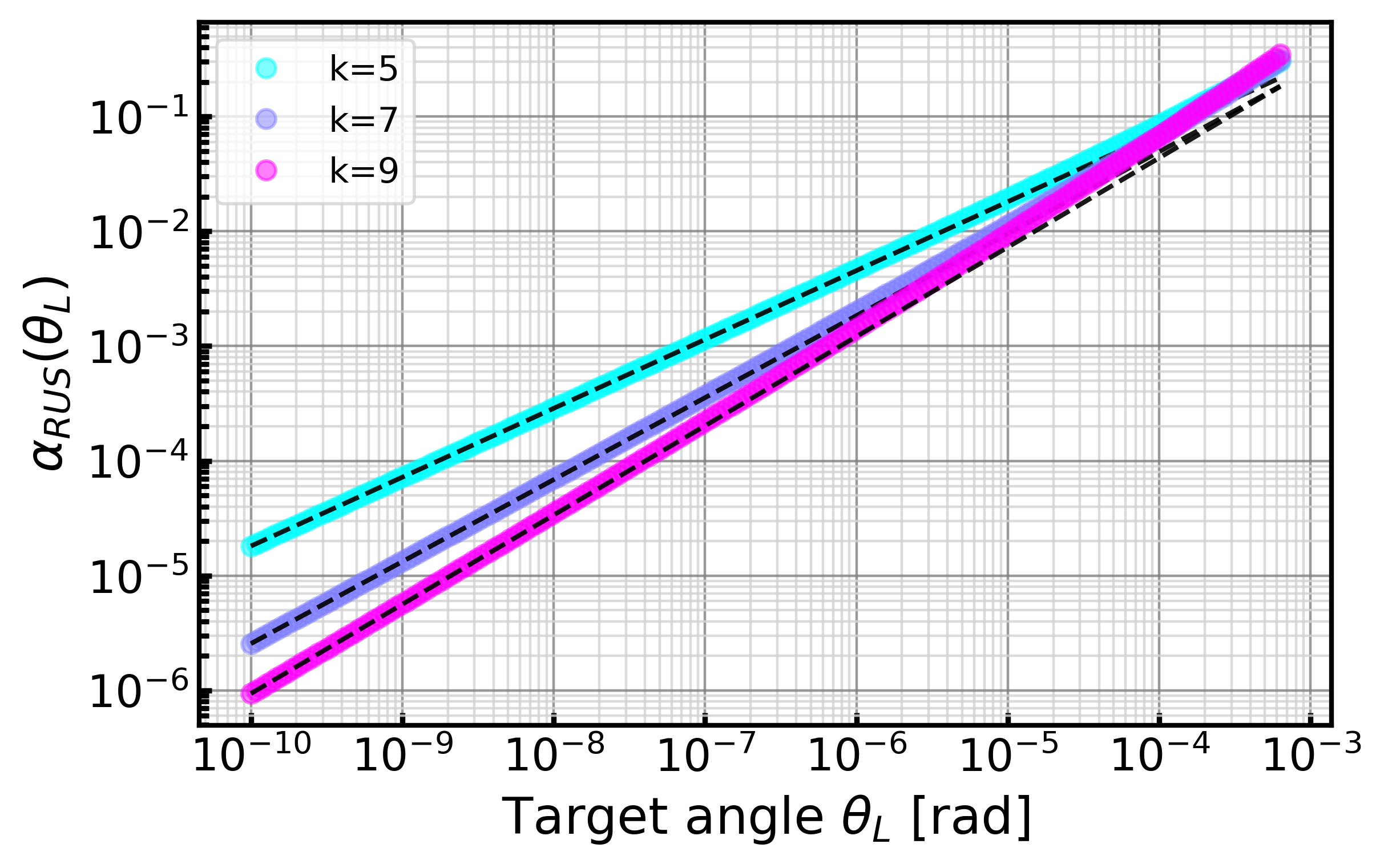}\\
  \end{tabular}
    \caption{(a) The RUS factor $\alpha_{\text{RUS}}$ and (b) the effective error rate $P_L$ of STAR-magic mutation with a fixed ratio $r=2^7=128$, i.e., $\theta_{\text{th}}=128\theta_L$. Here we again assume $p_{\text{ph}}=10^{-3}$ and $p_m=0$. (c) Asymptotic behavior of RUS factor $\alpha_{\text{RUS}}$ in the small-angle limit. The dotted lines denote the scaling expected from Theorem~\ref{thm:scaling}, namely, $\alpha_{\text{RUS}}(\theta_L)\propto\theta_L^{1-2/k}$.}
    \label{fig:error_rate_w_fixed_ratio}
\end{figure*}

\begin{figure*}
    \centering
  \begin{tabular}{lll}
{\normalsize (a)} & {\normalsize (b)} & {\normalsize (c)} \\
   \includegraphics[width=5.8cm]{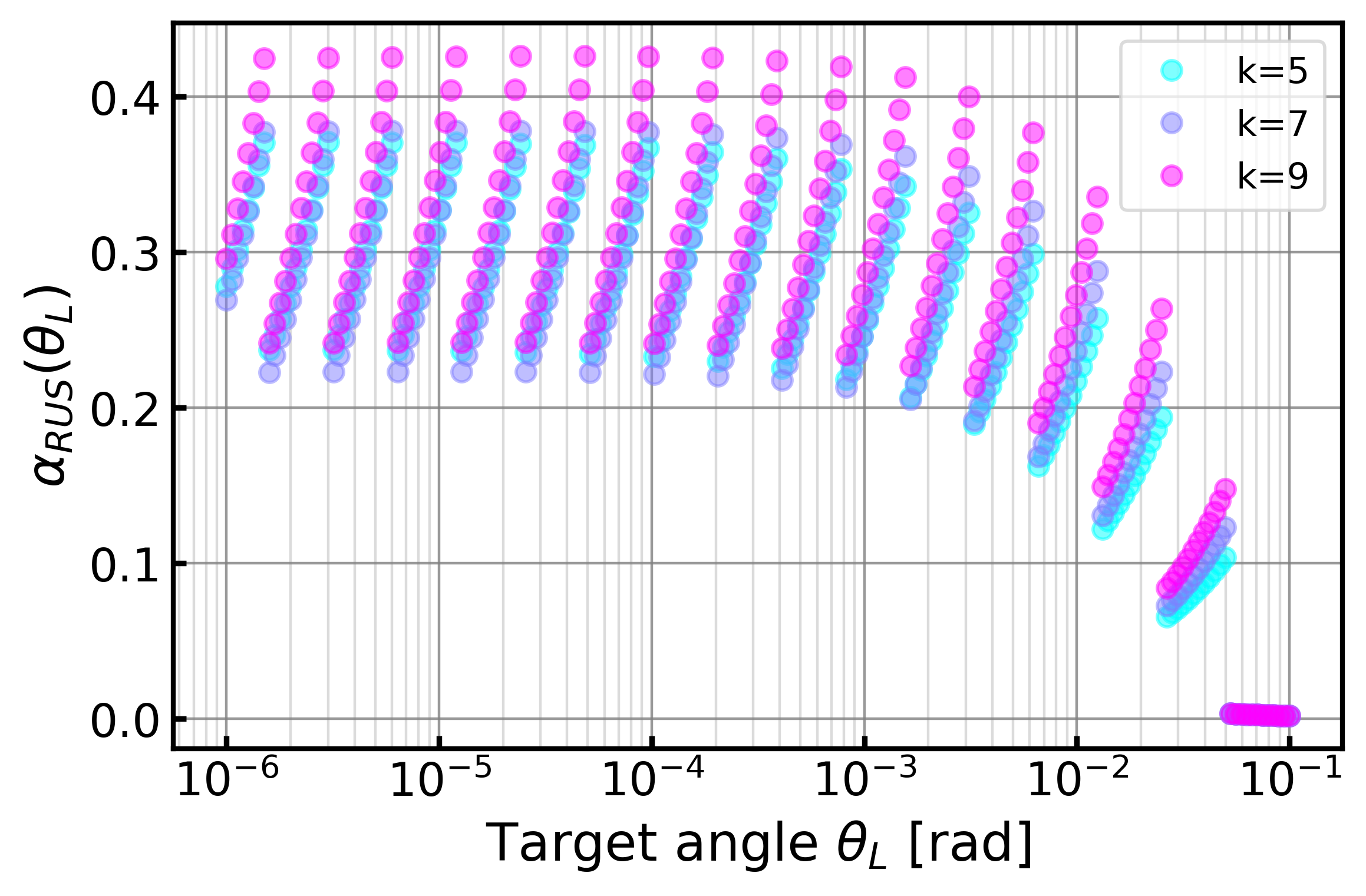}&
   \includegraphics[width=5.8cm]{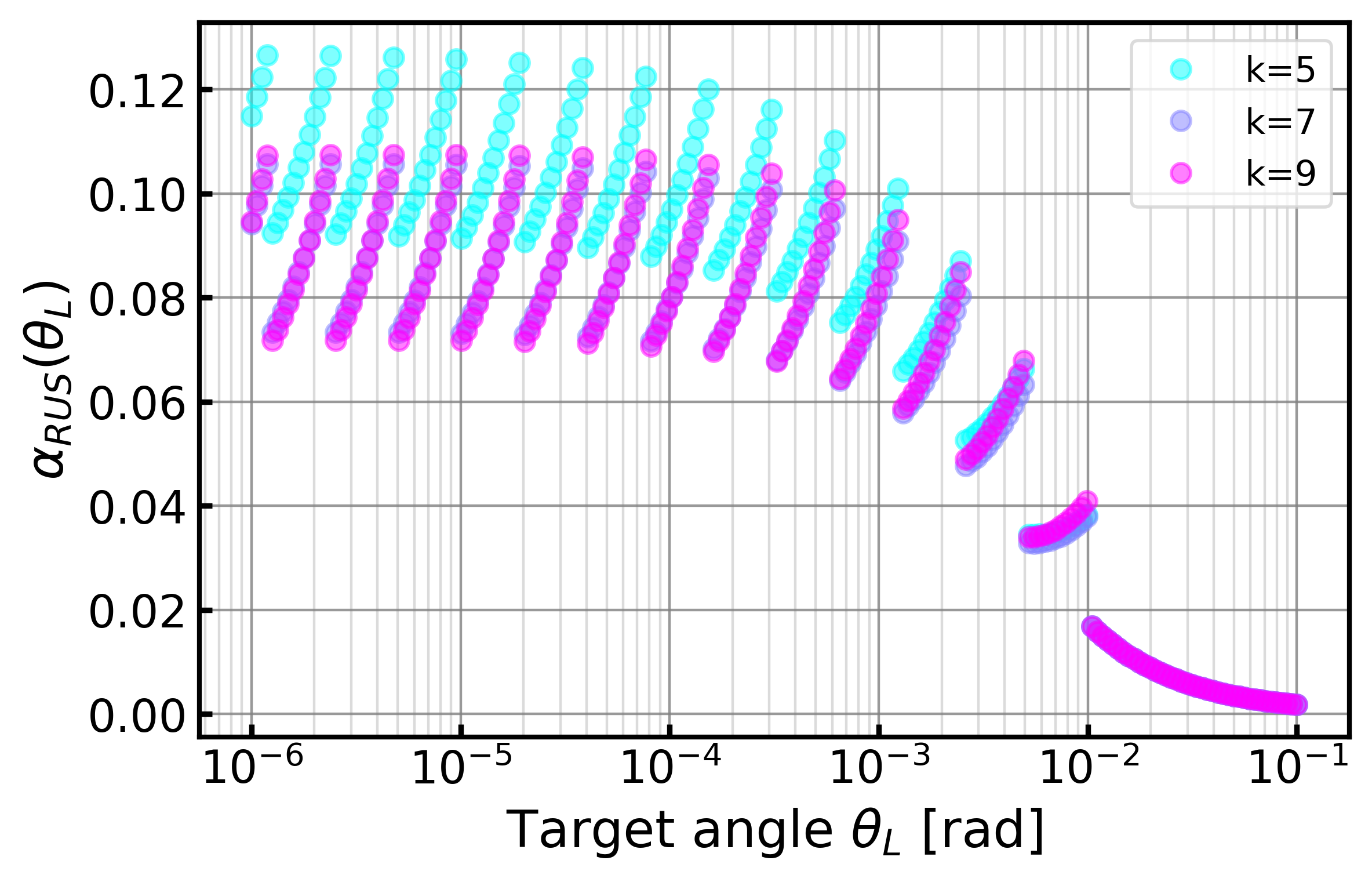}&
   \includegraphics[width=5.8cm]{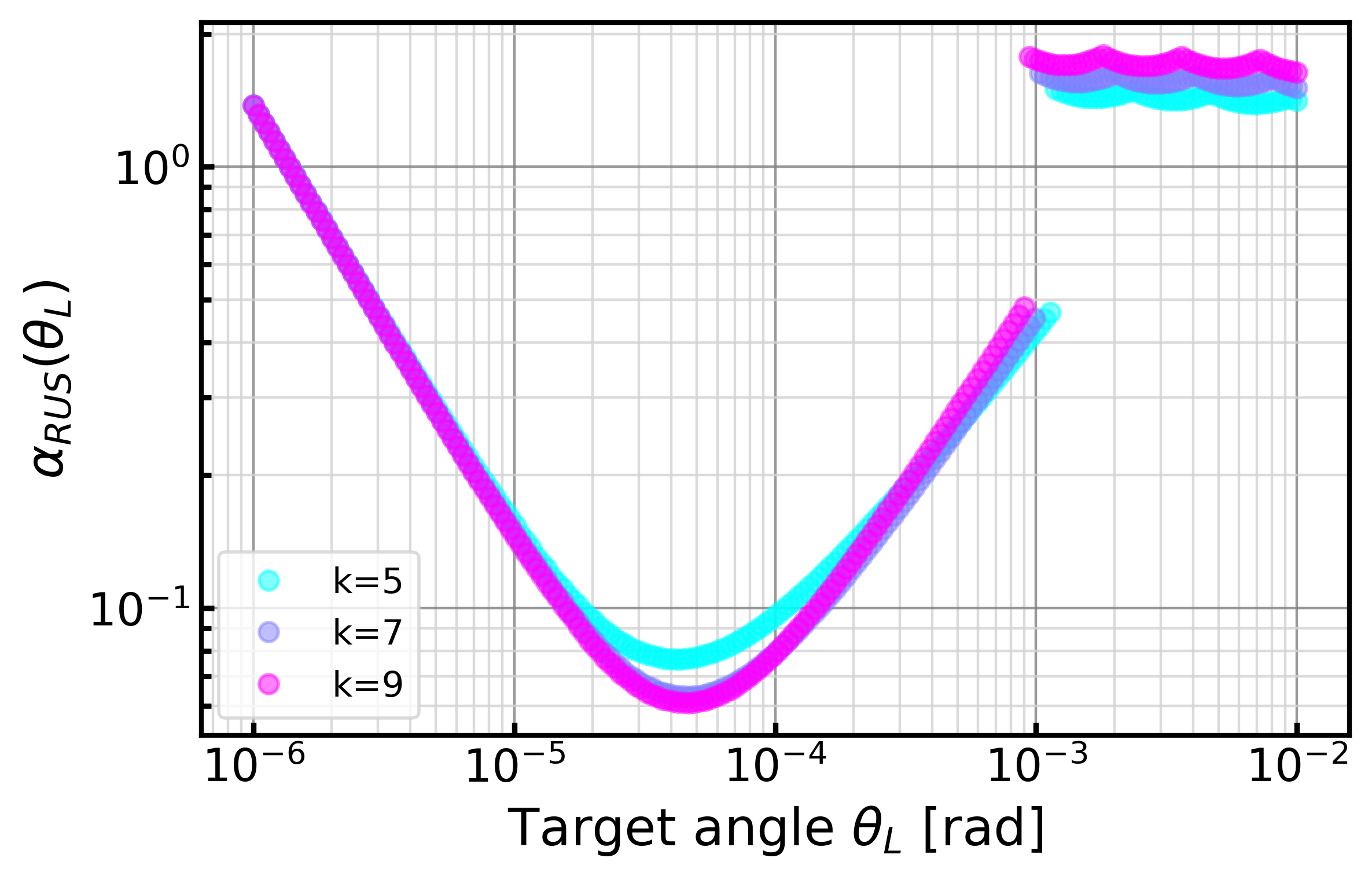}\\
  \end{tabular}
    \caption{The RUS factor $\alpha_{\text{RUS}}$ when the magic state has a finite error rate of $p_m=2\times10^{-9}$. (a,b) The RUS factor of STAR-magic mutation with a fixed threshold of $\theta_{\mathrm{th}}=0.05$ and $0.01$, respectively. (c) The RUS factor of STAR-magic mutation with a fixed ratio $r=2^7=128$, i.e., $\theta_{\text{th}}=128\theta_L$.}
    \label{fig:alpha_w_imperfect_magic_state}
\end{figure*}

\begin{figure}
    \centering
    \includegraphics[width=\linewidth]{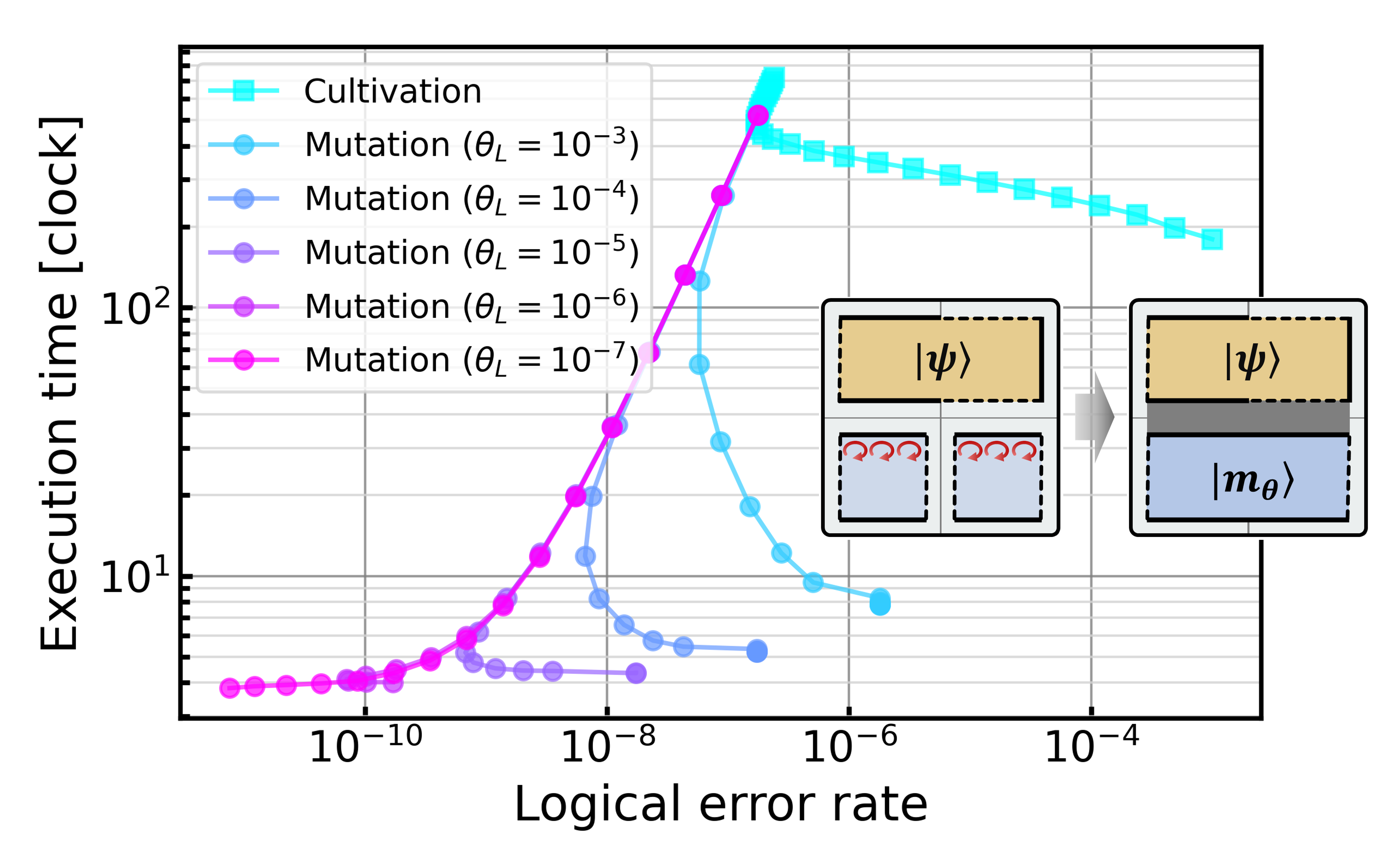}
    \caption{Tradeoff between logical error rate and execution time in STAR-magic mutation under $p_{\text{ph}}=10^{-3}$. We plotted the logical error rate and the execution time when we perform a Pauli-$Z$ rotation gate $R_{z,L}(\theta_L)$ with $\theta_L=10^{-k}$ $(k=3,4,\cdots,7)$ via the STAR-magic mutation, varying the value of threshold $\theta_{\text{th}}=2^n \theta_L$ $(n=0,1,2,\cdots, 15)$ (circle). In particular, the case of $n=0$ corresponds to the data point located at the upper end of each curve.
    For comparison, we also plotted the logical error rate and the execution time when performing $R_{z,L}(\theta_L)$ only with $T$-gate synthesis and magic state cultivation (square), where we varied the target accuracy in $T$-gate synthesis for each square point. The error rate of each $T$-gate is assumed to be $p_m=2\times10^{-9}$~\cite{Gidney2024cultivation} for both cases.
    In both cases, we assume the setup shown in the inset. Here, only two surface code patches are assigned for preparing a magic or resource state, and any single-qubit Pauli rotation gate is executed within a single clock. The execution time includes the latencies for state preparation and gate-teleportation.}
    \label{fig:tradeoff}
\end{figure}

\subsection{Numerical results}
\label{sec:numerical_results}

In this section, we present the numerical data of STAR-magic mutation.
First, Fig.~\ref{fig:error_rate_w_fixed_threshold} and Fig.~\ref{fig:error_rate_w_fixed_ratio} show the numerical results of STAR-magic mutation with a fixed threshold $\theta_{\text{th}}$ or a fixed ratio $r$, respectively, assuming the supply of perfect magic states ($p_m=0$). 
For convenience, here we define the RUS factor $\alpha_{\mathrm{RUS}}$ as 
\begin{equation}
    \alpha_{\mathrm{RUS}}(\theta_L)\equiv \frac{P_L(\theta_L)}{\theta_L p_{\mathrm{ph}}}\ \ \Leftrightarrow\ \ P_L(\theta_L)= \alpha_{\mathrm{RUS}}\theta_L p_{\mathrm{ph}}
\end{equation}
Then, considering the results in Theorem~\ref{thm:scaling}, this factor is expected to scale as follows:
\begin{equation}
    \alpha_{\mathrm{RUS}}(\theta_L) \simeq 
\left\{
\begin{array}{ll}
\mathcal{O}(\theta_L^{1-2/k}) & \ \ (r:\mathrm{fixed}; \ \theta_L\to 0) \\
\mathcal{O}(1) & \ \  (\theta_{\mathrm{th}}:\mathrm{fixed};\  \theta_L\to 0)
\end{array}
\right.
\end{equation}
As shown in Fig.~\ref{fig:error_rate_w_fixed_threshold} (a), the RUS factor in Ref.~\cite{Toshio2024} was given in the range of $1.5<\alpha_{\mathrm{RUS}}<1.8$ for $k=5,7,9$. 
The plotted results exhibit periodic behavior on a logarithmic axis because the value of $\alpha_{\mathrm{RUS}}$ depends strongly on the value of the angle at which the analog-digital switching occurs after the RUS angle is doubled repeatedly in the RUS process.
In contrast, Fig.~\ref{fig:error_rate_w_fixed_threshold} (b) and (c) demonstrate that STAR-magic mutation can achieve smaller RUS factor in the range of $0.2<\alpha_{\mathrm{RUS}}<0.4$ for $\theta_{\text{th}}=0.05$ or $0.05<\alpha_{\mathrm{RUS}}<0.11$ for $\theta_{\text{th}}=0.01$.
On the other hand, Fig.~\ref{fig:error_rate_w_fixed_ratio} (a) illustrates that the RUS factor decreases with target angle $\theta_L$ when the ratio $r$ is fixed. In particular, Fig.~\ref{fig:error_rate_w_fixed_ratio} (c) shows that the RUS factor scales $\alpha_{\text{RUS}}(\theta_L)\propto\theta_L^{1-2/k}$ in the small-angle limit.

When using MSC in the digital stage, the effective error rate of SMM is increased due to the finite error rate of magic states ($p_m\neq0$). In this case, Eq.~\eqref{eq:effective error rate} is modified as follows:
\begin{equation}
    P_L(\theta_L) = p_{\text{analog}} +\left( \frac{1}{2}\right)^{N_{\text{RUS}}} \left(\delta + p_m N_{\text{syn}}(\delta)\right),
\end{equation}
\begin{equation}
    p_{\text{analog}}\ \equiv\  \sum_{m=1}^{N_{\text{RUS}}} \left( \frac{1}{2}\right)^m P_L^{(m)}(\theta_L),
\end{equation}
where $\delta$ is the target accuracy of $T$-gate synthesis and $N_{\text{syn}}(\delta)$ is the number of magic states required for the synthesis. Using the standard approach of $T$-gate synthesis in Ref.~\cite{Ross2016}, $N_{\text{syn}}(\delta)$ is approximated as $3\log_2(1/\delta)$. In what follows, for simplicity, we set the value of $\delta$ as 
\begin{equation}
\label{eq:synthesis error}
    \delta = \max\left(p_m, \ 0.1\times 2^{N_{\text{RUS}}} \times p_{\text{analog}}
    \right),
\end{equation}
so that the synthesis error is always smaller than other types of errors. 
Although this value can be further fine-tuned, such tuning is expected to have little impact on the results of this paper.

In Fig.~\ref{fig:alpha_w_imperfect_magic_state}, we show the RUS factor in the case of $p_m=2\times10^{-9}$. Fig.~\ref{fig:alpha_w_imperfect_magic_state} (a) and (b) suggest that the value of $\alpha_{\text{RUS}}$ increases only slightly when $\theta_{\text{th}}$ is set to a sufficiently large value, except in the case of $\theta_L>\theta_{\text{th}}$. In contrast, Fig.~\ref{fig:alpha_w_imperfect_magic_state} (c) demonstrates that, given a fixed ratio $r$, the value of $\alpha_{\text{RUS}}$ actually begins to increase when the target angle falls below a certain angle. 
This is because the accumulated error in the digital stage surpasses that in the analog stage in the small-angle limit.

To maximize the performance of SMM, the threshold angle $\theta_{\text{th}}$ must be adjusted to an optimal value. Thereby, we can balance the tradeoff between the accumulated errors in the analog and digital stages, or 
between the execution time and the total error.
To illustrate this point, in Fig.~\ref{fig:tradeoff}, we plot the 2D plot of the logical error rate and the execution time of SMM. In this figure, we vary the value of the threshold angle as $\theta_{\text{th}}=2^n \theta_L$ $(n=0,1,2,\cdots, 15)$. The execution time includes the latencies for magic or resource state preparation, as well as for the gate-teleportation throughout the RUS process.
Fig.~\ref{fig:tradeoff} clearly illustrates that we can significantly reduce both the execution time and the error rate of logical rotation gates by tuning the value of $\theta_{\text{th}}$ properly.
For comparison, we also plot the data for the standard $T$-gate synthesis when magic states are generated by MSC with an error rate of $p_m = 2 \times 10^{-9}$ and a preparation time of $t_m = 10$ [clocks].
Remarkably, comparing these results, we can conclude that SMM can execute an analog rotation gate for $\theta_L<10^{-5}$ with an error rate two orders of magnitude smaller than the standard approach, while simultaneously reducing the execution time by two orders of magnitude. 
This is one of the key results of this paper.

\section{Novel early-FTQC architecture: STAR ver.~3}
\label{sec:STARv3}

As detailed in Sec.~\ref{sec:Preliminary}, the STAR architecture is originally based on the Clifford+$\phi$ gate set, which consists of logical Clifford gates and non-fault-tolerant analog rotation gates. 
The TMR protocol enables us to implement arbitrary logical rotation gates locally, eliminating the need for costly procedures such as MSD and $T$-gate synthesis. This results in a considerable reduction in the spacetime overhead of analog rotation gates compared to standard FTQC architectures~\cite{Litinski2019}.
One drawback of this framework is that even a $T$-gate must be implemented using these non-fault-tolerant approaches to avoid the additional footprint required for magic state factories.
While the fidelity of the TMR protocol is remarkably high for small-angle rotations, it deteriorates for large-angle rotation gates, especially $T$-gates.

However, the advent of MSC enables us to prepare a high-fidelity magic state locally, eliminating the need for large spatial resources dedicated to magic state distillation. In particular, it is compatible with two crucial features of the STAR architecture: the removal of magic state factories and the easy-to-parallelize implementation of rotation gates.
This complementarity and compatibility between STAR and MSC strongly motivate us to refine the basic concepts of STAR architecture.
In this section, we demonstrate how to utilize MSC and SMM to revamp the framework of previous STAR architectures~\cite{Akahoshi2023,Toshio2024}.

\begin{figure*}
    \centering
    \includegraphics[width=0.95\linewidth]{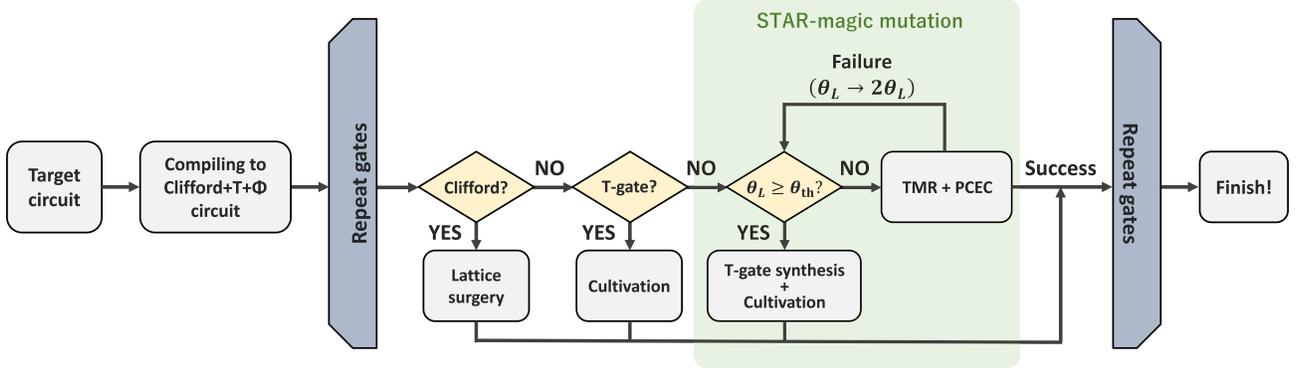}
    \caption{Flowchart describing how to execute quantum circuits on the STAR ver.~3. First, we compile the input circuit into the Clifford+T+$\phi$ gate set. Each gate in the compiled circuit is classified into the following four classes and performed with operations suited for each class: (1) Clifford gates are performed with standard lattice surgery (or transversal gate) techniques~\cite{Horsman2012, Litinski2019, Ismail2025}. (2) $T$-gates are performed via gate-teleportation by preparing a magic state with MSC (or MSD)~\cite{Gidney2024cultivation}. Here, we can use MSD, instead of MSC, to improve the fidelity of $T$-gate. (3) Analog rotation gates satisfying $\theta_L \geq \theta_{\text{th}}$ are performed by combining $T$-gate synthesis~\cite{Ross2016} and MSC. (4) Analog rotation gates satisfying $\theta_L < \theta_{\text{th}}$ are performed via gate-teleportation by preparing a resource state with the TMR protocol, followed by the post-processing for PCEC. This process succeeds with probability of $1/2$ and, otherwise, we perform a feedback rotation gate after updating the target angle from $\theta_L$ to $2\theta_L$.}
    \label{fig:flowchart}
\end{figure*}

\subsection{Definition}

We first give the formal definition of our proposed architecture, ``STAR ver.~3," as follows:

\begin{itemize}
    \item {\bf Partial fault-tolerance}: Quantum information is encoded on some error-correcting codes, and arbitrary Clifford operations are performed on it in a fault-tolerant manner, such as lattice surgery or transversal gate techniques.
    \item {\bf Clifford+$T$+$\phi$ gate set}: The input quantum circuit is first decomposed into Clifford operations, $T$-gates, and analog rotation gates. The ratio of $T$-gates to analog rotation gates is determined to minimize the total time overhead, including the sampling cost for error mitigation. 
    \item {\bf Fault-tolerant $T$-gate}: Each $T$-gate is implemented in a fault-tolerant manner by preparing a magic state via MSC or MSD.
    \item {\bf Fault-tolerant implementation of large-angle analog rotation gates}: When $\theta_L \geq\theta_{\text{th}}$, each analog rotation gate $\hat{R}_{P,L}(\theta_L)$ is approximated with standard Clifford+$T$ decompositions and synthesized with fault-tolerant $T$-gates. 
    \item {\bf Non-fault-tolerant implementation of small-angle analog rotation gates}: When $\theta_L < \theta_{\text{th}}$, each analog rotation gate $\hat{R}_{P,L}(\theta_L)$ is implemented using STAR-magic mutation.
    \item {\bf Error mitigation}: Any residual error in $T$-gates and analog rotation gates is mitigated by some error mitigation techniques (e.g., probabilistic error cancellation (PEC)~\cite{Temme2017,Endo2018}).
\end{itemize}
The flowchart in Fig.~\ref{fig:flowchart} illustrates how quantum circuits are executed on STAR ver.~3, following the above definition.

Adopting the Clifford+$T$+$\phi$ gate set enables the execution of more complex circuits than those in the previous STAR architectures~\cite{Akahoshi2023,Toshio2024} (see also Fig.~\ref{fig:bound}).
For example, typical block-encoding circuits are composed of a combination of $T$-gates (or Toffoli gates) and small-angle rotation gates~\cite{Babbush2018qubitization,Martyn2021}. Similarly, some modern Trotter-based algorithms periodically perform the fermionic fast Fourier transform~\cite{Kivlichan2020improved} or the single-particle basis transformation~\cite{Motta2021} to reduce the circuit depth when solving electronic structure problems. These oracles include many large-angle rotation gates and cause a significant contribution to the total circuit errors in STAR ver. 2~\cite{Toshio2024}. On the other hand, STAR ver.~3 can minimize the accumulated errors in these circuits by properly switching the methods for executing non-Clifford gates.

Another major advantage of STAR ver.~3 is the locality of magic and resource state preparation. In usual FTQC architectures based on MSD~\cite{Litinski2019magic}, magic states are generated in MSD factories, and each non-Clifford gate is sequentially performed by allocating ancilla regions between the factories and the target qubits for lattice surgery. 
Although we can perform some non-Clifford gates in parallel by properly compiling paths for lattice surgery~\cite{Beverland2022compiling, Hamada2025routing, LeBlond2025compiling}, this approach results in expanding the scale of the MSD factories, significantly increasing the spatial overhead for MSD.
In contrast, our architecture can perform any local non-Clifford gate in place by preparing magic and resource states near the target qubits.
This local state preparation does not require additional spatial overhead since we can use the ancilla region originally prepared for lattice surgery.
This locality allows us to perform multiple rotation gates in parallel, further accelerating quantum computations on STAR ver. 3.
For example, Ref.~\cite{Akahoshi2024} has previously investigated locality-aware parallel compilation on STAR ver.~2 and achieved an acceleration of over 10 times compared to naive serial compilation in the Trotter simulation.
Similar directions have also been investigated in other prior works~\cite{Hirano2025localityaware,Sethi2025RESCQ}.

\subsection{Error mitigation and theoretical bound on the feasible circuit size}

This section describes how to mitigate the residual errors in STAR-magic mutation or direct $T$-gates by utilizing the standard PEC techniques~\cite{Temme2017,Endo2018}.
As shown below, the simulatable circuit size on STAR ver.~3 is strictly restricted by the sampling cost for PEC. 

The application of PEC to FTQC architectures was previously investigated in detail in Refs.~\cite{Suzuki2022, Piveteau2021}. For example, we can generally describe a quantum channel for noisy $T$-gate, $\mathcal{N}_T$, with a stochastic
Pauli-$Z$ noise channel after twirling it by logical Clifford operations:
\begin{equation}
\label{eq:pauli-Z noise channel}
    \mathcal{N}_T = \mathcal{E}_T \circ \mathcal{R}_{\pi/8},\ \ \ 
    \mathcal{E}_T(\hat{\rho}) = (1-p_m)\hat{\rho} + p_m \hat{Z}\hat{\rho} \hat{Z}, 
\end{equation}
where $p_m$ describes an error rate due to the infidelity of the magic state.
We can explicitly construct the inverse map for this type of noise channel as 
\begin{equation}
\label{eq:inverse map}
    \mathcal{E}_T^{-1}(\hat{\rho}) \ =\ 
    \frac{1-p_m}{1-2p_m}\rhohat 
    -\frac{p_m}{1-2p_m}\zhat\rhohat \zhat.
\end{equation}
Then, by applying this operation after $\mathcal{N}_T$, we can obtain an ideal $T$-gate as follows:
\begin{equation}
\label{eq:noiseless T gate}
    \mathcal{R}_{\pi/8}= \mathcal{E}_T^{-1}\circ \mathcal{N}_T
    = \gamma_T \left[ (1-p_m)\cdot \mathcal{N}_T - p_m \mathcal{Z}\circ \mathcal{N}_T,
    \right]
\end{equation}
where $\gamma_T = (1-2p_m)^{-1}\simeq 1+2p_m$ and $\mathcal{Z}$ is a Pauli-$Z$ gate channel. 

However, the inverse channel $\mathcal{E}_T^{-1}$ cannot be implemented only by sampling some unitary gates probabilistically, since it includes a negative probability. In the PEC method~\cite{Temme2017,Endo2018}, we virtually realize this inverse channel by focusing only on the measurement outcomes. Using Eq.~\eqref{eq:noiseless T gate}, we can describes the noise-free expectation value of any observable $\hat{O}$ as 
\begin{equation}
\label{eq: expectation value}
    \ev*{\hat{O}}_{\mathcal{R}_{\pi/8}}\  \simeq \ \gamma_T\left((1-p_m) \ev*{\hat{O}}_{\mathcal{N}_T}  - p_m\ev*{\hat{O}}_{\mathcal{Z} \circ \mathcal{N}_T}\right),
\end{equation}
where $\ev*{\hat{O}}_{\mathcal{N}}=\tr[\hat{O}\mathcal{N}(\rhohat)]$ is the expectation value of $\hat{O}$ after applying a channel $\mathcal{N}$ to a target state. This equation indicates that, by randomly sampling the correcting operations $\mathcal{Z}$ with a probability of $p_m$ for each shot, we can estimate the noise-free expectation value of any observable $O$.
Now it is crucially important that the variance of the estimator in Eq.~\eqref{eq: expectation value} is amplified by the factor $\gamma_T^2\simeq1+4 p_m$, compared to the case without PEC.
Consequently, in the PEC, we require $\gamma^2_T$ times more samples to suppress the amplified statistical errors.

This PEC formalism can be applied to the gate-teleportation for analog rotation gates directly, since the noise channel for each RUS trial can be approximately described as stochastic Pauli-$Z$ noise in Eq.~\eqref{eq:canceled rotation gate for STARv3}. In this case, the parameter $p_m$ in Eq.~\eqref{eq:pauli-Z noise channel} is replaced with $2\sum_{j=1}^k\bar{q}_j\sin^2(\Delta_{j}) \simeq 2\bar{q}_1 \sin^2(\Delta_1) $ for each RUS trial. As detailed in Ref.~\cite{Toshio2024}, the total RUS process is described as a classically mixed channel composed of the random events where the RUS process finishes at different numbers of RUS trials. Considering the convex-linearity of $\gamma$ factor, we can determine the mitigation cost for the resulting rotation gate $\hat{R}_{P,L}(\theta_L)$ as 
\begin{equation}
\label{eq:gamma for analog rotation}
    \gamma_{\theta_L}^2 \simeq 1+ 4 P_L(\theta_L) = 1+4\alpha_{\text{RUS}}(\theta_L)\cdot\theta_L \cdot p_{\text{ph}},
\end{equation}
when assuming that we can neglect the mitigation cost for the synthesis error by setting $\delta$ sufficiently small. According to Ref.~\cite{Suzuki2022}, the synthesis error can also be mitigated using the PEC technique, but its mitigation cost has a more complex form compared to the case of simple Pauli-$Z$ errors.
Therefore, Eq.~\eqref{eq:gamma for analog rotation} is not a rigorous formula for $\gamma_{\theta_L}^2$. 
Nonetheless, in this paper, we assume that we can use the simple formula in Eq.~\eqref{eq:gamma for analog rotation} as a good approximation, since we set $\delta$ sufficiently small as in Eq.~\eqref{eq:synthesis error}.

Next, let us consider a quantum circuit composed of any number of Clifford gates, $N_T$ $T$-gates, and $N_R$ analog rotation gates with angles $\{\theta_i\}_{i=1,2,\cdots,N_R}$. 
For this circuit, the overall factor representing the total mitigation cost is calculated as follows: 
\begin{equation}
\label{eq:mitigation cost in generic case}
    \gamma_{\text{total}}^2 \ \equiv\ \prod_{i=1}^{N_T} \gamma_T^2 \times \prod_{i=1}^{N_R} \gamma_{\theta_i}^2\  \simeq \ e^{4P_{\text{total}}},
\end{equation}
\begin{equation}
    P_{\text{total}}\ \equiv \ N_T \cdot p_m + \sum_{i=1}^{N_R} \alpha_{\text{RUS}} (\theta_i)\cdot \theta_i \cdot p_{\text{ph}}.
\end{equation}
Thus, to avoid an exponential growth of sampling overhead, input quantum circuits are required to satisfy 
\begin{equation}
\label{eq:theoretical bound}
    P_{\text{total}}\ \lesssim \ 1.
\end{equation}
This inequality highlights a fundamental limitation of STAR ver.~3. 

As an example, consider the first-order Trotter simulation of a quantum system with Hamiltonian $\hat{H} = \sum_{k=1}^L c_k \hat{P}_k$.
In this case, the Trotter circuit has the following form: 
\begin{equation}
    e^{-iT\hat{H}} \simeq \left(\prod_{k=1}^L \hat{R}_{P_k}(\theta_k)
    \right)^M,
\end{equation}
where $\theta_k \equiv c_k T/M$. Then, the total error rate $P_{\text{total}}$ is explicitly given as 
\begin{equation}
    P_{\text{total}} = M\sum_{k=1}^{L} \alpha_{\text{RUS}} (\theta_k)\cdot \theta_k \cdot p_{\text{ph}} \leq \alpha_{\text{max}} \lambda T p_{\text{ph}},
\end{equation}
where $\lambda =\sum_k|c_k|$ is the L1-norm of the Hamiltonian, and we assume that $\alpha_{\text{RUS}} (\theta_k)\leq \alpha_{\text{max}}$ is satisfied for any $k$.
Then, the theoretical bound in Eq.~\eqref{eq:theoretical bound} results in the following requirements for the problem size:
\begin{equation}
    \lambda T \  \lesssim  \  \frac{1}{ \alpha_{\text{max}} p_{\text{ph}} }.
\end{equation}
This theoretical bound has already been derived in Ref.~\cite{Toshio2024}.
For example, assuming $d=21$ ($k=7$) and $\theta_{\text{th}}=0.01$, STAR-magic mutation leads to $\alpha_{\text{max}}\simeq 0.1$ as shown in Fig.~\ref{fig:alpha_w_imperfect_magic_state} (b).
Consequently, STAR ver.~3 can simulate quantum systems with $\lambda=100$ up to $T\lesssim100$ under $p_{\text{ph}}=10^{-3}$ without an exponential cost of error mitigation.

Finally, based on Eq.~\eqref{eq:theoretical bound}, we compare the capabilities of different quantum computing architectures, including previous STAR architectures and MSC-based FTQC architecture. To this end, let us consider a quantum circuit composed of any number of Clifford gates, $N_T$ $T$-gates, and $N_R$ analog rotation gates with a fixed angle $\theta_*$. In the original STAR architecture (STAR ver.~1)~\cite{Akahoshi2023}, any analog rotation gate is executed using [[4,1,1,2]] subsystem code-based injection protocol with an error rate of $2\times\frac{2}{15}p_{\text{ph}}$. Here, the factor of $2$ originates from the RUS process and does not apply to $T$-gates. Thus, the total error rate for this circuit is easily determined as follows:
\begin{equation}
    P_{\text{total}}^{\text{(v1)}}\ \simeq \  \frac{2}{15}\left(N_T + 2N_R\right)p_{\text{ph}}.
\end{equation}
Meanwhile, in STAR ver.~2~\cite{Toshio2024}, small-angle rotation gates are implemented using another approach based on the TMR protocol, which has an error rate of $P_L(\theta_L)=\alpha_{\text{RUS}}^{\text{(v2)}}\theta_Lp_{\text{ph}}$. Thus, the total error rate is calculated as 
\begin{equation}
    P_{\text{total}}^{\text{(v2)}}\ \simeq \  \left(\frac{2}{15}\cdot N_T +  1.6 \cdot  N_R  \theta_*\right)p_{\text{ph}},
\end{equation}
where we use the fact that $\alpha_{\text{RUS}}^{\text{(v2)}}\simeq 1.6$ for $k=7$ as shown in Fig.~\ref{fig:error_rate_w_fixed_threshold} (a) and assume that $T$-gate is performed via the [[4,1,1,2]] protocol. 
On the other hand, when assuming an FTQC architecture where any non-Clifford is implemented with Clifford+$T$ gate set and magic state is implemented via MSC, the total error rate is determined as follows:
\begin{equation}
    P_{\text{total}}^{\text{(cul)}} \ \simeq \ \left(N_T + N_R N_{\text{sys}}(\delta) \right) p_m +N_R \delta,
\end{equation}
where $\delta$ is the synthesis error and $N_{\text{sys}}$ is the number of $T$-gate required for implementing an analog rotation gate.

\begin{figure}
    \centering
    \includegraphics[width=1.0\linewidth]{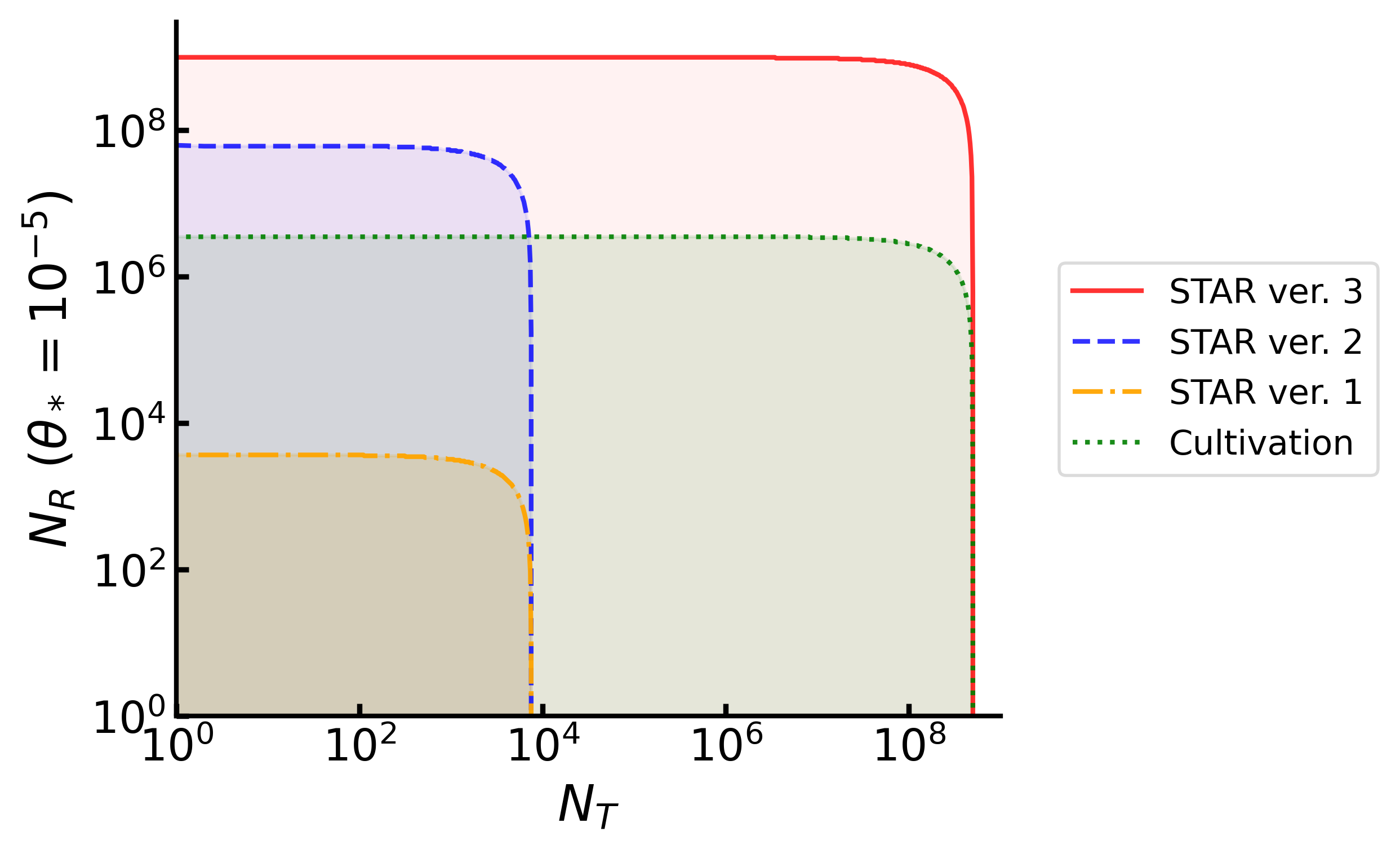}
    \caption{Feasible circuit sizes in different quantum computing architectures: STAR ver.~3 ({\bf red}), STAR ver.~2 ({\bf blue}), STAR ver.~1 ({\bf orange}), and MSC-based FTQC ({\bf green}). Here, $N_R$ and $N_T$ denote the total numbers of analog rotation gates with $\theta_*=10^{-5}$ and $T$-gates included in the circuit, respectively. The plotted curves are determined to satisfy $P_{\text{total}}=1$, assuming $p_{\text{ph}}=10^{-3}$ and $p_m=2\times10^{-9}$. In the FTQC setup, we set $\delta=p_m$ and assume that magic states are prepared with MSC and not distilled anymore to avoid an additional overhead of MSD.}
    \label{fig:bound}
\end{figure}

As discussed above, the theoretical bounds on feasible circuit sizes are roughly estimated as the gate counts satisfying $P_{\text{total}}=1$ for each quantum computing architecture.
Fig.~\ref{fig:bound} plots the numerical results for previous STAR architectures and MSC-based FTQC, as well as STAR ver.~3. This figure clearly illustrates that STAR ver.3 can cover a vastly broader class of quantum circuits than the previous STAR architectures.
Similarly, our architecture outperforms the MSC-based FTQC architecture for circuits where small-angle rotation gates are predominant. Notably, STAR ver.3 not only reduces the total error rate by over two orders of magnitude but also cuts the execution time of each rotation gate by two orders of magnitude, as shown in Fig.~\ref{fig:tradeoff}.

\section{Practical Applications}
\label{sec:practical applications}

In this section, we demonstrate the potential of STAR ver.~3 by providing detailed resource estimates for simulating the dynamics of quantum many-body systems, including molecules and spin systems. These simulations are crucial for understanding a range of non-equilibrium phenomena~\cite{Fauseweh2024review}, including chemical reactions~\cite{Lidar1999, Kassal2008} and laser-induced ultrafast dynamics~\cite{Magann2021, Chan2023grid, Kohler1995review,Cavalleri2018review, Ghimire2019review}, as well as for interpreting experimental data from various spectroscopic experiments, such as nuclear magnetic resonance (NMR)~\cite{Sels2020, Brien2022, Seetharam2023, Google2025NMR} and inelastic neutron scattering (INS)~\cite{Vilchez-Eestevez2025}.
In addition, as suggested by prior works~\cite{Childs2018speedup,Beverland2022assessing}, quantum dynamics simulations could be one of the earliest practical applications of FTQC, as they require significantly fewer space-time resources than other classically infeasible problems such as prime factoring and quantum chemistry.

Typical quantum algorithms for simulating quantum many-body dynamics, including Trotterization~\cite{Trotter1959, Suzuki1990, Suzuki1991, Lloyd1996, Abrams1997} and various randomized algorithms~\cite{Campbell2019,Chakraborty2024, Kiumi2024}, require a large number of small-angle rotation gates. This property makes those applications an ideal target of STAR ver. 3, which is designed to execute these gates efficiently without costly procedures such as $T$-gate synthesis and magic state distillation. 

In this paper, we analyze the spacetime cost of simulating quantum many-body dynamics based on a latest randomized algorithm called the {\it Time Evolution by Probabilistic Angle Interpolation} (TE-PAI) algorithm~\cite{Kiumi2024}. This algorithm has two major advantages for our purpose: (i) The circuit depth and sampling overhead of TE-PAI depend only on the evolution time $T$ and target Hamiltonian's L1-norm $\lambda\equiv ||H||_1$, rather than the detailed structure of target Hamiltonians. This is in contrast with the Trotter-based approach, where we need to calculate a theoretical bound of the Trotter norm for each target Hamiltonian to determine the optimal size of the Trotter step.
The simplicity of TE-PAI makes it easy to estimate the spacetime costs of simulating a broad range of many-body systems in a unified framework. 
Additionally, the TE-PAI algorithm is suited for simulating molecular systems because their Hamiltonians include a huge number of terms due to long-range Coulomb interactions.
(ii) The TE-PAI circuits are constructed from a gate set composed of Clifford gates and an analog rotation gate with a unique angle $\Delta$. In actual quantum devices, fixing the rotation angle will significantly reduce the calibration cost of physical rotation gates. This point will be crucial in realistic applications of the STAR architecture.

In what follows, we first overview the TE-PAI algorithm in Sec.~\ref{sec:TE-PAI}. Then, we provide the detailed resource analyses for simulating quantum many-body dynamics in Sec.~\ref{sec:estimation}.

\subsection{TE-PAI: Exact quantum dynamics simulation by sampling random circuits}
\label{sec:TE-PAI}

The TE-PAI algorithm effectively achieves an exact time evolution of quantum many-body systems by sampling random quantum circuits. This algorithm constructs an unbiased estimator for the entire time-evolution superoperator, enabling the estimation of expectation values of time-evolved observables by sampling the output of quantum circuits. It is also compatible with advanced measurement techniques, such as classical shadows~\cite{Huang2020measurement} and Pauli grouping techniques~\cite{Crawford2021,Jena2019}.

TE-PAI builds on the so-called Probabilistic Angle Interpolation (PAI) technique~\cite{Koczor2024}. For illustration, consider a time-independent Hamiltonian $\hat{H} = \sum_{k=1}^L c_k \hat{P}_k$, where $c_k$ are real coefficients and $\hat{P}_k$ are Pauli strings. The first-order Trotter-Suzuki formula approximates its time evolution operator as 
\begin{equation}
\label{eq: Trotter for TE-PAI}
    U \equiv e^{-i\hat{H}T} \approx \left( \prod_{k=1}^L e^{-ic_k \hat{P}_k \frac{T}{N}} \right)^N = \prod_{j=1}^N \left(\prod_{k=1}^L \hat{R}_{P_k}(\theta_{k}) \right),    
\end{equation}
where we introduce $\theta_{k} = c_k T/N$. 
In TE-PAI, we decompose the superoperator representation $\mathcal{R}_{P_k}(\theta_{k})$ of each rotation gate $\hat{R}_{P_k}(\theta_{kj})$ as 
\begin{equation}
\mathcal{R}_{P_k}(\theta_{k}) = \gamma_1(|\theta_{k}|) \mathcal{A} + \gamma_2(|\theta_{k}|) \mathcal{B}_{k} + \gamma_3(|\theta_{k}|) \mathcal{C}_k,
\end{equation}
where $\mathcal{A}$, $\mathcal{B}_{k}$, and $\mathcal{C}_k$ denote the superoperator representation of the following gate sets: 
\begin{equation}
    \hat{A} = \hat{I}, \ \ \ \hat{B}_{k} = \hat{R}_{P_k}(\text{sign}(\theta_{k}) \Delta), \ \ \ \hat{C}_k = \hat{R}_{P_k}(\pi/2).
\end{equation}
Here, we can choose the value of $\Delta$ arbitrarily and the coefficients $\gamma_l(|\theta_{k}|)$ $(l=1,2,3)$ are explicitly determined as specific trigonometric functions~\cite{Kiumi2024}. However, as explained below, it is preferable to determine $\Delta$ to minimize the total computational cost of TE-PAI.

The TE-PAI algorithm replaces each superoperator of rotation gate, $\mathcal{R}_{P_k}(\theta_{kj})$ in Eq.~\eqref{eq: Trotter for TE-PAI}, with its unbiased estimator
\begin{equation}
\hat{\mathcal{R}}_{P_k}(\theta_{k}) = ||\gamma(|\theta_{k}|)||_1 \cdot \text{sign}[\gamma_l(|\theta_{k}|)] \cdot \hat{\mathcal{D}}_l,
\end{equation}
where $\hat{\mathcal{D}}_l\in \{ \mathcal{A}, \mathcal{B}_{k}, \mathcal{C}_k  \}$ denotes a randomly selected superoperator according to the probabilities $p_l = |\gamma_l(\theta_{k})| / ||\gamma(\theta_{k})||_1$ for $l\in \{1,2,3\}$.
This procedure allows for the construction of an unbiased estimator for the entire time evolution superoperator $\mathcal{U}=\prod_{j=1}^N \left(\prod_{k=1}^L \mathcal{R}_{P_k}(\theta_{k}) \right)$:
\begin{equation}
\label{eq:TE-PAI circuit}
\hat{\mathcal{U}} = \prod_{j=1}^N \left(\prod_{k=1}^L \hat{\mathcal{R}}_{P_k}(\theta_{k}) \right).
\end{equation}
Specifically, this estimator satisfies $E[\hat{\mathcal{U}}] = \mathcal{U}$ when averaged over the sampling in $\hat{\mathcal{R}}_{P_k}$.

Theorem 1 in Ref.~\cite{Kiumi2024} suggests that the expected number of non-trivial gates in $\hat{\mathcal{U}}$ can be approximated as
\begin{equation}
    N_{\text{gate}}=\csc(2\Delta) (3 - \cos (2\Delta)) \lambda T \ \geq \  2\sqrt{2} \lambda  T
\end{equation}
in the limit of $N\to\infty$.
This linear scaling with time $T$ and the Hamiltonian norm $\lambda\equiv \sum_k|c_k|$ demonstrates the efficiency of TE-PAI. 
However, in practice, the rotation angle $\Delta$ needs to scale as $\mathcal{O}(1/\lambda T)$ as discussed below, to avoid an exponential increase in the sample complexity. 

\begin{table*}
 \begin{center}
   \caption{Examples of target quantum many-body systems for our resource estimation. Here, ``RFIC" and ``TFIM" denote the random-field Ising chain model and the transverse-field Ising model, respectively. $[\text{2Fe-2S}]$ and $[\text{4Fe-4S}]$ denote iron-sulfur clusters containing 2 or 4 transition metal atoms~\cite{Lee2023evaluating}.
   $N_L$ and $\lambda$ are the number of logical qubits and Hamiltonian's L1-norm of the target quantum systems. The L1-norms for molecules are minimized using orbital optimization~\cite{Koridon2021} and block-invariant symmetry shift (BLISS)~\cite{Loaiza2023BLISS,Patel2025BRISS}.}
   \label{tab: L1-norm}
  \begin{tabular}{p{4cm}p{3cm}p{3cm}p{3cm}} \hline
   System & Actice space & $N_L$ & $\lambda$ \\ \hline
   $N$-sites RFIC & --- & $N$ & $(3J+h/2)N$ \\ 
   $(L\times L)$-sites TFIM & --- & $L^2$ & $(2J+h)L^2$ \\
   $(L\times L)$-sites Hubbard & --- & $2L^2$ & $(4t+U/4)L^2$ \\ 
   $[\text{2Fe-2S}]$~\cite{Lee2023evaluating}  & (30e, 20o) & 40 & 38.2 [Hartree] \\
   $[\text{4Fe-4S}]$~\cite{Lee2023evaluating} & (54e, 36o) & 72 & 137.8 [Hartree] \\
   FeMoco (S=0)~\cite{Reiher2017} & (54e, 54o) & 108 & 308.2 [Hartree] \\ 
   FeMoco (S=3/2)~\cite{Li2019FeMoCo} & (113e, 76o) & 152 & 512.5 [Hartree] \\ 
   \hline
  \end{tabular}
 \end{center}
\end{table*}

To calculate the total computational cost of TE-PAI, we need to determine the required number of shots $N_s$ to achieve a precision $\epsilon$ in estimating time-evolved expectation values.
Ref.~\cite{Kiumi2024} suggests that $N_s$ is upper bounded by $\gamma_N^2/\epsilon^2$, where $\gamma_N = \prod_{j=1}^N \prod_{k=1}^L ||\gamma(|\theta_{k}|)||_1$. Considering the limit of $N\to\infty$, this overhead can be approximated as
\begin{equation}
\gamma_\infty^2 \equiv\lim_{N\to\infty}\gamma_N^2 = \exp\left( 2 \lambda T \tan\left(\Delta\right) \right).
\end{equation}
TE-PAI can control the tradeoff between circuit depth and measurement overhead. 
Specifically, by setting the rotation angle as $\Delta =\arctan\left(\frac{Q}{2\lambda T}\right) \simeq Q/2\lambda T$, TE-PAI requires a constant overhead of $\gamma_\infty^2=e^Q$ and a single-shot gate count of $N_{\text{gate}} =2(\lambda T)^2/Q + Q$. Here, $Q$ is a tunable parameter that balances the sampling cost and the gate count in the TE-PAI algorithm.
In conclusion, the total gate count for TE-PAI is determined as $(2(\lambda T)^2/Q + Q)e^Q/\epsilon^2$.

The TE-PAI circuits in Eq.~\eqref{eq:TE-PAI circuit} are highly suitable for the STAR architecture in the sense that they are composed of any Pauli-string operations and a unique rotation gate with a fixed small angle $\Delta$. Notably, fixing the rotation angle to a unique value will significantly alleviate the cost for the calibration of physical-level rotation angles.

\subsection{Resource estimation on STAR ver.~3: Simulation of quantum many-body dynamics}
\label{sec:estimation}

This section provides detailed resource analyses for simulating quantum many-body dynamics on STAR ver.~3 using the TE-PAI algorithm.
As shown in the previous subsection, the computational costs of TE-PAI are uniquely determined by specifying the evolution time $T$ and the Hamiltonian's L1-norm $\lambda$ of target systems. 
For instance, Table.~\ref{tab: L1-norm} shows a list of L1-norm values for several quantum systems, including typical lattice models and molecules.

Our analyses of circuit runtimes on STAR ver.~3 rely on the sequential Pauli-based computation in Ref.~\cite{Litinski2019}. This scheme employs the elimination of all Clifford gates from the input
circuit, resulting in a compiled circuit consisting only of analog rotation gates and Pauli-string measurements. Especially for TE-PAI, the compiled circuit results in a long sequence of $\Delta$-angle rotation gates with various Pauli-string axes. These operations can be efficiently implemented via STAR-magic mutation on STAR ver.~3.

In this paper, we assume the fast block layout~\cite{Litinski2019}, where a single RUS trial for any analog rotation gates is implemented within a single clock via lattice surgery. In this setup, STAR-magic mutation is executed within $C_{\text{smm}}\simeq 3$ [clock] for $\theta_{\text{th}}=2^6\cdot \theta_L$, assuming a sufficient supply rate of magic and resource states.
Consequently, the single-shot runtime of TE-PAI is estimated as $N_{\text{gate}}C_{\text{smm}} =(2(\lambda T)^2/Q + Q)C_{\text{smm}}$ [clock]. 
Strictly speaking, $N_{\text{gate}}$ includes not only the analog rotation gate but also Pauli-string operations, and so, the actual count of analog rotation gates is lower than $N_{\text{gate}}$ [clock] in Pauli-based computation.
However, we easily find that the TE-PAI circuit is basically dominated by analog rotations rather than Pauli-string operations, and thus, we estimate the total clock as mentioned above.

\begin{figure*}
    \centering
  \begin{tabular}{lll}
{\normalsize (a) Number of physical qubits} & {\normalsize (b) Single-shot runtime [sec]} & {\normalsize (c) Total runtime [day]}  \\
   \includegraphics[width=5.8cm]{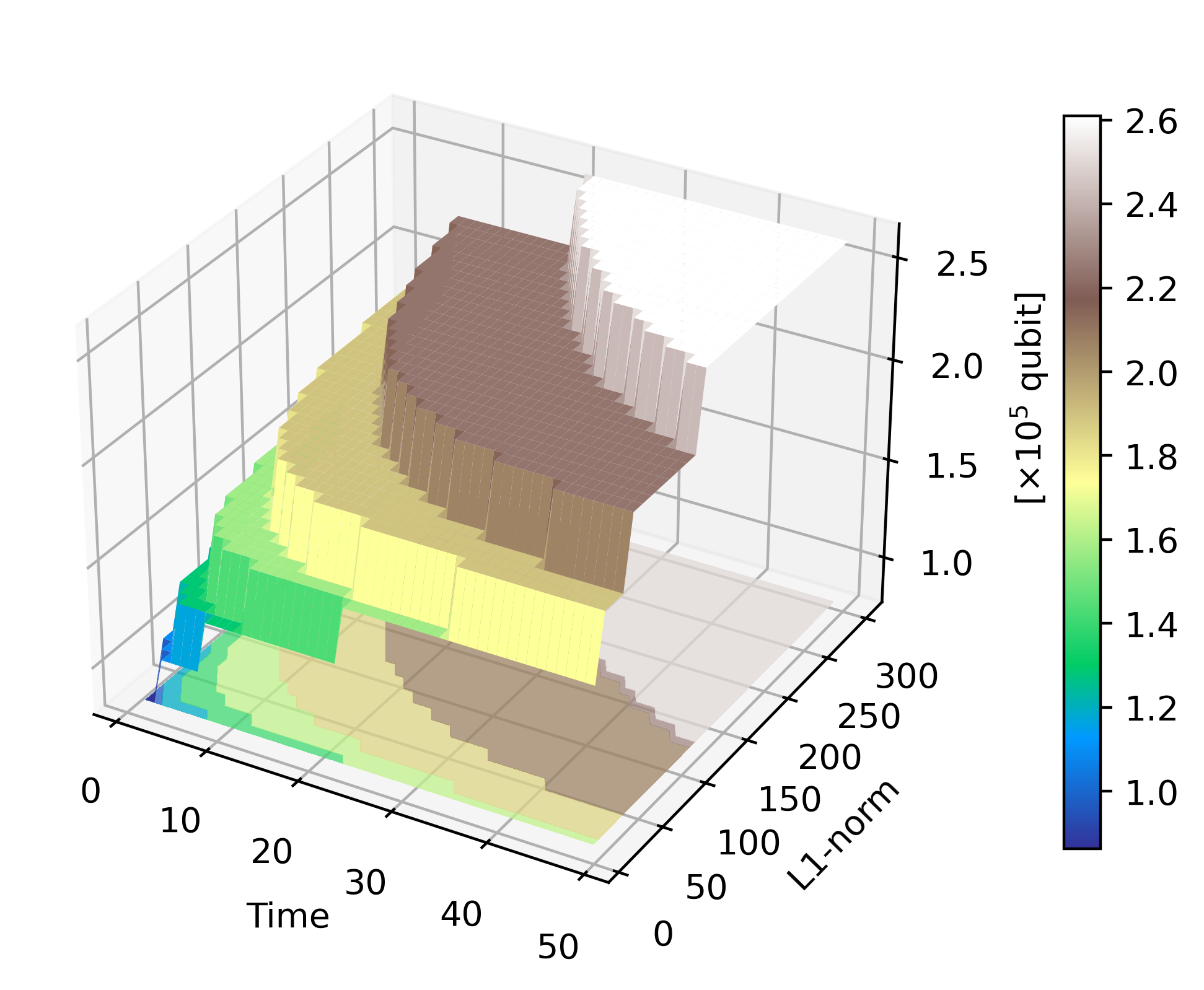}&
   \includegraphics[width=5.8cm]{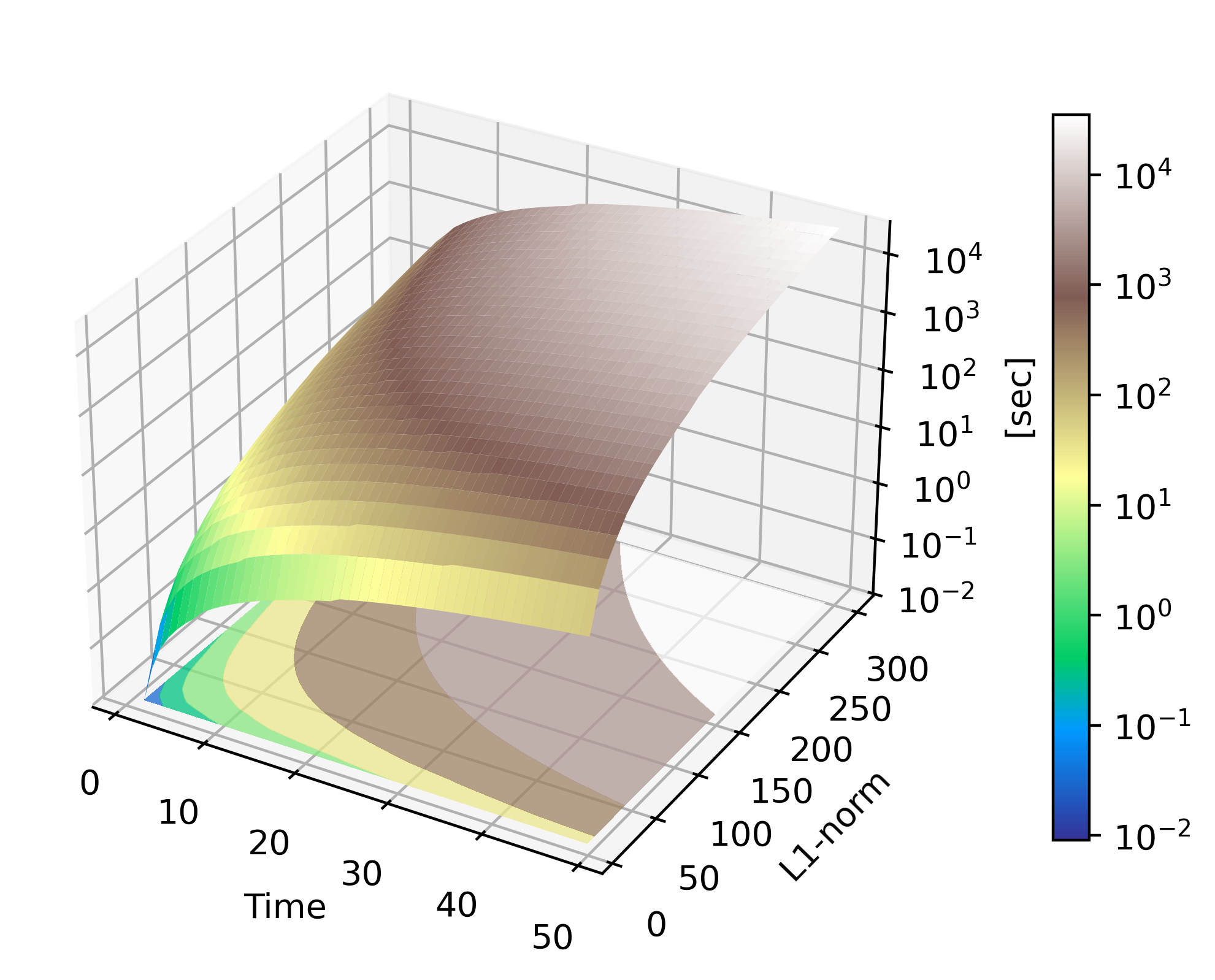}&
   \includegraphics[width=5.8cm]{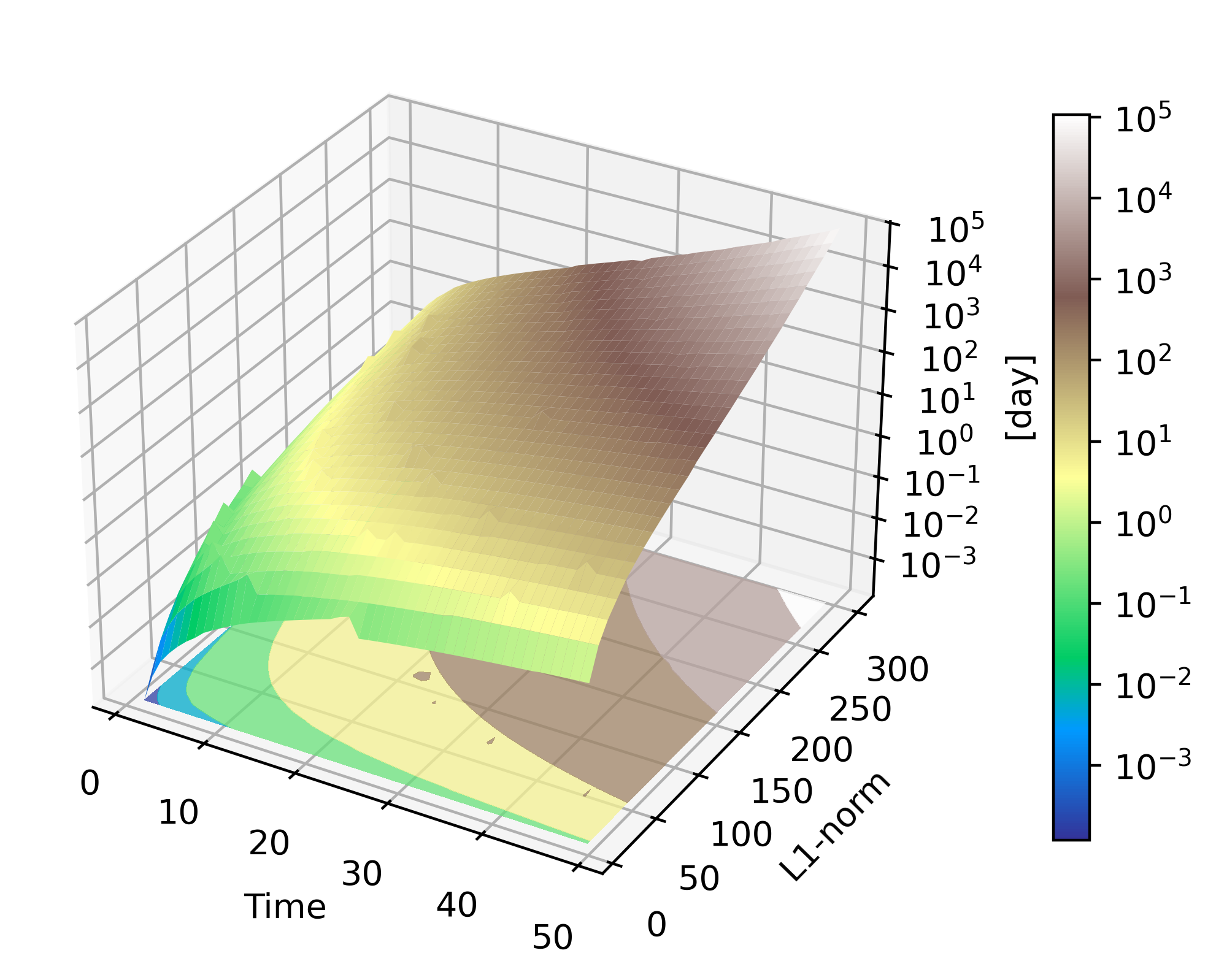}\\
  \end{tabular}
    \caption{Spacetime costs of TE-PAI simulation on STAR ver.~3: (a) Number of physical qubits that include those for data patches and ancillary space for lattice surgery and state preparation, (b) single-shot runtime of TE-PAI circuit, and (c) total runtime including the sampling and mitigation overhead for achieving a specific accuracy $\epsilon=0.05$ in estimating time-evolved expectation values. In these plots, we assume $p_{\text{ph}}=10^{-3}$ and analyze the dynamics simulation up to $T\in [1, 50]$ for virtual molecules with 72 spin orbitals and a L1-norm $\lambda\in [10, 300]$. Here, $T$ and $\lambda$ are assumed in the atomic units (a.u.) and $T\simeq41.3$ [a.u.] corresponds to 1 fs (= $1\times 10^{-15}$ sec).}
    \label{fig:TE-PAI for molecules}
\end{figure*}

Next, let us interpret the clock scale into a real-time scale.
On a current superconducting qubit chip~\cite{Arute2019,Google2023suppressing}, a single round of syndrome extraction takes less than 1 $\mu$s.
Therefore, we assume that a single code cycle takes 1 $\mu$s, as in previous related works~\cite{Yoshioka2022hunting,Babbush2018qubitization,Kivlichan2020improved}. This means that the single-shot runtime of TE-PAI is $(2(\lambda T)^2/Q + Q)C_{\text{smm}}d$ [$\mu$s], where $d$ is the code distance of the surface codes used for this simulation.
For similar reasons, we can readily estimate the total computational time as $(2(\lambda T)^2/Q + Q)C_{\text{smm}}de^Q e^{4P_{\text{total}}}/\epsilon^2$ [$\mu$s], where the factor $ e^{4P_{\text{total}}}$ is the mitigation cost for PEC. 

To complete our resource analyses, we must determine the optimal value of code distance $d$ for our task. 
According to Ref.~\cite{Fowler2018}, using a minimum-weight perfect matching decoder under the circuit-level noise model, the logical error rate per code cycle is approximated as
\begin{equation}
    p_L(p_{\text{ph}},d) = 0.1\times(100p_{\text{ph}})^{(d+1)/2}.
\end{equation}
 for the rotated surface code. The optimal value of the code distance $d$ is determined to satisfy $p_L(p_{\text{ph}},d)^{-1}\gg dN_{\text{gate}}C_{\text{smm}}N_{\text{patch}}$,
so that errors in the Clifford operations are sufficiently suppressed.
Here $N_{\text{patch}}$ is the total number of code patches in our layout.
We note that this requirement is slightly pessimistic because some of the ancilla patches are not activated during each gate operation.
Following prior studies~\cite{Toshio2024,Babbush2018qubitization}, in this study, we will determine the code distance $d$ to satisfy
\begin{equation}
\label{eq:code distance condition}
    p_L(p_{\text{ph}},d)^{-1}\  \geq \ 100 \times dN_{\text{gate}}C_{\text{smm}}N_{\text{patch}}.
\end{equation}
Here we introduce the factor of $100$ to ensure that logical errors do not significantly impact our estimate of the total execution time. Introducing this factor ensures that the total Clifford errors in the entire circuit become less than 1\%. These residual Clifford errors could be mitigated with almost negligible overhead using the probabilistic error cancellation technique~\cite{Suzuki2022, Huggins2025FLASQ}.

Finally, we specify the value of $N_{\text{patch}}$ in the patch layout for STAR ver.~3.
As shown in Ref.~\cite{Litinski2019}, the fast block layout requires $2N_L+\sqrt{8N_L}+1$ patches, which includes $N_L$ data qubits and ancillary space for lattice surgery. Furthermore, we add 10 additional patches allocated for non-Clifford state preparation to the fast block layout, ensuring a sufficient supply rate of magic and resource states for STAR-magic mutation under $t_m = 10$ [clocks]. In conclusion, our layout requires $N_{\text{patch}}=2N_L+\sqrt{8N_L}+11$. By applying this formula to Eq.~\eqref{eq:code distance condition}, we can determine the code distance $d$ for each target system.

In Fig.~\ref{fig:TE-PAI for molecules}, we present theoretical estimates of the space-time costs of TE-PAI simulations on STAR ver.~3.
We consider model molecules with 72 spin orbitals and an $L_1$-norm $\lambda\in[10,300]$, enabling comparison with the $[\text{4Fe-4S}]$ cluster~\cite{Lee2023evaluating}.
Here, we set the parameters as $T\in[1, 50]$, $p_{\text{ph}}=10^{-3}$, $Q=1$, and $\epsilon=0.05$ (see also Appendix.~\ref{appendix:Hubbard} for the results of the Hubbard model). 
The Hilbert space of these molecules far exceeds what is exactly tractable with classical computation, necessitating classical heuristics such as the density matrix renormalisation group (DMRG)~\cite{Baiardi2020review}. These methods generally lack rigorous error guarantees; in this respect, TE-PAI simulations on quantum computers offer a clear advantage over classical approaches.

Fig.~\ref{fig:TE-PAI for molecules} (a) indicates that real-time dynamics simulations for large molecular systems, such as enzyme active centers like [4Fe-4S], become feasible on STAR ver.~3 with a superconducting quantum processor unit (QPU) comprising from $1.0$ to $2.6\times10^5$ physical qubits.
This contrasts with prior FTQC studies~\cite{Low2025amplification, Lee2021}, which typically require a larger-scale QPU with several millions of physical qubits, which are largely consumed by spatial overheads from MSD and ancilla logical qubits for block encoding.
Notably, a recent technical review of superconducting quantum devices~\cite{Mohseni2025} cautions that, once the target scale exceeds a few hundred thousand physical qubits, it may no longer be feasible to accommodate the full quantum device within a single dilution refrigerator. At that scale, distributed quantum computing---enabled by quantum interconnects between dilution refrigerators---may become unavoidable, thereby requiring major advances in quantum networking technologies.
Accordingly, our achievement in reducing the required qubit count is particularly valuable: it may help avoid a severe scaling bottleneck and bring forward the practical realization of early-FTQC devices.

However, Fig.\ref{fig:TE-PAI for molecules} (c) suggests that to keep the total runtime within a week, the value of $\lambda T$ must be lower than around $1.5\times10^3$, which corresponds to $T \lesssim 10$ [a.u.] for the $[\text{4Fe-4S}]$ cluster ($\lambda=137.8$ [Hartree]).
This substantial time overhead may raise a serious doubt about whether STAR ver.~3 can handle large-scale computational tasks required for practical quantum advantages.
Here we emphasize that the resource analyses presented in this work are preliminary and based on a simplified approach, not yet fully optimized with respect to specific quantum tasks or compilation strategies. We anticipate that significant reductions in time overhead and improvements in the tractable problem size could be achieved by leveraging the high parallelism of rotation gates or more advanced quantum algorithms that exhibit an optimal scaling like the Heisenberg limit.
In fact, our parallel work by Kanasugi {\it et al}.~\cite{Kanasugi2026} has demonstrated that ground-state energy estimation for biologically-relevant molecules like [4Fe-4S] can be executed on STAR ver.~3 within a realistic runtime by using a partially randomized quantum phase estimation~\cite{Gunther2025partiallyrandomized} coupled with a specialized Hamiltonian optimization technique, assuming parallel sampling with a few QPUs.

Another quantum task relevant to our analyses is real-time simulation of quantum non-adiabatic dynamics, such as vibronic effects~\cite{Motlagh2025,Loaiza2025spectroscopy} or laser-induced ultrafast charge migration~\cite{Schriber2019, Frahm2019, Chan2023grid} in photochemistry. In these situations, time-evolved quantum states are far from equilibrium state and relevant to broader Hilbert subspaces including high-energy eigenstates, making its simulation more classically challenging than ground-state energy estimation.
Nonetheless, given that powerful classical methods exist for simulating real-time dynamics, such as the time-dependent DMRG~\cite{Baiardi2020review}, claiming a quantum advantage for these tasks would require a more detailed comparative analysis in the future.

\section{Conclusion}

In this work, we have introduced STAR-magic mutation, a novel scheme for implementing arbitrary analog rotation gates with an effective error rate of $\mathcal{O}(\theta_L^{2(1-\Theta(1/d))}p_{\text{ph}})$, requiring ancillary space for only a single surface code patch. 
This scheme is based on an adaptive switching between the TMR protocol and MSC-based $T$-gate synthesis during the RUS process for gate teleportation. Specifically, we utilize the TMR protocol for $\theta_{\text{RUS}}<\theta_{\text{th}}$ and otherwise, a standard $T$-gate synthesis by using magic states generated via MSC or MSD.

Furthermore, we have proposed a partially fault-tolerant quantum computing architecture called STAR ver.~3. This architecture decomposes an input circuit into a gate sequence comprising Clifford+$T$+$\phi$ gate set, enabling efficient and parallelizable execution of analog rotation gates with STAR-magic mutation and digital rotation gates with MSC. We have clarified the theoretical bound of feasible circuit size on our architecture, which is determined by error mitigation cost and described with a L1-norm of the target Hamiltonian. Our analysis of the spacetime costs for simulating quantum many-body dynamics of lattice models and molecules using the TE-PAI algorithm demonstrates that our early-FTQC architecture can simulate the dynamics of various many-body Hamiltonians of a classically intractable size without approximation, assuming a realistic physical error rate of $p=10^{-3}$. This represents a significant improvement over previous studies on STAR architecture~\cite{Akahoshi2023,Toshio2024, Akahoshi2024}, which required an optimistic value of physical error rate, $p=10^{-4}$, to solve similar tasks.

To close this paper, we summarize the important issues that have not been addressed in this study. These could provide interesting avenues for future research. 
(i) Firstly, to minimize the circuit runtime on STAR ver.~3, it is desirable to develop an optimal compilation technique that fully leverages the locality and parallelism of STAR-magic mutation.
In Sec.~\ref{sec:estimation}, we only focus on the spacetime cost of many-body dynamics simulations under the sequential Pauli-based computation in Ref.~\cite{Litinski2019}. However, as discussed in Ref.~\cite{Akahoshi2024}, our architecture will unlock its full potential when executing quantum circuits comprising parallelizable rotation gates, such as the Trotter circuits.
(ii) Secondly, integrating several recent proposals~\cite{Zhang2025transversal, Zeng2025star} into STAR ver.~3 is a promising avenue for future research. 
Ref.~\cite{Zhang2025transversal} has proposed the {\it multi-level transversal injection} method, which recursively performs the TMR protocol on higher-level concatenated QEC codes by using the magic state pumping technique. This method enables reducing the fidelity of the prepared resource state to an arbitrary high value by increasing the level of injections. Meanwhile, Ref.~\cite{Zeng2025star} has suggested that the infidelity of the TMR protocol can be reduced from $\mathcal{O}(\theta_L^{2(1-1/k)}p_{\text{ph}})$ to $\mathcal{O}(\theta_L^{2(1-1/k)}p_{\text{ph}}^2)$ by leveraging some noise structures in realistic hardware setups. These techniques can be integrated with our proposals to overcome the theoretical bound on the feasible circuit size on STAR ver.~3.
(iii) Thirdly, our error mitigation strategies require the development of efficient tomography techniques for prepared resource or magic states. In particular, the PCEC protocol introduced in Sec.~\ref{sec:higher-order} employs the information on the error parameters $\{ \bar{q}_j \}_{j=1,2,\cdots,k}$ to construct a probabilistic inverse rotation channel.
After that, we also need the information on the leading residual error of $2\bar{q}_1 \sin^2(\Delta_1)$ to perform the usual PEC protocol. If a circuit-level error model is specified, these parameters can be efficiently determined via Clifford circuit simulations in Ref.~\cite{Choi2023,Toshio2024}. However, in a realistic setup, we do not know the exact error model of quantum devices.
Developing efficient experimental methods to evaluate these parameters is crucial for realizing the STAR ver.~3 on real hardware devices.
(iv) Finally, it is also crucial to consider the impact of control errors on the performance of our scheme and develop more refined error mitigation strategies for them. As discussed in Sec.~\ref{sec:control errors}, the STAR architecture is fragile to a specific type of control errors. To overcome this limitation, we need to develop an efficient method for real-time calibration of logical-level rotation angles.

\section{Acknowledgement}

We are grateful to thank Yutaka Hirano, Norifumi Matsumoto, Shinichiro Yamano, and Yutaro Akahoshi for fruitful discussions. K. F. is supported by MEXT Quantum Leap Flagship Program (MEXT Q-LEAP)
Grant No. JPMXS0120319794, JST COI-NEXT Grant No. JPMJPF2014, and JST
Moonshot R\&D Grant No. JPMJMS2061.

\section{Author contributions}

R. T. proposed the concepts of STAR-magic mutation and STAR ver.~3, formulated the theoretical details, performed the numerical simulations for all figures, analyzed the data, and wrote the original draft of this paper. 
S. K. contributed to the technical discussion on the application of STAR ver.~3 in Sec.~\ref{sec:practical applications} and provided the numerical data in Table.~\ref{tab: L1-norm} by implementing the BLISS and orbital optimization. 
J. F., H. O., and S. S. were instrumental in supervising the project, securing necessary resources, and overseeing project administration.
K. F. provided technical supervision for this work and contributed to the conceptualization and the interpretation of the numerical results.
All authors participated in the review of the manuscript.

\appendix

\begin{figure*}
    \centering
  \begin{tabular}{lll}
{\normalsize (a) Number of physical qubits} & {\normalsize (b) Single-shot runtime [sec]} & {\normalsize (c) Total runtime [day]}  \\
   \includegraphics[width=5.8cm]{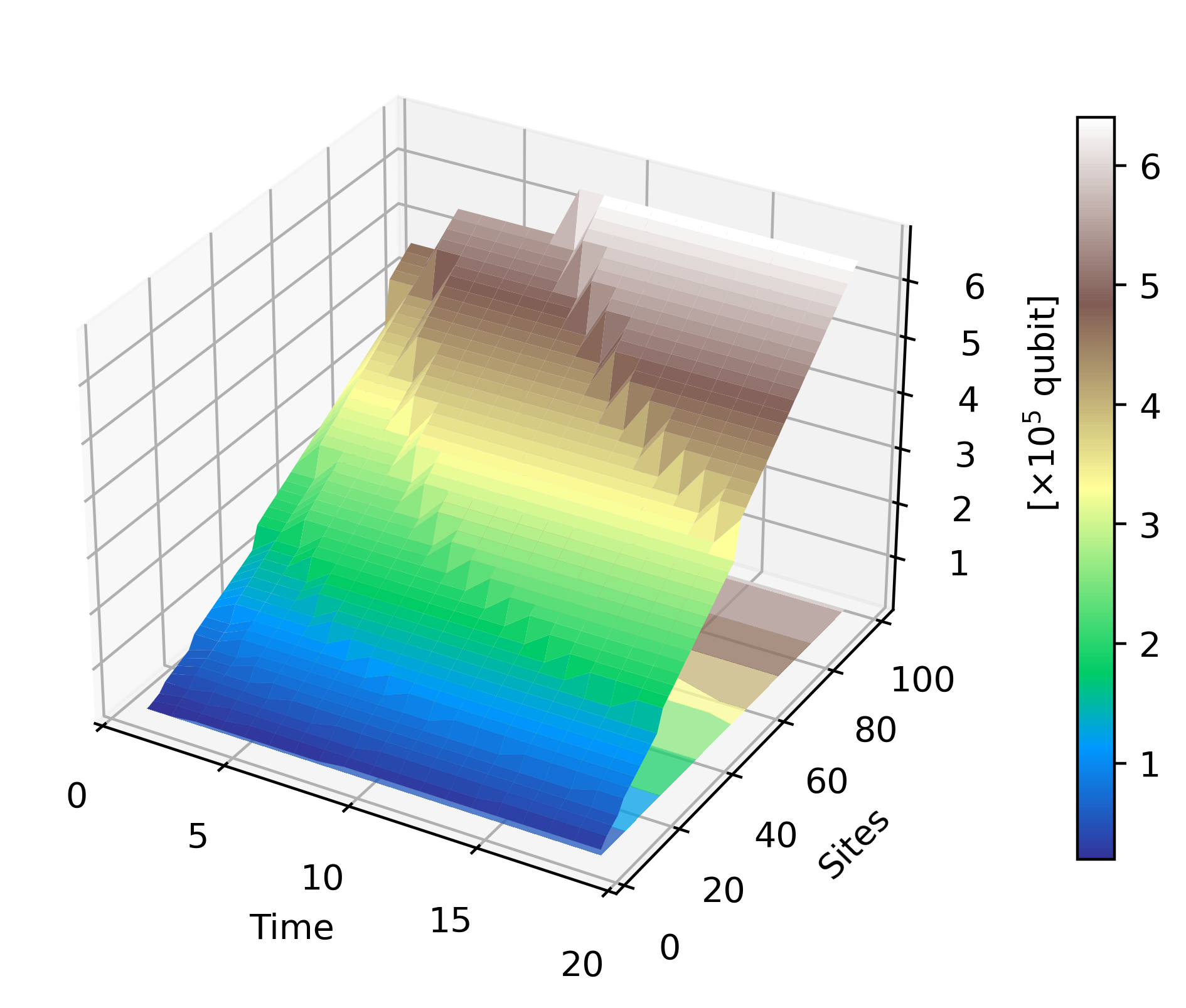}&
   \includegraphics[width=5.8cm]{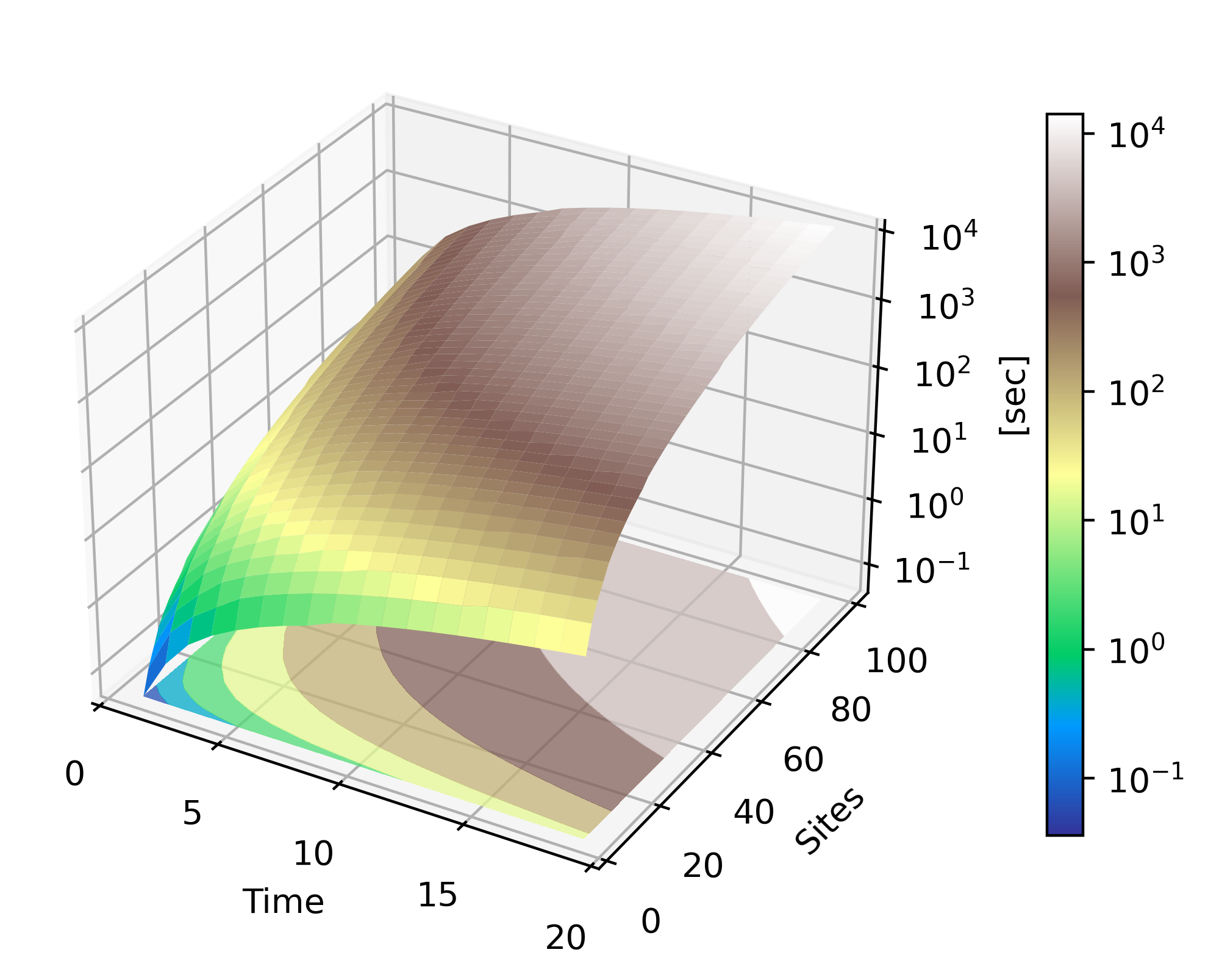}&
   \includegraphics[width=5.8cm]{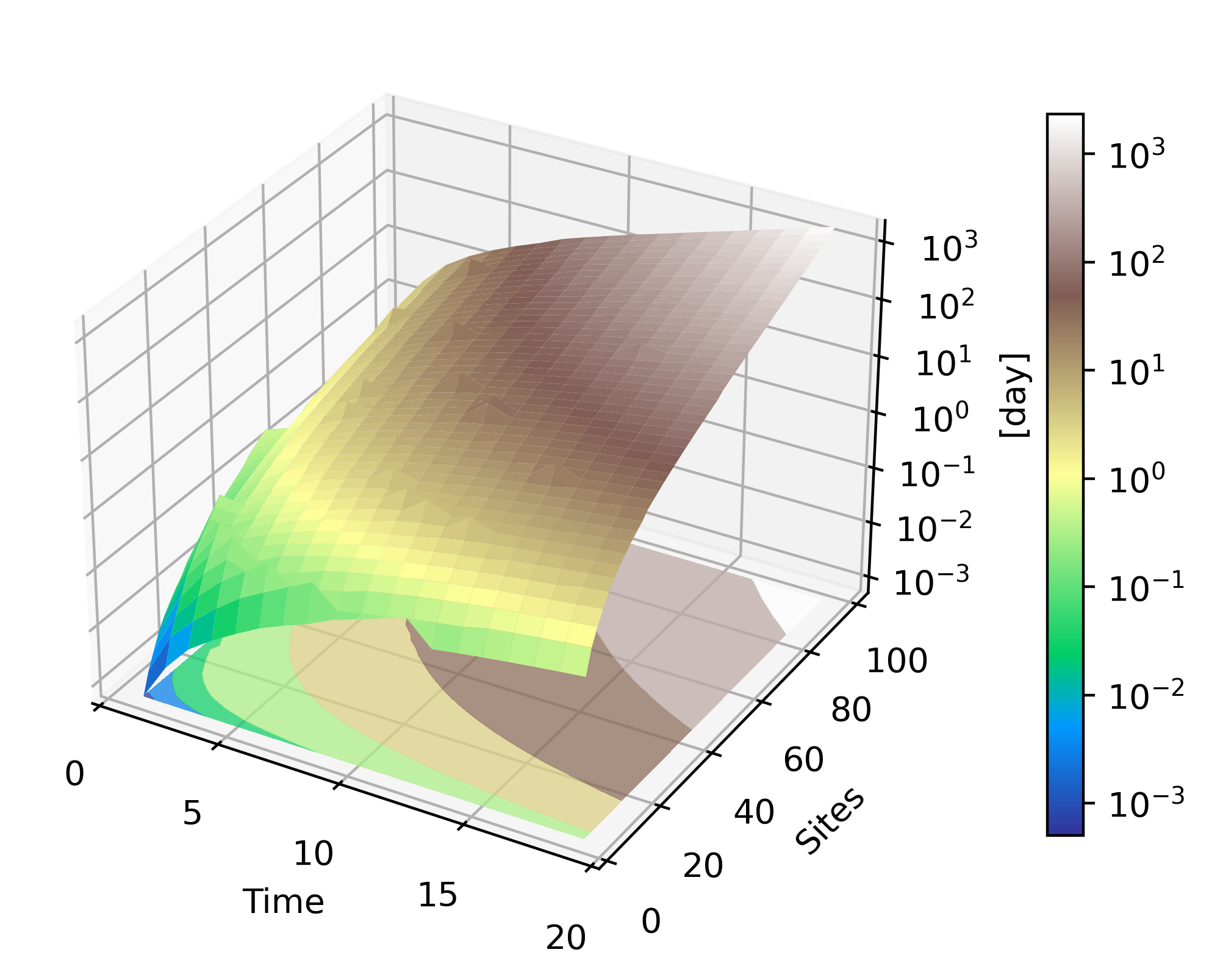}\\
  \end{tabular}
    \caption{Spacetime costs of TE-PAI simulation on STAR ver.~3 for the $N$-sites 2D Hubbard model ($t=1$, $U=4$): (a) Number of Physical qubits, (b) single-shot runtime of TE-PAI circuit, and (c) total runtime including the sampling and mitigation overhead for achieving a specific accuracy $\epsilon=0.05$ in estimating time-evolved expectation values. In these plots, we assume $p_{\text{ph}}=10^{-3}$ and analyze the dynamics simulation up to $T\in [1, 20]$. The number of sites $N$ is varied within a range of $[4,100]$.}
    \label{fig:TE-PAI for hubbard}
\end{figure*}

\section{The spacetime costs of TE-PAI simulations of the 2D Hubbard model}
\label{appendix:Hubbard}

This section presents supplementary data on the spacetime cost of TE-PAI dynamics simulations for the 2D Hubbard model. The Hamiltonian of the 2D Hubbard model is defined as 
\begin{equation}
\label{eq:Usual Hubbard Hamiltonian}
    \hat{\mathcal{H}} = -t \sum_{\langle i,j \rangle, \sigma} (\hat{c}^{\dag}_{i, \sigma}\hat{c}_{j, \sigma} + {\rm h.c.}) + U \sum_{i} \hat{n}_{i, \uparrow} \hat{n}_{i, \downarrow}, 
\end{equation}
where $\hat{c}^{(\dag)}_{i, \sigma} \ (\sigma = \uparrow, \downarrow)$ are fermionic creation (annihilation) operators, and $\hat{n}_{i, \sigma} = \hat{c}^{\dag}_{i, \sigma} \hat{c}_{i, \sigma}$ is the number operator for the fermionic mode. 
The notation $\langle i,j \rangle$ indicates any pair of adjacent sites on the 2D square lattice. 
The parameters $t$ and $U$ denote the strength of the hopping and on-site Coulomb interaction, respectively.

By performing the Jordan-Wigner transformation and constant energy shift, we can represent the 2D Hubbard model in the form of a linear combination of Pauli string operators as follows:
\begin{equation}
\label{eq:Spin Hamitonian}
\begin{aligned}
      \hat{\mathcal{H}} &= -\frac{t}{2} \sum_{\langle i,j \rangle,\sigma} (\hat{X}_{i,\sigma}\hat{Z}^{\leftrightarrow}_{i,j,\sigma} \hat{X}_{j,\sigma} + \hat{Y}_{i,\sigma} \hat{Z}^{\leftrightarrow}_{i,j,\sigma}\hat{Y}_{j,\sigma})\\
      &\qquad\qquad\qquad\qquad+ \frac{U}{4} \sum_{i} \hat{Z}_{i,\uparrow}\hat{Z}_{i,\downarrow},  
\end{aligned}
\end{equation}
where $\hat{Z}^{\leftrightarrow}_{i,j,\sigma} = \prod_{k = i+1}^{j-1} \hat{Z}_{k,\sigma}$
is the so-called Jordan-Wigner string, which is needed to preserve the appropriate commutation relations between fermionic creation and
annihilation operators.

In Fig.~\ref{fig:TE-PAI for hubbard}, we show the spacetime costs of TE-PAI dynamics simulations for the $N$-sites 2D Hubbard model, assuming the parameters $t=1$, $U=4$ $T\in[1, 20]$, $p_{\text{ph}}=10^{-3}$, $Q=1$, and $\epsilon=0.05$. 
The number of sites $N$ is varied within the range of $[4, 100]$, and the number of logical qubits is given as $N_L=2N$.
These plots indicate that we can simulate the real-time dynamics of the 2D Hubbard model with $N\leq100$ using less than $6\times10^5$ physical qubits, but the total runtime exceeds a weak for $NT\gtrsim 300$.

To alleviate this significant overhead, we must seek a more refined approach beyond TE-PAI.
For simple periodic models such as the Hubbard model, we can exploit the high parallelism of the Trotter circuit. In fact, several efficient Trotter-based approaches have already been proposed in Refs.~\cite{Kivlichan2020improved, Campbell2021early}. Notably, Ref.~\cite{Akahoshi2024} has shown that the simulation cost for the 2D Hubbard model on STAR ver.~2 can be significantly reduced by utilizing the fermionic swap gate and the parallel compilation of analog rotation gates.
Therefore, we believe that we can further reduce the spacetime costs on STAR ver.~3 for simulating lattice models by exploring more sophisticated analyses similar to that in Ref.~\cite{Akahoshi2024}.

\clearpage

\bibliography{citation}

\begin{thebibliography}{121}
\bibitem [{com()}]{comment}%
  \BibitemOpen
  \href@noop {} {}\bibinfo {note} {Please note that, throughout this paper, we use a notation of rotation gates that is slightly different from the one used commonly, i.e., $\hat{R}_{z,i}'(\theta)=e^{-i\theta\hat{Z}_i/2}$ to make simpler our formula obtained below.}\BibitemShut {Stop}%
\end{thebibliography}

\end{document}